\pdfoutput=1

\documentclass{jpp} 

\usepackage[UKenglish]{babel}
\usepackage[utf8]{inputenc}
\usepackage[T1]{fontenc}

\usepackage{amsmath}
\usepackage{bm}
\usepackage{mathrsfs}
\usepackage{amssymb}

\usepackage{graphicx}
\usepackage{svg}

\usepackage{multirow}

\usepackage{setspace} 

\usepackage{enumitem}

\usepackage{authblk}

\usepackage[colorlinks=true, allcolors=blue]{hyperref}
\usepackage{natbib}
\usepackage{cleveref}

\crefname{equation}{}{}
\crefname{figure}{Figure}{Figures}
\crefname{table}{Table}{Tables}
\crefname{section}{Section}{Sections}
\crefname{appendix}{Appendix}{Appendices}

\Crefname{equation}{}{}
\Crefname{figure}{Figures}{Figures}
\Crefname{table}{Table}{Tables}
\Crefname{section}{Section}{Sections}
\Crefname{appendix}{Appendix}{Appendices}

\newcommand{\quotemarks}[1]{``#1''}

\def\examplepulse{\ensuremath{\#90339}}

\newcommand{\addcomma}[1]{
  \ifx\relax#1\relax {}
  \else {#1,}
  \fi}

\newcommand{\psipos}[1]{\ensuremath{\psi_{\addcomma{#1}\mathrm{Pos}}}}
\newcommand{\psitop}[1]{\ensuremath{\psi_{\addcomma{#1}\mathrm{Top}}}}
\newcommand{\psiwid}[1]{\ensuremath{w_{\addcomma{#1}\mathrm{Ped}}}}
\newcommand{\coreslope}[1]{\ensuremath{s_{{\addcomma{#1}}\mathrm{Core}}}}
\newcommand{\pedheight}[1]{\ensuremath{h_{{\addcomma{#1}}\mathrm{Ped}}}}

\def\Tesep {\ensuremath{T_{e}^{\mathrm{Sep}}}}
\def\Nesep {\ensuremath{n_{e}^{\mathrm{Sep}}}}
\def\Tetop {\ensuremath{T_{e}^{\mathrm{Ped}}}}
\def\TetopNe{\ensuremath{n_{e}\left(\psitop{T_e}\right)}}
\def\Netop {\ensuremath{n_{e}^{\mathrm{Ped}}}}
\def\NetopTe{\ensuremath{T_{e}\left(\psitop{n_e}\right)}}
\def\psisteep{\ensuremath{\psi_{\mathrm{Steep}}}}
\def\psisep{\ensuremath{\psi_{\mathrm{Sep}}}}

\def\Rsep{\ensuremath{R^\mathrm{Sep}}}

\def\psirenorm{\ensuremath{X_{\mathrm{Ren}}}}

\def\Iplasma{\ensuremath {I_{\mathrm{Plasma}}}}
\def\Psep{\ensuremath {P_{\mathrm{Sep}}}}
\def\vthe{\ensuremath {v_{\mathrm{th},e}}}
\def\Qegb{\ensuremath {Q_{e,\mathrm{gB}}}}
\def\Qeturb{\ensuremath {Q_{e,\mathrm{Turb}}}}
\def\fuelrate{\ensuremath{\Gamma_{D}}}
\def\betapol{\ensuremath{\beta_\theta}}

\def\fgw{\ensuremath{f_{\mathrm{Gw}}}}
\def\ngw{\ensuremath{n_{\mathrm{Gw}}}}

\def\etacr{\ensuremath{\eta_{e,\mathrm{crit}}}}
\def\fitcoeff{\ensuremath{\etacr}}

\newif\ifvisiblechanges

\visiblechangestrue

\newcommand{\der}[2]{\ensuremath{\frac{\mathrm{d}#1}{\mathrm{d}#2}}}
\newcommand{\dertxt}[2]{\ensuremath{\mathrm{d}#1/\mathrm{d}#2}}

\def\LNe{\ensuremath{L_{n_e}}}
\def\LTe{\ensuremath{L_{T_e}}}
\def\RLNe{\ensuremath{R/\LNe}}
\def\RLTe{\ensuremath{R/\LTe}}

\def\fit {\texttt{fit} }
\def\raw {\texttt{raw} }
\def\mtanh {\texttt{mtanh} }

\newlist{predictions}{itemize}{1}
\setlist[predictions]{
	leftmargin=1cm,
}

\newlist{concl}{itemize}{1}
\setlist[concl]{
	leftmargin=0.375cm,
}

\shorttitle{Electron-temperature reconstructions for JET-ILW pedestals}
\shortauthor{L.-P. Turica et al.}

\title{Reconstructions of electron-temperature profiles from EUROfusion Pedestal Database using turbulence models and machine learning}
\author[1,2,3]{L.-P. Turica\footnote{\email{leonard-petru.turica@physics.ox.ac.uk}}}
\author[2]{A. R. Field}
\author[4]{L. Frassinetti}
\author[1,5]{A. A. Schekochihin}
\author[a]{\mbox{JET Contributors}}
\author[b]{the EUROfusion Tokamak Exploitation Team}

\affil[1]{\small Rudolf Peierls Centre for Theoretical Physics, University of Oxford, Oxford, OX13PU, UK}
\affil[2]{United Kingdom Atomic Energy Authority, Culham Campus, Abingdon, OX14 3DB, UK} 
\affil[3]{University College, Oxford, OX1 4BH, UK}
\affil[4]{KTH Royal Institute of Technology, Stockholm SE-100 44, Sweden}
\affil[5]{Merton College, Oxford, OX1 4JD, UK}
\affil[a]{see the author list of \citet{JETTeam_Mailloux_2022_full}} 
\affil[b]{see the author list of \citet{EUROFusion_Joffrin_2024_full}} 

\begin{document}
	\begin{titlepage}
		\maketitle
		
		\normalsize
		
		\pagenumbering{gobble}
		\begin{abstract}

			This study makes use of plasma-profile data from the EUROfusion pedestal database \citep{Frassinetti_Database}, focusing on the electron-temperature and electron-density profiles in the edge region of H-mode ELMy JET ITER-Like-Wall (ILW) pulses. We make systematic predictions of the electron-temperature pedestal, taking engineering parameters of the plasma pulses and the density profiles as inputs.
			We first present a machine-learning algorithm which, given more inputs than theory-based modelling and $80\%$ of the database as training data, is able to reconstruct the remaining $20\%$ of electron-temperature profiles within $20\%$ of the experimental values, giving accurate estimates of the temperature-pedestal widths and locations. We find that the most consequential engineering parameters for such predictions are the magnetic field strength, the particle fuelling rate, the plasma current, and the strike-point configuration. This result confirms the conceptual possibility of accurate prediction via models that rely on large databases. 
			Next, taking a simple theoretical approach that assumes a definite local relationship between the electron-density ($R/L_{n_e}$) and electron-temperature ($R/L_{T_e}$) gradients, we find that a range of power-law scalings $R/L_{T_e}=A\left(R/L_{n_e}\right)^\alpha$ with $\alpha\approx 0.4$ correctly capture the behaviour of the electron-temperature in the steep-gradient region. When the constants $A$ and $\alpha$ are fit independently for each pedestal, we obtain a one-to-one correlation between $A$ and $\alpha$, which also holds for JET-C pedestals. For $\alpha = 1$, $A \equiv \eta_e=L_{n_e}/L_{T_e}$, which is a parameter known to govern the turbulence saturation in the standard picture of slab-ETG modes. Determining the measured $\eta_e$ across the region between the density-pedestal top and the separatrix yields a distribution of values that lie considerably above the known slab-ETG linear stability threshold, implying either a non-linear threshold shift or a measurably supercritical saturated turbulent state.
			Simulations of gyrokinetic turbulence in steep-gradient regions of the pedestals of JET and other tokamaks suggest that simple scalings exists between the pedestal turbulent heat flux and the local values of the gradient scale-lengths $R/L_{T_e}$ and $R/L_{n_e}$. We provide parameters for these scalings that best reconstruct JET-ILW pedestals. The inclusion of more experimental parameters is necessary for such heat-flux models to match the accuracy of our machine-learning results.
			
		\end{abstract}
	
	\ifvisiblechanges

		
		\tableofcontents
		
	\fi

	\end{titlepage}
	
	\clearpage

\section{Introduction} \label{sec:intro}
\setcounter{page}{1}
\pagenumbering{arabic}

The operation of tokamaks with a divertor can result in an edge-transport barrier (ETB), which occurs once the heating power exceeds a certain threshold. The appearance of this ETB marks the onset of the high-confinement mode [the H-mode \citep{Wagner_Hmode_1982} -- as opposed to a lack of an ETB in the L-mode \citep{Solano_LH_2022}]. This barrier results in enhanced confinement and correspondingly higher plasma temperatures and densities in the core region, which are desirable for the optimisation of fusion power. As a result, it is the H-mode that is also planned to be achieved in future tokamaks such as ITER.

The ETB is located just inside the edge of the confined plasma, which is identified by the last closed flux surface (LCFS) bounded by the separatrix. The ETB is characterised by high temperature and density gradients, so the plasma profiles appear to be \quotemarks{raised up} upon a pedestal. The pedestal exhibits a stability limit, which constrains the pressure gradient (via unstable ballooning modes) and the bootstrap current (via unstable peeling modes) \citep{Connor_Wilson_PeelBalloon}. When this peeling-ballooning threshold is crossed, the equilibrium becomes subject to magnetohydrodynamic (MHD) instabilities that are thought to trigger Edge-Localised Modes (ELMs). Peeling and ballooning instabilities extend radially on scales larger than the typical pedestal width \citep{Snyder_PeelBalloon_2004}, and therefore the stability of these modes imposes a \quotemarks{global} condition on the pedestal's width and height. Such stability conditions are used by the semi-empirical EPED model \citep{Snyder_EPED_2009} in order to predict the pedestal width and height in fusion devices.

Alongside the MHD limitations on the plasma equilibrium, the pedestal profile is also believed to be determined by the transport caused by turbulence saturated in these steep-gradient regions. We assume that the action of this turbulence is local, i.e., that the gradients of the temperature and density are sources free energy that linearly destabilise turbulent modes on micro-scales (much smaller than the extent of the pedestal), and that the non-linear saturation of these unstable modes determines the effective transport at each location in the pedestal.

The exact relationship between the transport properties of such turbulence and the underlying equilibrium is still largely unknown. Here, we approach this question using data-driven methods, focusing on the pedestal structure. Previous studies have found that a distinctive feature of pedestals in several machines is a general radial shift between the density-pedestal location and the  temperature-pedestal location: the steep-temperature-gradient region occurs further inside the last closed flux surface (LCFS) than the steep-density-gradient region. This has been found on JET, ASDEX, and DIII-D by \citet{Frassinetti_Database, Frassinetti_2021,Stefanikova_PedShift,Dunne_PedShift_ASDEX,Wang_PedShift_D3D}. Of these works, \citet{Frassinetti_Database, Frassinetti_2021} used the EUROfusion database, which contains the pedestal characteristics (position, height, width) of over 2000 JET-C and JET-ILW Type-I ELMy H-mode pulses, alongside their engineering and magnetic-equilibrium parameters. The pedestal profile data are obtained by compiling high-resolution Thompson-scattering (HRTS) measurements of density and temperature pedestals, which are in near-steady state. Because of the slow evolution timescale of the plasma profiles compared to that of turbulent fluctuations, the transport properties of the pedestal plasma ought to be set \quotemarks{quasistatically} by a saturated state of the local turbulence.

Here we will make use of the pedestal profile data for pulses in the EUROfusion database in order to identify correlations between the electron-temperature ($T_e$) and electron-density ($n_e$) profiles of $1251$ JET-ILW pedestals and the corresponding pulse parameters. These pedestal profiles constitute information with which it should be possible to characterise the nature of edge transport in JET-ILW and the effect of this transport on confinement. In order to do this, we will reconstruct electron-temperature profiles using machine learning (ML) or a number of \quotemarks{physics-based} turbulence models, and we will use the quality of these reconstructions to determine the accuracy of each method.

\Cref{sec:geometry} provides an overview of the pedestal profiles and the functional fits that represent these profiles. Formal definitions of the pedestal height, width, and position are given in \cref{ssec:mtanh}; we identify how these values correlate with the $T_e$ and $n_e$ profiles and the corresponding gradients in \cref{ssec:exampleGradients}. In \cref{ssec:locations}, we define reference positions at characteristic locations across the pedestal (the locations of the density-pedestal top, temperature-pedestal top, steep-gradient point, and separatrix); in \cref{ssec:psiRenorm}, we describe a method to align these locations across the database. \Cref{sec:database} explains our usage of the contents of the EUROfusion database. We justify the choice of the $1251$-pulse subset of the database used in our analysis in \cref{ssec:dataSelection}, and then outline the database parameters and their values in \cref{ssec:params}. \Cref{appendix:parameters} defines these parameters formally.

Our first method of reconstructing pedestal $T_e$ profiles, presented in \cref{sec:neuralnets}, uses an ML approach. This is a natural step to take in order to use non-linear correlations in the multidimensional parameter-space of the EUROfusion Pedestal Database. We will conclude that it is indeed possible to infer electron-temperature profiles statistically, taking the electron-density profiles and various pulse parameters as inputs. The ML method fully exploits the information in the pedestal database. Therefore, it gives us an effective upper bound for the accuracy of such a pedestal prediction in our database using ML, physics-based models, or otherwise.

We build two different families of models, which generate \quotemarks{global} and \quotemarks{local} predictions. Global predictions, presented in \cref{ssec:mlFullProf}, reconstruct the full temperature-pedestal profile using the knowledge of the full density-pedestal profile and different groups of database parameters. Local models, presented in \cref{ssec:mlLocalVal}, assemble the temperature profile from predictions of the local electron temperature, using the corresponding local electron density and various database parameters. The local approach allows us to mimic a local-transport model and compare the outcome to the global predictions of \cref{ssec:mlFullProf}. In \cref{ssec:mlImportantPars}, we scan systematically through all combinations of database parameters and identify those that result in most accurate predictions. This exhaustive approach of training a different network for each parameter combination reveals the most important parameters for a successful local or global prediction. \cref{appendix:neuralnets} describes the numerical methods employed for out ML models.

As an alternative to ML, we turn to physics-based turbulence models to reproduce the electron-temperature pedestal. \Cref{sec:gradients} depicts the loci of normalised gradients of the electron density ($\RLNe$) and electron temperature ($\RLTe$) over the pedestal region. With this information, we discuss the relevance of linear instabilities and sources of turbulence in this physical regime. The plasma gradients are found to lie in a regime unstable to electron-temperature-gradient (ETG) modes. In particular, the steep-gradient pedestal turbulence is thought to be dominated by slab-ETG modes \citep{Field_Stiff}, which determine the nature of local turbulence and corresponding transport, as we will discuss in \cref{ssec:gradientLitReview}.

In \cref{sec:etae}, we investigate the experimental distribution of $\eta_e \equiv \LNe/\LTe$, a parameter that governs the linear instability of slab-ETG modes. In \cref{ssec:etaDist}, we find that in steep-gradient region of the pedestal, the mean value of this parameter is $\approx 2$. This mean value is consistent with past findings, however, we find a considerable spread in the values of $\eta_e$. 
In \cref{ssec:etaPredictions}, by assuming that the turbulence in the pedestal region can be described as \quotemarks{pinned} to a nominal value of $\eta_e$, we reconstruct electron-temperature profiles by numerically inverting and integrating the relationship between the temperature and density gradients, with the density gradient as an input (as described in \cref{appendix:predictionMethods}). Doing this by minimising the discrepancy between the reconstructed and the experimental profile in the steep-gradient region yields an \quotemarks{effective} $\eta_e = 1.9$ across the database. If, instead, the nominal $\eta_e$ is allowed to vary between pulses, then this optimal pulse-dependent  $\eta_e$ agrees well with the measured local $\eta_e$ in the steep-gradient region of the pedestal.

In \cref{sec:alpha}, we complicate the model by using the experimental distribution of $\RLNe$ and $\RLTe$ and an assumed relation between them of the form $\RLTe = A (\RLNe)^{\alpha}$ to reconstruct the $T_e$ pedestal profiles, with $A$ and $\alpha$ being fitting constants. \Cref{ssec:alphaDist} highlights that the exponent $\alpha$ maintains values below $1.5$ across the pedestal when calculated locally.  Such an approach can indeed produce correct plasma profiles, since any value of $\alpha \neq 1$ allows the gradient ratio $\eta_e$ to adjust over the span of the pedestal. \Cref{ssec:alphaPredictions} identifies an effective value of $\alpha \approx 0.4$ required for an accurate full-database reproduction, however, most such reconstructions of the $T_e$ profiles do not produce a pedestal, instead growing exponentially towards the core and failing to reproduce the correct gradient inside the temperature-pedestal top. If $A$ and $\alpha$ are fit independently for each pulse, a clear relationship between $A$ and $\alpha$ emerges, meaning that identifying either of these parameters for each pulse is sufficient for excellent reconstructions of the pedestal profile up to the position of the electron-temperature top.

Having found in \cref{sec:neuralnets} that the power transported through the separatrix is a crucial parameter, we use the power balance in the pedestal and the knowledge of turbulent transport for more accurate reconstructions of $T_e$ profiles in \cref{sec:heatFluxModels}. In \cref{ssec:etgTurb}, we discuss recent results on modelling electron-channel transport in the pedestal by \citet{Guttenfelder_2021,Hatch_RedModels,Hatch_modeling_2024,Chapman-Oplopoiou_2022,Farcas_transportEq_2024}. The models of turbulent transport by \citet{Guttenfelder_2021} and \citet{Chapman-Oplopoiou_2022} represent a physical picture of ETG transport: the electron-channel turbulent heat flux is thought to increase with the difference between the actual gradient ratio $\eta_e$ and some non-linear critical value of it, $\etacr$ \citep{Field_Stiff}.{ More general models of heat transport, for which the physical picture is more uncertain, should still be considered because of the existence in the pedestal of ETG modes with low parallel wavenumbers (toroidal ETG) and other electron-scale instabilities such as micro-tearing modes (MTMs) \citep{Parisi_ETGdominance,Told_MTM, Leppin_Complex}. These other modes can alter the saturation of pedestal turbulence, and could therefore heavily affect the resulting heat fluxes.}
No comprehensive comparison between the predictions of various heat-flux scaling laws and experimental data has been carried out before on the EUROfusion database. We carry this out in \cref{ssec:heatFluxPredictions} by fitting the free parameters of the models proposed by \citet{Chapman-Oplopoiou_2022} and \citet{Hatch_RedModels}. We obtain models that yield far better pedestal reconstructions than those of \cref{sec:etae,sec:alpha}, however not better than the ML results of \cref{sec:neuralnets}, as evidenced in \cref{appendix:heatFluxFits}. We also discuss the differences in the analytical forms of the heat-flux models and the consequences these differences have on the reconstructed profiles. We find that more recent heat-flux models, which also account for the local properties of the local magnetic field \citep{Hatch_modeling_2024,Farcas_transportEq_2024}, comply with the requirements for good $T_e$ profile reconstructions, showing promise for a study which has access to magnetic-field equilibrium reconstructions.

We provide a summary and a discussion of our results in \cref{sec:conclusions}.

	\section{Pedestal Profiles} \label{sec:geometry}

For the purpose of this analysis, the electron temperature $T_e$ and density $n_e$ are measured for Type-I ELMy JET-ILW pulses using the high-resolution Thompson-scattering (HRTS) diagnostic \citep{Pasqualotto_HRTS}. The HRTS laser path crosses the magnetic flux surfaces on the device midplane, from the outboard side through the centre of the plasma core, and back out. The HRTS diagnostic uses a laser beam fired at $20\mathrm{Hz}$ throughout the duration of the pulse. The scattered light from the beam is measured at different locations by detectors \citep{Pasqualotto_HRTS}; this information is then used to determine the electron temperature and density at these locations \citep{Frassinetti_MeasurementMethods}. The profiles that occur within the last $20\%$ of each inter-ELM period are then selected manually. Only inter-ELM periods longer than $2\tau_{E}$ are used, where the profiles of temperature and density reach near-steady-state. Here, $\tau_{E} = W/P_{\mathrm{Loss}}$ is the energy-confinement time, with $W$ the plasma energy and $P_{\mathrm{Loss}}$ the total power loss. This data selection obtains profiles close to the peeling-ballooning stability threshold, which is reached many times over the duration of a pulse. Once this threshold is crossed, an ELM is triggered. More information regarding Type-I ELMs and their relevance for the pedestal profiles can be found in \cref{ssec:dataSelection}.

The profiles obtained from the HRTS diagnostic will be referred to as $\raw$ profiles (scatter points in \Cref{fig:Gradients}a,b). These $\raw$ profiles were used by \citet{Frassinetti_Database} to determine the deconvolved $\mtanh$ fits used in our analysis (which we introduce in \cref{ssec:mtanh}). These $\raw$ profiles will be used in our analysis to estimate the experimental confidence intervals of the data for the machine-learning methods described in \cref{sec:neuralnets}, avoiding the systematic biases of the $\mtanh$ fits.

\subsection{Radial Coordinates}\label{ssec:radialcoord}
The $\raw$ profiles and other calculated plasma profiles are cast as functions of the radial coordinate: either $R$ (the radial distance from the axis of JET-ILW) or $\psi_N$ (the poloidal-magnetic-flux label). The normalisation convention of $\psi_N$ and the radial shift applied to each pulse will be described below.

\begin{figure}
	\centering
	\includegraphics[width=\textwidth]{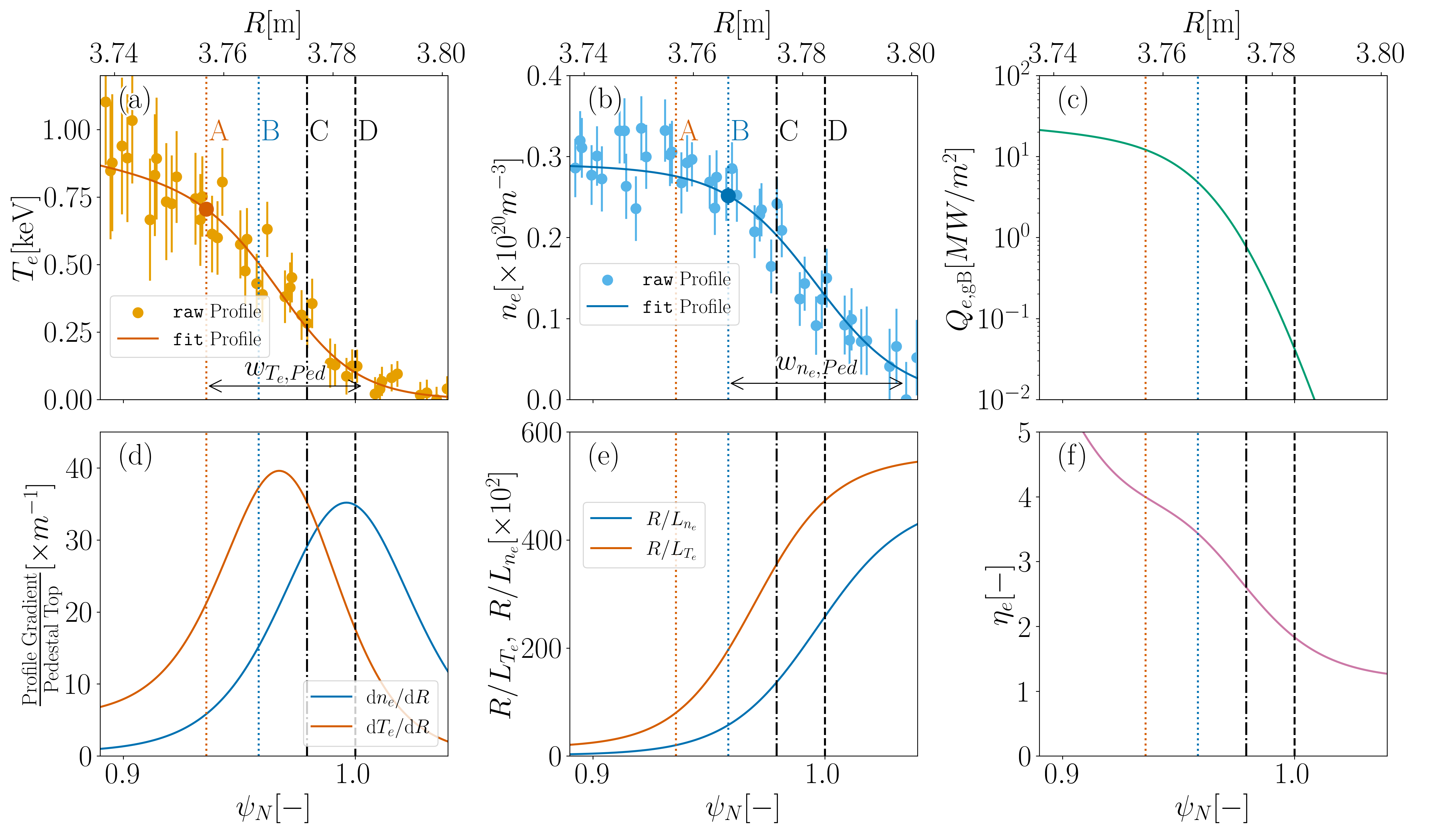}
	\caption{\centering { 
			The $\fit$ (lines) and $\raw$ (dots) profiles for pulse $\examplepulse$: 
			(a) the electron density $n_e$ and 
			(b) the electron temperature $T_e$, with the pedestal-top points highlighted; 
			(c) the gyro-Bohm heat flux $\Qegb$;
			(d) the normalised radial gradients of the profiles, $(\dertxt{T_e}{R} )/ \Tetop$ and $(\dertxt{n_e}{R} )/ \Netop$, which are the gradients of a normalised \mtanh fit with height $1$ at the pedestal top; 
			(e) the \RLTe and \RLNe  profiles; (f) the $\eta_e \equiv \LNe / \LTe$ profile. 
			The horizontal axes represent the normalised poloidal-magnetic-flux coordinate $\psi_N$ (lower) and the radial position $R$ (upper).
			These profiles are plotted alongside the four characteristic pedestal locations introduced in \cref{ssec:locations} (vertical lines): A is the $T_e$ top position $\psitop{T_e}$, B is the $n_e$ top position $\psitop{n_e}$, C is the steep-gradient point $\psisteep$, D is the separatrix position $\psisep$. The steep-gradient region is bounded by points B and D, and the exact middle of this region is marked by $\psisteep$ \mbox{(point C)}. We define the core as the locations inside of point A, and the SOL as those outside of D. The horizontal arrows in panels (a) and (b) represent the radial extent of the pedestal region for $T_e$ and $n_e$.
	}}
	\label{fig:Gradients}
\end{figure}

Due to uncertainties in the exact underlying equilibrium and in the location of the separatrix, the HRTS profiles are shifted radially to pin them at a specific prescribed value of the electron temperature at the separatrix, 
\begin{equation}
	T_e(\Rsep)\equiv \Tesep,
\end{equation} a relationship used to determine the location of the separatrix $\Rsep$. The same separatrix location applies to both the electron-temperature and density profiles. The value of $\Tesep$ is estimated to be $100\;\mathrm{eV}$ in JET-ILW using the two-point model for heat flux, as described in Chapter 5.2 of \citet{Stangeby_SOLbook_2000}. The two-point model is a simplified model that describes the transport of heat along the scrape-off layer (SOL) between a point on the divertor and a point in the outboard plasma midplane. In this context, the model treats the SOL as being in a dominantly conductive (as opposed to convective) regime, which is consistent with the usage of stationary inter-ELM plasma profiles. This model yields a scaling of the SOL heat flux $Q_e \propto T_e^{7/2}(\Rsep)$, which implies that $\Tesep$ is very insensitive to changes in $Q_e$. EDGE2D simulations consistently show that $\Tesep$ varies in the range $80-110\;\mathrm{eV}$ \citep{Simpson_Separatrix}. As shown by \citet{Frassinetti_Database}, this uncertainty of the value of $\Tesep$ has a negligible effect on the radial location of the separatrix as a result of the steepness of the profiles around $R = \Rsep$. This can also be generally inferred from the shapes of the electron-temperature profiles shown in \cref{fig:Gradients}(a).

The poloidal-magnetic-flux label $\psi_N$ is a normalised poloidal-magnetic-flux coordinate \nolinebreak $\psi(R)$. It is defined to be the poloidal magnetic flux subtended between the magnetic axis and the radial position $R$ and obtained by reconstructing the equilibrium magnetic field using the magnetohydrodynamic code HELENA \citep{Huysmans_HELENA_1999}. For this, HELENA is provided with the plasma profiles and the shape of the plasma boundary reconstructed using EFIT \citep{Lao_EFIT_1985}. Then $\psi_N$ is obtained by normalising $\psi(R)$ so that $\psi_N = 1.0$ at the radial location of the separatrix, viz., 
\begin{equation}	
	\psi_N = \frac{\psi(R)}{\psi(\Rsep)} .
\end{equation}

\subsection{\mtanh Fits} \label{ssec:mtanh}
The pedestal profiles that are the subject of our analysis are functions fit to the measured \raw profiles. The general form of the fits is
\begin{equation}\label{eq:mtanh}
	\mathtt{mtanh}(x,\pedheight{},\coreslope{}) = \frac{\pedheight{}}{2}\left[\frac{(1+x \; \coreslope{} )e^x - e^{-x}}{e^x+e^{-x}}+1 \right],
\end{equation}
where $x = (\psipos{}- \psi_N)/\psiwid{} $, and all the free parameters will be defined below. Such \mtanh fits are commonly used in many studies of H-mode physics, and thus they offer the advantage of a generalised convention for pedestal profiles, while being differentiable fits with relatively few free parameters. These free parameters also represent formal definitions of several important properties of the pedestal. Namely, these are the profile value at the top of the pedestal $\pedheight{}$ ($= \Tetop$ and $=\Netop$ for the temperature and density, respectively), the position of the middle of the pedestal $\psipos{}$, the width of the pedestal $\psiwid{}$ ($=\psiwid{T_e}$ and $=\psiwid{n_e}$), and the position of the pedestal top $\psitop{} = \psipos{} - \psiwid{}/2$ ($= \psitop{T_e}$ and $= \psitop{n_e}$). The core slope to which the profile tends at $\psi_N \ll \psitop{}$ is determined by $\pedheight{} \coreslope{} / 2 \psiwid{}$, parametrised by $\coreslope{}$. 
The profiles derived using an \mtanh fit, as per \cref{eq:mtanh}, will be referred to as $\fit$ profiles, and they represent the nominal plasma profiles used for the EUROfusion database \citep{Frassinetti_Database}.

Importantly, the fits are performed taking into account the finite radial resolution of the HRTS diagnostic, as described by \cite{Frassinetti_MeasurementMethods}. Therefore, each \fit profile is computed from the \raw profile using deconvolved fits. For this, the \fit electron-density profile is obtained by minimising the residual between the \raw $n_e$ values and the \fit $n_e$ profile, convolved with the spatial instrument function of the HRTS diagnostic. Then, the \fit electron-temperature profile is obtained by minimising the residual between the \raw $T_e$ values and the \fit $T_e$ profile weighed by the \fit $n_e$ profile and convolved with the instrument function of the HRTS diagnostic. This is because the density profiles affect the inference of the temperature from the HRTS data. The instrument function of HRTS represents a \quotemarks{scattering kernel} with a full-width-half-maximum of approximately $11\;\mathrm{mm}$ in the pedestal region. The scatter is caused by the HRTS diagnostic imaging a finite volume of the plasma around each measurement location, so the variations of the profile across this volume must be taken into account.

\subsection{Gradients}\label{ssec:exampleGradients}
Fitting the $\raw$ profiles to a differentiable function allows for gradients to be meaningfully calculated. The gradients are an essential ingredient for studying the local quasi-equilibrium that the plasma settles to during the inter-ELM period. They must be related to the heat flux crossing the pedestal because they provide the drive for the turbulence that controls transport across the pedestal. It is important to acknowledge that the choice of an \mtanh fit is a source of systematic bias in estimating the gradients of the profiles, however, this fit has been used to characterise the properties of pedestal shapes across devices for more than 20 years, which helps make our study easily generalisable to other databases.

The gradients of the $T_e$ and $n_e$ profiles are calculated with respect to the radial coordinate $R$. For any quantity $A(R)$, we define 
\begin{equation}
	L_A = A(R) \left(\der{A}{R}\right)^{-1}
\end{equation} as the corresponding gradient scale length.

\cref{fig:Gradients} is an overview of pulse $\examplepulse$: the plasma profiles (a,b), the value of the gyro-Bohm heat flux [shown in (c) and defined below], the profile gradients (d,e), and the gradient-length-scale ratio $\eta_e = \LNe / \LTe$ (f). The radial gradients $\dertxt{T_e}{R}$ and $\dertxt{n_e}{R}$ in \cref{fig:Gradients}(d) are normalised to the pedestal-top values $\Tetop$ and $\Netop$, respectively. The most important trends manifest in \cref{fig:Gradients} are the increase by approximately two orders of magnitude of the normalised gradients $\RLTe$ and $\RLNe$ towards the separatrix, and the typical values of $\eta_e = \mathcal{O}(1) - \mathcal{O}(10)$. The finite values of $\RLTe$ and $\RLNe$ outside the separatrix region is a result of the $\mtanh$ convention that all $\fit$ profiles tend to $0$ in the SOL. These tendencies of the gradients of pulse $\examplepulse$ are qualitatively representative of the behaviour of most pulses across the database.

The gyro-Bohm heat flux is defined as 
\begin{equation} \label{eq:QeGB}
	\Qegb = n_e T_e \vthe \frac{\rho_e^2 }{R^2},
\end{equation}
where $\vthe = \sqrt{2 T_e / m_e}$ is the electron thermal velocity, $\rho_e = m_e \vthe/eB$ is the electron gyroradius, $m_e$ is the electron mass, $e$ is the electron charge, and $B$ is the magnitude of the magnetic field. The gyro-Bohm heat flux \cref{eq:QeGB} maintains values of $\mathcal{O}(\mathrm{MW}/\mathrm{m}^2)$ across the profiles inside of the half-way point between the density-pedestal top and and separatrix (defined as the \quotemarks{steep-gradient point} in \cref{ssec:locations}); $\Qegb$ decreases rapidly towards the separatrix, approaching $0$ together with the $T_e$ and $n_e$ profiles.

\subsection{Pedestal Regions and Relevant Locations} \label{ssec:locations}
It is useful to define formally some important radial locations and regions for our profiles. Using the parameters of the $\mtanh$ fits in $\psi_N$ coordinates, the pedestal-top positions of $T_e$ and $n_e$ are $\psitop{T_e}$ and $\psitop{n_e}$, respectively. These pedestal-top positions are defined as 
\begin{equation}
	\psitop{} = \psipos{} - \frac{\psiwid{}}{2}.
\end{equation}
We refer to the region $\psi_N < \psitop{T_e}$ as the core of the plasma. This definition with respect to $\psitop{T_e}$  holds across the database as a result of the relative shift between the tops of the $n_e$ and $T_e$ pedestals \citep{Frassinetti_2021}: the top of the density pedestal is radially further out than the top of the temperature pedestal. Hence, the low temperature gradient in the regions within $\psitop{T_e}$ guarantee low electron-density gradients. We will also use this pedestal-shift property in \cref{ssec:psiRenorm}: the separatrix position will be conventionally placed at $\psi_N=\psisep=1.0$ such that $T_e = 100 \; \mathrm{eV}$ as per the normalisation described in \cref{ssec:radialcoord}, and, consequently, the region of $\psi_N>1.0$ is referred to as the Scrape-Off Layer (SOL).

The region between the separatrix and $\psitop{T_e}$ is the pedestal. Because most of the pulses obey $\psitop{T_e}<\psitop{n_e}<1.0$, the interval between $\psitop{n_e}$ and $1.0$ will be referred to as the steep-gradient region, since both the electron-density and temperature gradients are very large there. In the interval between $\psitop{T_e}$ and $\psitop{n_e}$, the electron-density gradient is much smaller, which justifies making a qualitative distinction between these two neighbouring regions of the pedestal. We identify this distinction again in \cref{sec:gradients}, and we will discuss the physical difference between the two regions in the Sections thereafter.

The profile values at the top of the $T_e$ pedestal are $\Tetop$ and $\TetopNe$; at the top of the $n_e$ pedestal they are $\NetopTe$ and $\Netop$. These pedestal-top points [large dots in \Cref{fig:Gradients}(a,b)] are the locations where the gradient scale lengths change from the high-gradient regime of the pedestal to the low-gradient regime characteristic of the core. As a representative location in the steep-gradient region, we define the \quotemarks{steep-gradient point} located at 
\begin{equation}
	\psisteep = \frac{\psitop{n_e}+1}{2} ,
\end{equation}
i.e., half-way between the electron-density top and the separatrix. Further out from the steep-gradient point, towards (or beyond) the separatrix, the uncertainties of $\RLNe$ and $\RLTe$ become comparable to the values of the gradients themselves, as obtained by uncorrelated variation of the $\mtanh$ parameters within their confidence intervals \citep{Field_Stiff}. This is because the normalisation of the gradients against the asymptotically vanishing values of $n_e$ and $T_e$ in the SOL results in diverging errors. We shall, therefore, use the steep-gradient point as indicative of the pedestal's steep-gradient region.

\subsection{Rescaled Radial Coordinate}\label{ssec:psiRenorm}

In order to group together qualitatively similar regimes of the pedestal, we use the landmarks that we defined above, i.e., the separatrix position $\psisep$, the density-pedestal-top position $\psitop{n_e}$, and the temperature-pedestal-top position $\psitop{T_e}$, to define a new \quotemarks{renormalised} radial coordinate $\psirenorm$. This will map each point in the density-top and steep-gradient regions to a value $\psirenorm$ that depends on its position relative to these landmarks. 

Namely, a  point whose flux label is $\psi_N \in \left[ \min\left(\psitop{T_e},\psitop{n_e}\right), 1 \right]$ will have the radial coordinate $\psirenorm \in \left[ 0, 2 \right]$ defined by
\begin{equation}\label{eq:psirenorm}
\psirenorm \left(\psi_N\right) = 
\begin{cases} 
	\dfrac{\psi_N - 1}{\psitop{n_e} - 1} & \text{if } 1 \geq \psi_N \geq \psitop{n_e}, \\
	1 + \dfrac{\psi_N - \psitop{n_e}}{\psitop{T_e} - \psitop{n_e}} & \text{if } \psitop{T_e} \geq \psi_N > \psitop{n_e}.
\end{cases}
\end{equation}
This is a piecewise linear transformation that maps $\psisep$ to $0$, $\psitop{n_e}$ to $1$, $\psitop{T_e}$ to $2$, and $\psisteep$ to $0.5$. Because of the relative shift between the pedestal-top locations in the database \citep{Frassinetti_Database}, all but $22$ pedestals obey the pedestal-shift condition $(\psitop{n_e} > \psitop{T_e})$ in the subset relevant to this study, as per \cref{ssec:dataSelection}. For the pedestals that do not obey this condition, the map \cref{eq:psirenorm} will place the entire pedestal profile in the $\left[ 0, 1 \right]$ interval. The map \cref{eq:psirenorm} will be used in \cref{sec:etae,sec:alpha} to uncover the overall structure of pedestal gradients.

More complex functional forms of $\psirenorm$ that maintain the separation of  $\psisep$, $\psitop{n_e}$, and $\psitop{T_e}$, but are also, e.g., differentiable at $\psirenorm  = 1$, offer little advantage while being more cumbersome.

\section{Database and Parameters} \label{sec:database}

The database used for this study is the EUROfusion pedestal database described in \citet{Frassinetti_Database}. This database contains the parameters of over $2000$ ELMy H-mode pulses from JET-C (carbon wall) and JET-ILW (ITER-like wall, with beryllium main chamber and tungsten divertor plates). The pulse parameters that we will use are the fit coefficients of \mtanh functions as described in \cref{ssec:mtanh}. We narrow the scope of the database to a subset of $1251$ pulses, based on considerations of the plasma equilibrium, as explained below.

\begin{figure}
	\centering
	\includegraphics[width=\textwidth]{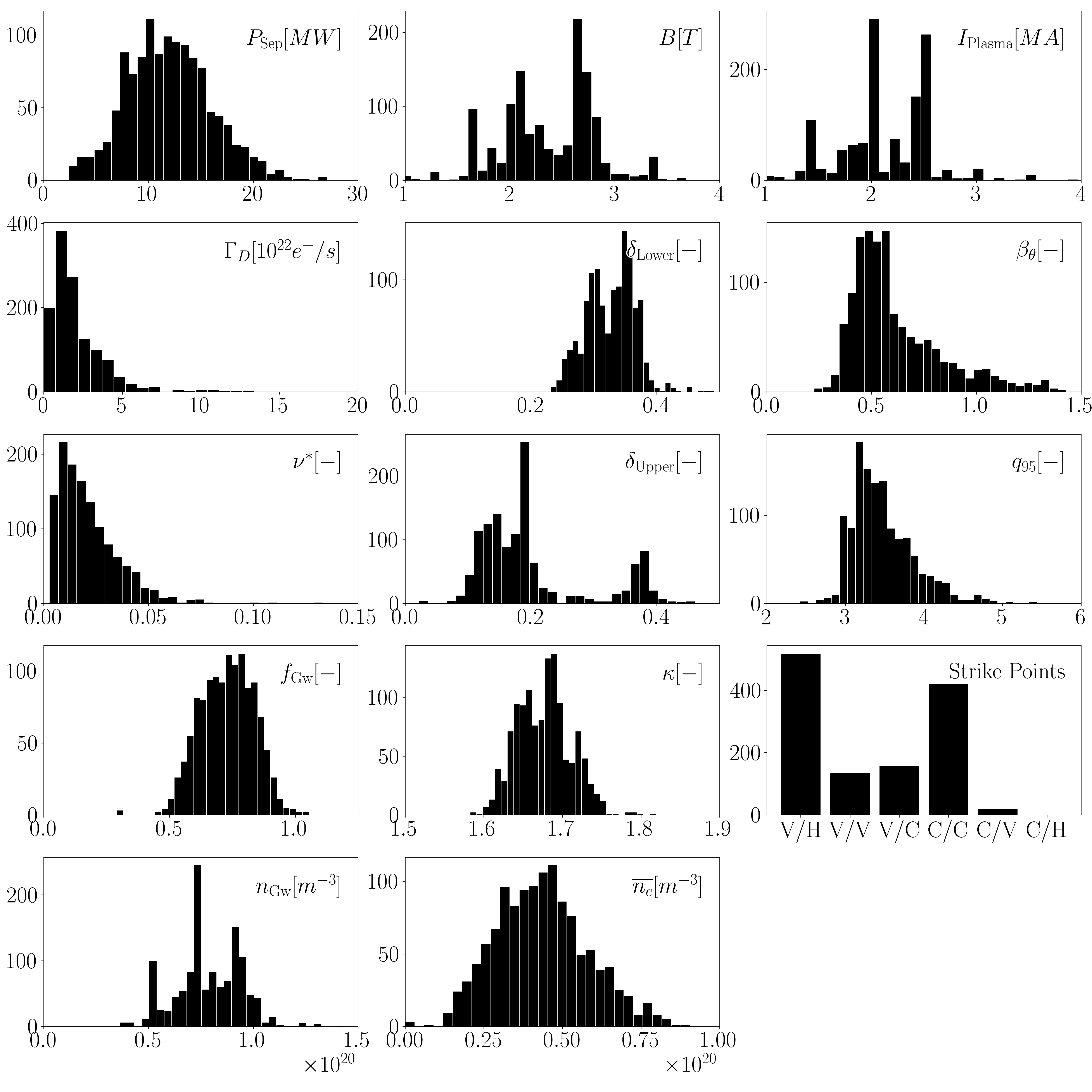}
	\caption{\centering {Histograms of the distributions of the engineering and equilibrium parameters described in \cref{sappendix:params}. The number of pulses with each strike-point configuration is also shown. The strike points are formatted as \quotemarks{inner/outer}, as explained in \cref{sappendix:params}.}}
	\label{fig:ParamHists}
\end{figure}

\subsection{Data Selection and ELMs} \label{ssec:dataSelection}

In order to contextualise the data representing our H-mode pulses, it is useful to understand how the pedestals in the database are measured and how they relate to the stability of ELMs.

The pedestal profiles are constrained by the peeling-ballooning MHD stability limit, which depends on the pressure gradient and the bootstrap current \citep{Connor_Wilson_PeelBalloon}. ELMs are thought to be triggered when a plasma profile strays past this stability limit, and they result in a sudden reduction of the energy density on spatial scales larger than the size of the pedestal \citep{Snyder_PeelBalloon_2004}. An \quotemarks{ELM cycle} contains the ELM itself and the recovery period of the plasma profile until the occurrence of the next ELM. The classification of ELMs depends on their frequency, associated energy loss, and on whether the pedestal is peeling-unstable or ballooning-unstable \citep{Connor_Wilson_ELM_Types, Murari_jetC_ELMs_2014}. \quotemarks{Giant} Type-I ELMs are the ones occurring throughout the pulses in our database.

The EUROfusion pedestal database contains data from Type-I ELMy H-mode pulses in JET, and the pedestal profiles are taken from measurements from the last $20\%$ of each ELM cycle, as described in \cref{sec:geometry}. This choice ensures that the pedestal profiles are in near-steady state over the course of the measurement, and that they represent optimal operation of the plasma, just below the ELM stability threshold.

While the EUROfusion Pedestal Database contains a wide range of pulses, we will narrow the scope in order to limit the variation of the underlying MHD instability of the plasma. We will focus on deuterium-only plasmas (so the effective ion mass will be $M_{\mathrm{eff}}=2$ atomic units). Furthermore, methods of ELM control or mitigation affect the equilibrium of the plasma, meaning that plasma pressure is reduced to a sub-critical, ELM-stable regime. Therefore, we chose pulses that had no methods of ELM control or mitigation applied, such as impurity seeding (added impurities producing a regime with high-frequency, low-amplitude ELMs), pellet injection (a method of plasma fuelling inside the edge-transport barrier, triggering ELMs), or vertical kicks (vertical displacements of the plasma magnetohydrodynamic equilibrium, also triggering ELMs).

This means that, for the current analysis, we will work with a subset of the database containing $1251$ pulses.

\subsection{Parameter Values} \label{ssec:params}

\begin{figure}
	\centering
	\includegraphics[width=\textwidth]{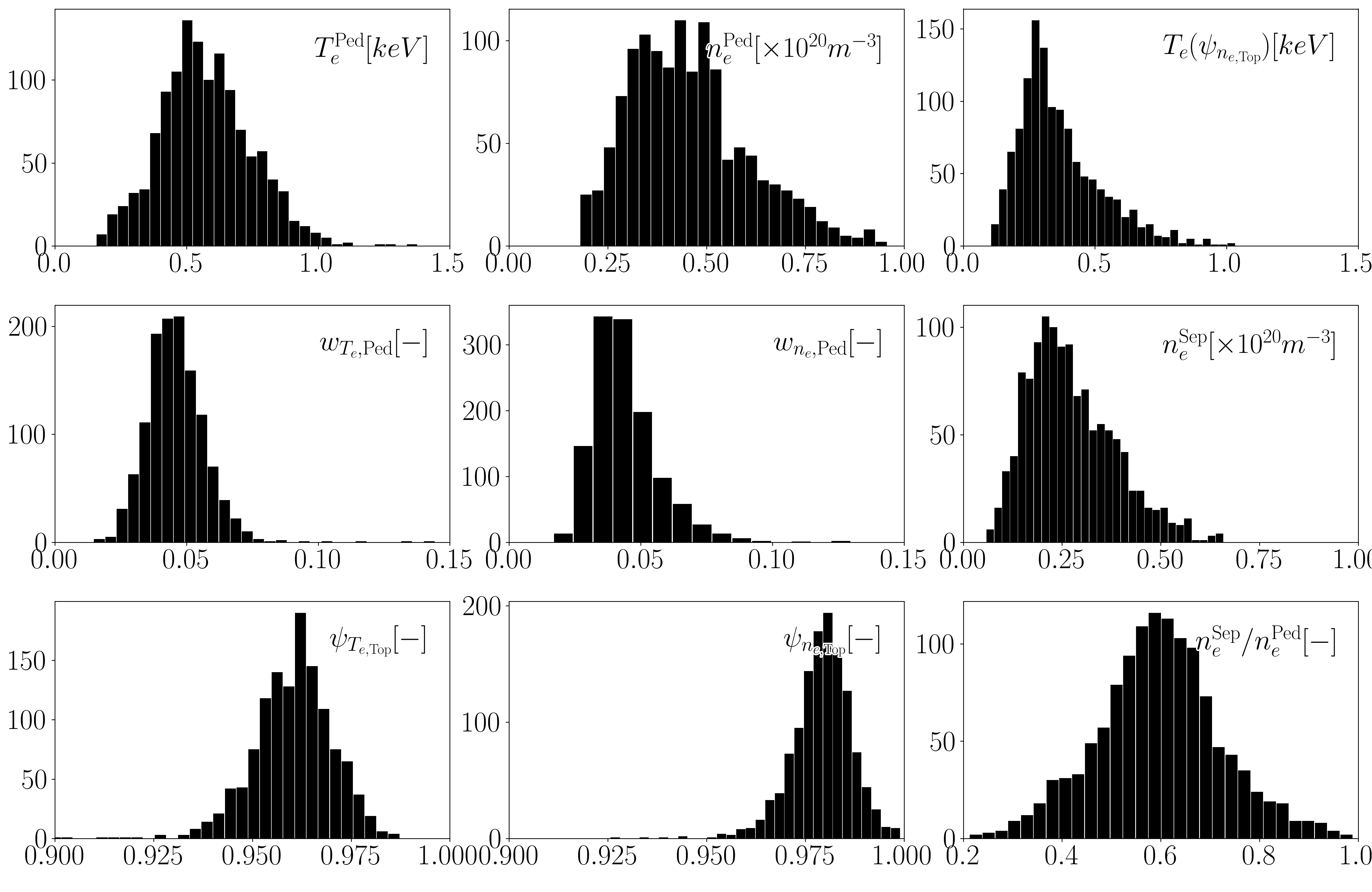}
	\caption{\centering {Histograms of the distributions of relevant $T_e$ and $n_e$ profile values. Their definitions are given in \cref{ssec:mtanh}.}}
	\label{fig:ProfValsHists}
\end{figure}

The database also contains a wide range of experimental parameters, so we limit ourselves to a physically relevant subset of engineering parameters and magnetic-equilibrium-reconstruction parameters. The list of these parameters is provided in \cref{sappendix:params} and an overview of their distribution is given in \cref{fig:ParamHists}. 

Because of the wide scatter in the values of these parameters, we anticipate the necessity for data augmentation (\cref{sappendix:augmentation}) for effective training of $T_e$-predicting neural networks in \cref{sec:neuralnets}. We note that the poloidal beta, $\beta_{\theta}$, and the electron-electron collisionality, $\nu^*$, are physically important in determining the nature of local turbulence in the pedestal, however the calculation of these parameters directly references $\Tetop$, as per \citet{Frassinetti_Database}. This becomes an issue as the neural networks of \cref{sec:neuralnets} manage to trivially invert the calculation of these parameters. This results in excellent $T_e$ predictions with a priori knowledge of the $T_e$ pedestal. The inclusion of such information into our neural networks defeats the point of our study, although they are worth mentioning amongst the parameters that are usually considered important for characterising the physical environment of a pedestal.

Some of these parameters will also be used in \cref{sec:heatFluxModels,appendix:heatFluxFits} for predictions using heat-flux scalings and for the calculation of $\Qegb$.

For our analysis of the plasma profiles themselves, we will make use of the $\raw$ and $\fit$ profiles described in \cref{sec:geometry}. These profiles are not normally a part of the EUROfusion database, so we sourced them directly from processed pulse files (PPFs) for each pulse. We have access to the electron-density and temperature profiles and the corresponding $R$ and $\psi_N$ for each pulse, as described in \cref{sec:geometry}. 

The distribution of $\mtanh$ parameters (defined in \cref{ssec:mtanh}) and other relevant values of the $T_e$ and $n_e$ profiles are shown in \cref{fig:ProfValsHists}. The value of the temperature at the density-pedestal top, $T_e(\psitop{n_e})$, will be an important measure of the accuracy of our temperature-pedestal reconstruction over the steep-gradient region. The density at the separatrix $\Nesep = n_e(\psi_N = 1.0)$, and the ratio  $\Nesep / \Netop$ between the separatrix density and the pedestal-top density are strongly correlated with temperature-pedestal parameters, as anticipated in \citet{Field_Stiff} and will be confirmed in \cref{fig:correlationMatrix}.

	\section{Neural Network Predictions} \label{sec:neuralnets}

The high-dimensional parameter space of the EUROfusion Pedestal Database together with the pulse profiles present an opportunity for machine learning (ML) to identify complex, non-linear correlations. Before attempting to use physics-based models, either stemming from theoretical considerations or from numerical experiments (\crefrange{sec:gradients}{sec:heatFluxModels}), we will leverage neural networks to predict $T_e$ profiles using $n_e$ profiles and selected database parameters. 

This section details the prediction of $T_e$ profiles using various input parameter groups, enabling us to quantify the contributions of specific engineering and equilibrium-reconstruction parameters to the accuracy of predictions. We will prove that such a prediction is indeed possible, and find the highest accuracy to which $T_e$ profiles can be reconstructed using an ML interpolation method over our data, effectively establishing also the maximal standard of accuracy of such reconstruction by any method. We will also explore the difference between predictions of $T_e$ carried out across the full pedestal profiles vs. ones over narrow windows of the pedestal profile, the latter of which mimic predictions using some model of local turbulence. Finally, we will find the pulse parameters in our database that are most important for the correct prediction of such $T_e$ profiles.

\def\tick {\ensuremath{\times}} 
\begin{table}
	\centering
	
	\begin{tabular}{c|c|c|c|c|c|c|c|c|}
		\multirow{2}{*}{\textbf{Parameters}} & \multicolumn{4}{c|}{\textbf{Full-Pedestal Groups}}& \multicolumn{4}{c|}{\textbf{Local-Value Groups}} \\ 
						 & All	 &No-EFIT& EFIT  & None  & All   & No-$R$&No-EFIT& None  \\ [3pt]
		$\Psep$		 	 & \tick & \tick &       &       & \tick & \tick & \tick &       \\ 
		$\fuelrate$ 	 & \tick & \tick &       &       & \tick & \tick & \tick &       \\ 
		$\ngw$ 			 & \tick & \tick &       &       & \tick & \tick & \tick &       \\ 
		$\overline{n_e}$ & \tick & \tick &       &       & \tick & \tick & \tick &       \\ 
		$\Iplasma$		 & \tick & \tick & \tick &       & \tick & \tick & \tick &       \\ 
		$B$ 			 & \tick & \tick & \tick &       & \tick & \tick & \tick &       \\ 
		Strike Points 	 & \tick & \tick & \tick &       & \tick & \tick &  	 &       \\ 
		Upper $\delta$ 	 & \tick &       & \tick &       & \tick & \tick &  	 &       \\ 
		Lower $\delta$ 	 & \tick &       & \tick &       & \tick & \tick &  	 &       \\ 
		$q_{95}$		 & \tick &       & \tick &       & \tick & \tick &  	 &       \\ 
		$\kappa$ 		 & \tick &       & \tick &       & \tick & \tick &  	 &       \\ 
		$R$ 			 & \tick & \tick & \tick & \tick & \tick &       & \tick & \tick \\ 
		$n_e$ 			 & \tick & \tick & \tick & \tick & \tick & \tick & \tick & \tick \\ 
		\RLNe 			 & 	 	 & 		 &  	 & 		 & \tick & \tick & \tick & \tick \\ 
	\end{tabular}
	
	\caption{\centering  {The parameter groups and the respective included parameters for the neural networks informed of the full $n_e$ pedestal profile (\cref{ssec:mlFullProf}) and of the local values of the $n_e$ profile (\cref{ssec:mlLocalVal}).}}
	\label{table:mlPars}
\end{table}



\begin{figure}
	\centering
	\includegraphics[width=\textwidth]{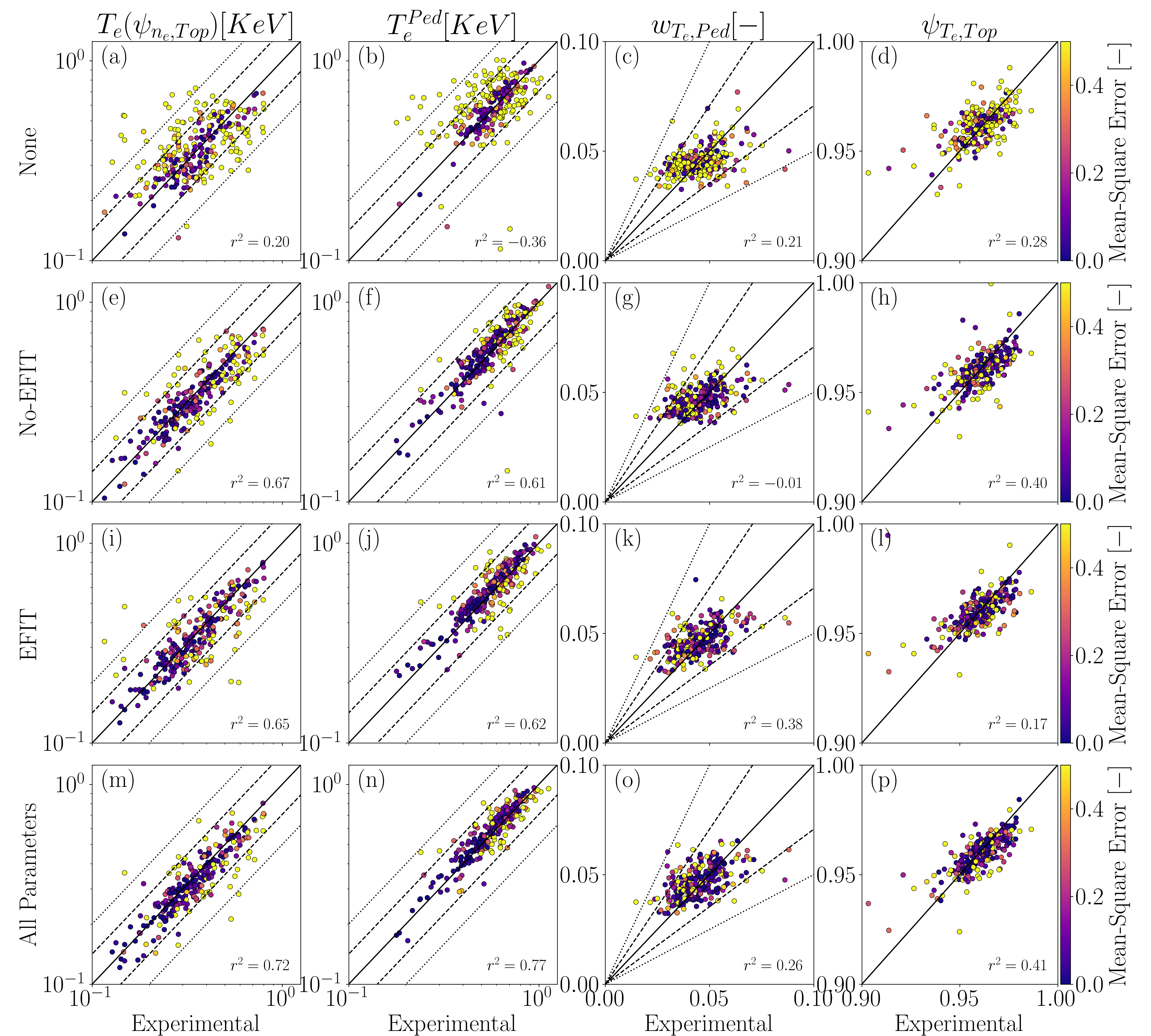}
	\caption{\centering{Database predictions using neural networks informed of the full density pedestal. These correspond to each of the four groups of parameters listed in \cref{table:mlPars}: (a-d) no database parameters used as inputs, with the exception of $R$ and $n_e$; (e-h) only engineering parameters alongside $R$ and $n_e$; (i-l) only the parameters relevant to the magnetohydrodynamic-equilibrium reconstruction, alongside $R$ and $n_e$; (m-p) all of the aforementioned information used. The colour bar represents the mean-square error between the predicted $T_e$ and the experimental \fit profile. For each pedestal prediction, a new \mtanh fit is performed in order to establish the \ensuremath{T_e}-pedestal parameters. Comparisons of the prediction accuracy of these $\mtanh$ parameters are provided (a,e,i,m) for $T_e$ at the inner edge of the steep-gradient region; (b,f,j,n) for $T_e$ at the top of the temperature-pedestal; (c,g,k,o) for temperature-pedestal widths; (d,h,l,p) for the temperature-pedestal locations. The solid lines represent the correct prediction, the dashed lines represent a discrepancy of a factor of \ensuremath{\sqrt{2}}, and the dotted lines represent a discrepancy of a factor of \ensuremath{2}. The text on the bottom right of each plot signifies the respective $r^2$ coefficient of determination, with a value of $1$ indicating a perfect prediction and lower values indicating less accurate predictions. }}
	\label{fig:mlFullProfPred}
\end{figure}

In \cref{sappendix:preprocessing}, we explain the form of the data used in training our networks. Importantly, we will use $T_e$, $n_e$, and $\RLNe$ profiles from the radial interval $R = [3.7, 3.85] \; \mathrm {m}$, sampled at evenly distributed locations in $R$. Once a new $T_e$ profile is computed, a new $\mtanh$ fit is constructed in order to obtain the pedestal height, width, and top position. These are used as metrics of agreement with the experimental profile (for which they are also derived from a deconvolved $\mtanh$ fit, as explained in \cref{sec:geometry}).

A baseline requirement for using ML for such predictions is that the database parameter space is populated densely enough and that experimental uncertainties are low enough in order to be able to infer correct trends from the data. We deal with the issues of parameter-space sparsity and experimental uncertainty by using data augmentation in order to obtain robust methods to predict the $T_e$ profiles. Methods and criteria for data augmentation are explained in \cref{sappendix:augmentation}. 

Our choices of the model loss, the train-test split, and network architecture are described in \cref{sappendix:normAndLoss,sappendix:trainTestSplit,sappendix:modelArchitecture}. In particular, for all predictions, the database is split into a training set of $80\%$ and a testing set of $20\%$ of the database. The same split is maintained for consistency across all the presented results, and all our results will be shown for a testing set that is unseen during training.

Initial tests revealed that the electron collisionality $\nu^*$ and poloidal beta $\beta_\theta$ artificially improved accuracy of predictions because they included exact information about the $T_e$ pedestal. These parameters were originally obtained using a specific reference to $\Tetop$ \citep{Frassinetti_Database}. Therefore, we have excluded these parameters to avoid circular reasoning and maintain the integrity of the predictive task. Similarly, $\psi_N$ is omitted altogether because its normalisation explicitly demands that $\Tesep = 100\mathrm{eV}$.

\begin{figure}
	\centering
	\includegraphics[width=\textwidth]{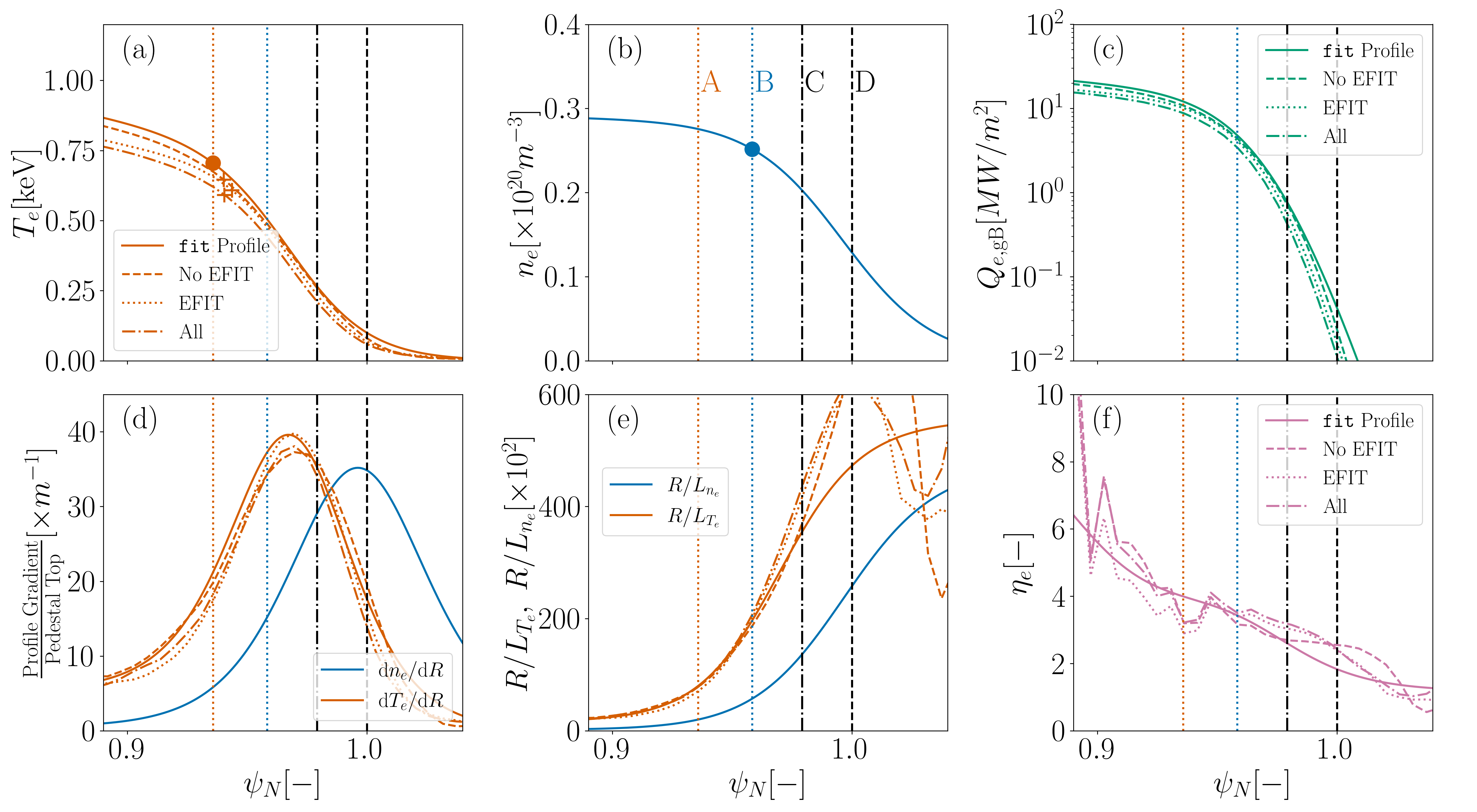}
	\caption{\centering { 
			Pulse $\examplepulse$ $\fit$ profiles compared with neural-network predictions of the $T_e$ profile over the full pedestal. The layout is the same as in \cref{fig:Gradients}. The $\;\fit$ profiles are shown as solid lines, while the predictions using different parameter groups listed in \cref{table:mlPars} are shown as dash-dotted (\quotemarks{All Parameters}), dashed (\quotemarks{No-EFIT}), and dotted (\quotemarks{EFIT}) lines. The vertical lines and dots represent the landmarks of the $\fit$ profiles (see \cref{ssec:locations,fig:Gradients}), while the crosses represent the pedestal top of the $T_e$ predictions.
	}}
	\label{fig:mlExampleVsFullProf}
\end{figure}

\subsection{Full-Profile Electron-Temperature Predictions}\label{ssec:mlFullProf}

We first make predictions of the full temperature-pedestal profile using the information from the full density-pedestal profile and various database parameters. Randomly-located radial cropping is used for the $n_e$ profile during augmentation to reduce bias, ensuring that networks remain independent of specific radial data points while preserving overall profile features. The random crop keeps $80\%$ of the pedestal information, i.e., $L$ is close to the full pedestal extent as detailed in \cref{fig:randomCrop} of \cref{sappendix:augmentation}.

Four groups of parameters are defined for this comparison (found under the heading \quotemarks{Full-Pedestal Groups} in \cref{table:mlPars}). \cref{fig:mlFullProfPred} illustrates the prediction results across these parameter groups, showing the \mtanh parameters that describe the reconstructed $T_e$ pedestals for the $251$ pulses in the test set. The results in \cref{fig:mlFullProfPred} correspond to the groups of parameters listed in \cref{table:mlPars} as follows:
\begin{predictions}
	\item[(a-d): ] \quotemarks{None} -- only the radial coordinate and $n_e$ profile, with no additional pulse parameters;
	\item[(e-h): ] \quotemarks{No-EFIT} -- $R$ and $n_e$ profiles and engineering parameters only, excluding magnetic-equilibrium reconstruction parameters;
	\item[(i-l): ] \quotemarks{EFIT} -- $R$ and $n_e$ profiles and magnetic-equilibrium reconstruction parameters, excluding engineering parameters;
	\item[(m-p): ] \quotemarks{All Parameters} -- all available engineering and EFIT parameters alongside $R$ and $n_e$ profiles.
\end{predictions}
   
The \quotemarks{None} group yielded largely random predictions of the $T_e$ pedestal's width $\psiwid{T_e}$ and top $\Tetop$, demonstrating insufficient predictive power without parameter input. 

The \quotemarks{No-EFIT} group performed significantly better, with marked improvements in the predictions of $T_e(\psitop{n_e})$, $\Tetop$, and $\psitop{T_e}$, suggesting that engineering parameters are highly relevant to the $T_e$ profile. 

The \quotemarks{EFIT} group showed enhanced accuracy of the prediction of the pedestal width $\psiwid{T_e}$, and comparable performance in predicting $T_e(\psitop{n_e})$ and $\Tetop$ as the \quotemarks{No-EFIT} group. However, this group predicts the pedestal-top location $\psitop{T_e}$ more poorly compared to the \quotemarks{No-EFIT} group.

The \quotemarks{All Parameters} group provided the most accurate predictions, achieving close alignment with experimental profiles across all our metrics for pedestal amplitudes and position. The prediction of $\psiwid{T_e}$ also improved, compared to the \quotemarks{None} and \quotemarks{No-EFIT} groups, underscoring the complementary roles of EFIT and engineering parameters in predicting $T_e$ profiles. 

Examples of these predictions  are shown in \cref{fig:mlExampleVsFullProf} for pulse $\examplepulse$. The outputs of the models using the \quotemarks{All Parameters}, \quotemarks{No-EFIT}, and \quotemarks{EFIT} groups closely follow the experimental $T_e$ and $\RLTe$ profiles.

\subsection{Local Electron-Temperature Predictions}\label{ssec:mlLocalVal}

We further explore the $T_e$ predictions based on local values of the density $n_e$ and normalised density gradient $\RLNe$ within a narrow radial window. This approach emulates predictions using physical models that use local parameters within the pedestal. Here, neural networks are trained to predict $T_e$ values over narrow radial intervals, using the density information at these positions. The radial extent used at each location is $L = 2.5\%$ of the sampled radial interval (\cref{sappendix:augmentation}). By choosing many randomly positioned samples, the full $T_e$ profile is then assembled from local predictions.

Four parameter groups are defined under the heading \quotemarks{Local-Value Groups} in \cref{table:mlPars}. The resulting \mtanh parameters of these reconstructions are given in \cref{fig:mlBigCropPred}, arranged as follows:

\begin{figure}
	\centering
	\includegraphics[width=\textwidth]{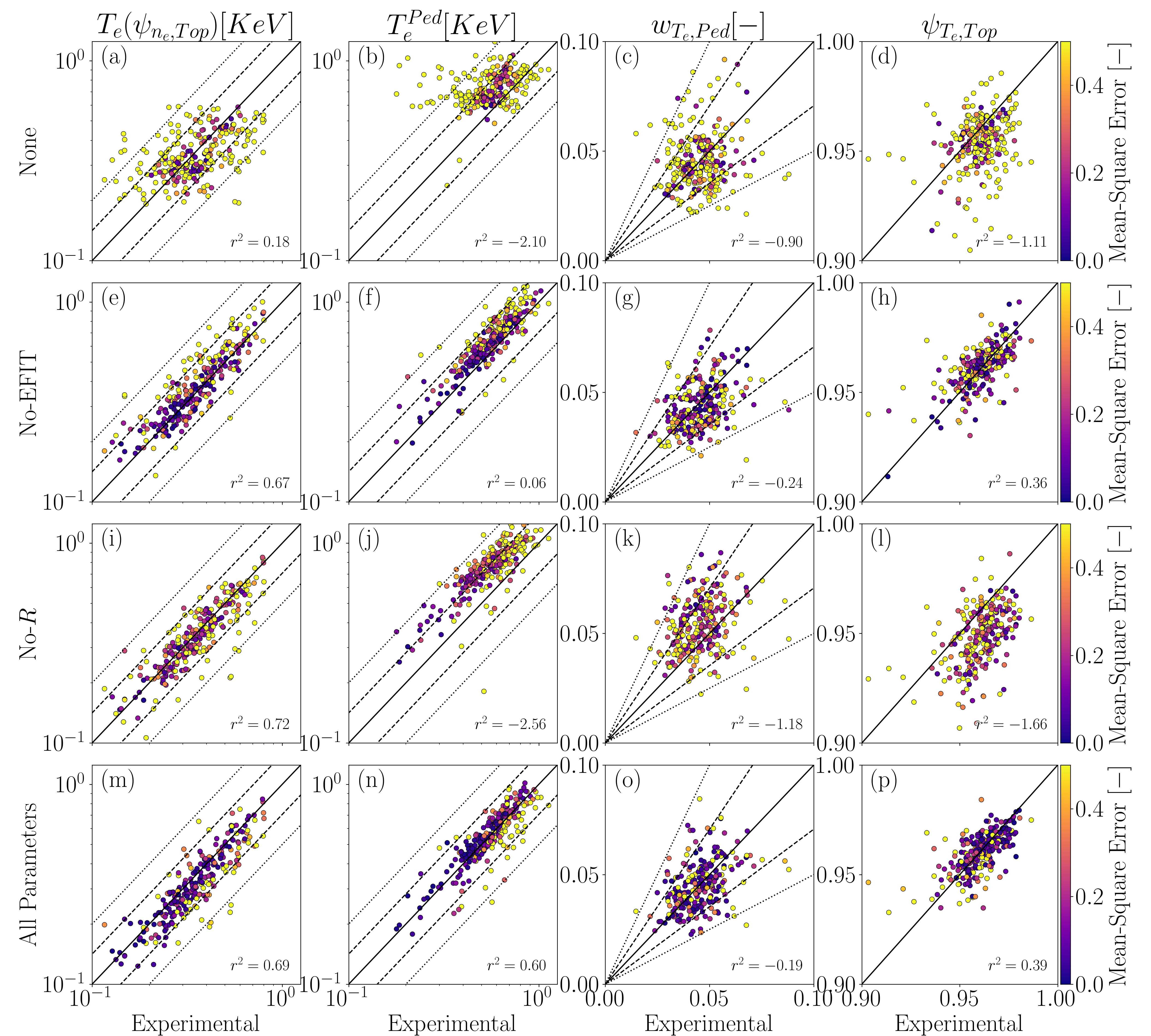}
	\caption{\centering  {Database predictions using neural networks informed of the local values of $n_e$ \nolinebreak and \nolinebreak$\RLNe$. These correspond to each of the four groups of parameters in \cref{table:mlPars}: (a-d) no database parameters used as inputs, with the exception of $R$ and $n_e$; (e-h) only engineering parameters alongside $R$ and $n_e$; (i-l) all available parameters and the $n_e$ values, omitting the radial location \nolinebreak$R$; (m-p) all available parameters used, alongside $R$ and $n_e$. The colour bar represents the mean-square error between the predicted $T_e$ and the experimental \fit profile. The plots follow the same layout as in \cref{fig:mlFullProfPred}.}}
	\label{fig:mlBigCropPred}
\end{figure}

\begin{predictions}
	\item[(a-d): ] \quotemarks{None} -- only the radial coordinate $R$, $n_e$, and $\RLNe$ profiles, with no additional pulse parameters;
	\item[(e-h): ] \quotemarks{No-EFIT} -- engineering parameters and $R$, $n_e$, and $\RLNe$ profiles, excluding magnetic-equilibrium reconstruction parameters;
	\item[(i-l): ] \quotemarks{No-$R$} -- all parameters, $n_e$, and $\RLNe$, but excluding the radial coordinate $R$;
	\item[(m-p): ] \quotemarks{All Parameters} -- all available engineering and EFIT parameters, and profiles including the radial coordinate $R$.
\end{predictions}

\begin{figure}
	\centering
	\includegraphics[width=\textwidth]{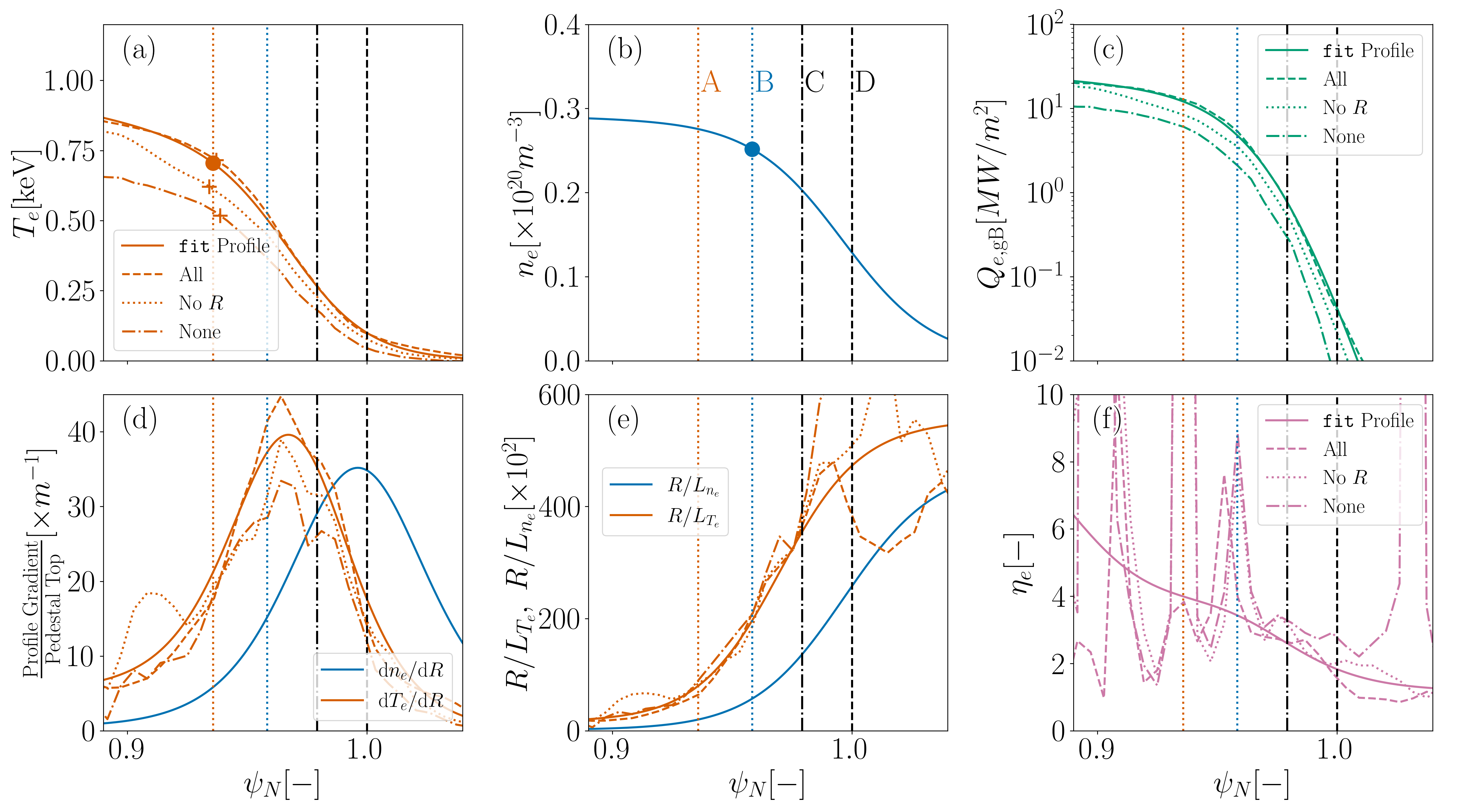}
	\caption{\centering { 
			Pulse $\examplepulse$ $\fit$ profiles, compared with neural-network local predictions of the $T_e$ values. The layout is the same as in \cref{fig:Gradients,fig:mlExampleVsFullProf}. The $\fit$ profiles are shown as solid lines, the predictions with different parameter groups listed in \cref{table:mlPars} are shown as dash-dotted (\quotemarks{None}), dashed (\quotemarks{All Parameters}), and dotted (\quotemarks{No-$R$}) lines. The vertical lines and dots represent the landmarks of the $\fit$ profiles (see \cref{ssec:locations,fig:Gradients}), the crosses represent the pedestal top of the $T_e$ predictions. The full predicted $T_e$ profiles are a result of assembling many local predictions of $T_e$ values at different locations.}}
	\label{fig:mlExampleVsBigCrop}
\end{figure}

The \quotemarks{None} group yields near-constant pedestal, reflecting the failure of local $n_e$ values alone in capturing pedestal variation over our database. This is a case where insufficient information is provided, resulting in the network yielding an output with little variation.

The \quotemarks{No-EFIT} group enhances predictions of $T_e(\psitop{n_e})$, $\Tetop$, and $\psitop{T_e}$ with a strong correlation with the experimental values, although the predictions of the pedestal width remain inaccurate.

The \quotemarks{No-$R$} group demonstrates the critical role of the radial position $R$ for precise pedestal predictions; without $R$, $\Tetop$ is over-predicted by a factor of about $1.5$, and the pedestal shapes are predicted poorly, indicating that $R$ could act as a proxy for larger-scale spatial effects. This means that the local shape of the density profile alone is not sufficient to predict how tall the $T_e$ profile must be in that location.

The \quotemarks{All Parameters} group achieves high accuracy across all metrics, comparable with the results using the full-pedestal profile in \cref{ssec:mlFullProf}. The pedestal width, however, is reproduced with a large inaccuracy, suggesting that $\psiwid{T_e}$ is linked to global properties of the $n_e$ profile. This link was also noticed by \citet{Saarelma_density_2024,Saarelma_2023}, who found $\psiwid{n_e}$ and $\Netop$ to be related to the $T_e$ profile based on the penetration of neutrals and their ionisation through the pedestal region.

Examples of these predictions for $\examplepulse$ are presented in \cref{fig:mlExampleVsBigCrop}, showing the effectiveness of local predictions with full parameter inclusion, and a high scatter in the temperature gradient of the reassembled profiles. Overall, the local profile predictions provide slightly poorer pedestal reconstructions than the global ones of \cref{ssec:mlFullProf}, and these results are generally far more easily degraded by omitting information.

\subsection{Assessment of Most Important Database Parameters} \label{ssec:mlImportantPars}

\begin{figure}
	\centering
	\includegraphics[width=\textwidth]{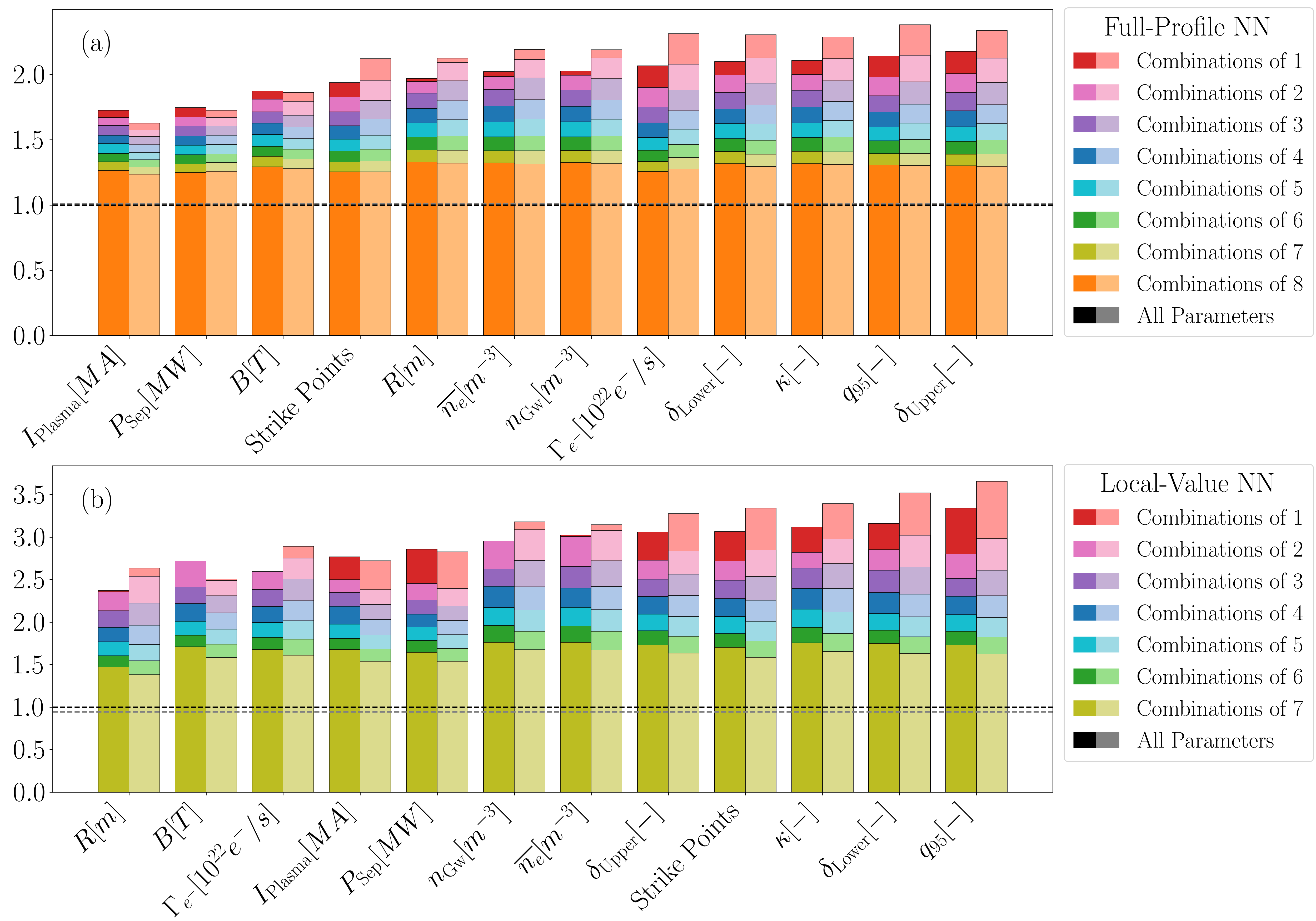}
	\caption{\centering (a) Normalised mean-square error of neural-network predictions of (a) full $T_e$ profiles and (b) local $T_e$ profiles, averaged over combinations of $N$ parameters that include each given parameter (horizontal axis). The bright bars indicate the test-set errors, while the pale bars represent the training-set errors. The red bars show the error when only the single parameter marked on the horizontal axis is used. The pink bars display the average error when the parameter is used in two-parameter combinations, with further bars showing averages across higher combinations. The parameters are arranged left to right in ascending order of the single-parameter test error, demonstrating the relative importance of each parameter for accurate full-profile predictions. }
	\label{fig:paramScan}
\end{figure}

For a quantitative assessment of parameter importance, we measure and compare the prediction accuracy of $T_e$ profiles using various parameter combinations as inputs. We scan sequentially through combinations of $N$ parameters from \cref{table:mlPars} and identify the ones that correlate most with the success of the $T_e$ pedestal reconstruction.

Because of the difference in the sensitivity to parameter inclusion and in the performance of full-profile and local-value predictions, we scan parameters for each case. The results of these parameter scans are presented in \cref{fig:paramScan}, where we rank the performance of networks that use $N$ parameters as inputs, with $N$ indicated by different colours, as shown in the legends. The diagrams present average performance for each parameter on the horizontal axis, when $N-1$ other parameters are included. Our metric for the performance of each network is the total mean-square error across all pulses, normalised to the error obtained from the respective \quotemarks{All Parameter} groups from \cref{ssec:mlFullProf,ssec:mlLocalVal}. The error for any given network is defined as
\begin{equation}
	\mathcal{E} = \sum_{\mathrm{Pulses}} \sum_{i}\left[T_{\fit}(R_i) - T_{\mathrm{Pred}}(R_i)\right]^2,
\end{equation}
where $\{R_i\}$ denote the evenly spaced radial locations where the profiles are sampled numerically, $T_{\fit}(R_i)$ is the experimental fit profile, and $T_{\mathrm{Pred}}(R_i)$ is the predicted value at each location. Bright colours represent the sum over the test pulses, pale colours represent the training pulses.

For full-profile predictions (\cref{fig:paramScan}a), for $N \geq 2$, the deuterium gas fuelling rate $\fuelrate$ becomes one of the highest-performance parameters in spite of its low performance in the $N=1$ group. Overall, the inclusion of $\fuelrate$, the separatrix loss power $\Psep$, the plasma current $\Iplasma$, and the strike-point configuration has proved essential for maximising the accuracy of neural networks with $N\geq 4$. Notably, in the full-profile models, the magnetic-equilibrium parameters have less impact than the engineering parameters.

For local predictions (\cref{fig:paramScan}b), the radial coordinate $R$ has proved to be the most important parameter across all metrics used, consistent with the findings of \cref{ssec:mlLocalVal}. The inclusion of $R$ and the magnitude of the magnetic field $B$ on the axis has turned out to be the most effective 2-parameter choice, however, for $N\geq 3$, $\Psep$, $\Iplasma$, and the strike-point configuration have proved to be the best-performing set.

Overall, these scans of parameter combinations suggest that the $\Psep$, $\Iplasma$, $\fuelrate$, and the strike points are the parameters strongest correlated to the correct prediction of the $T_e$ pedestal shapes. The difference between the \quotemarks{All Parameter} groups in the full-profile and local predictions suggests that successful predictions rely on both global and local plasma characteristics. These global characteristics are contained in the full geometry of the $n_e$ profile, and omitting this information degrades the quality of predictions. 

To conclude, the ML approach has proved quite successful in reconstructing the temperature pedestal given the density profile alongside other available parameters. Each of the $8$ neural networks presented in \cref{ssec:mlFullProf,ssec:mlLocalVal} correspond to entries 39-46 in \cref{table:FitCoeff}, where their performances are compared to the other results in this study, stressing the success of such methods. Besides the practical importance of this success for modelling endeavours aiming to help design and operate future devices, it also suggests that the search for a physics-based model of local transport is not a hopeless enterprise.

	\section{Temperature and Density Gradients}\label{sec:gradients}

\begin{figure}
	\centering
	\includegraphics[width=\textwidth]{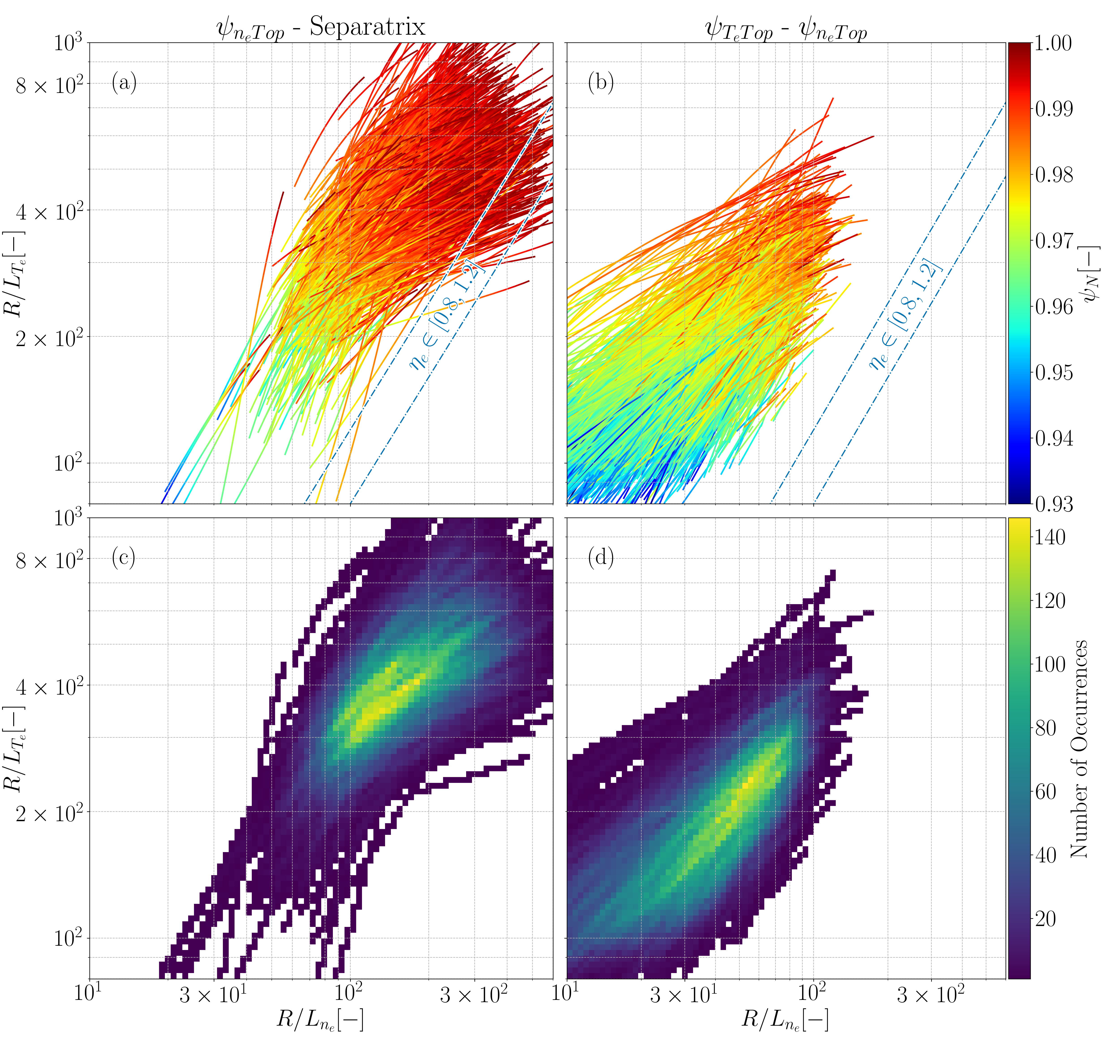}
	\caption{\centering {Loci of the data from all pulses in terms of $\left(\RLNe, \RLTe\right)$ coordinates: (a) and (c) contain the subsets of each profile spanning the steep-gradient region; (b) and (d) contain profiles between the $T_e$-pedestal top and the $n_e$-pedestal top; (a) and (b) contain all pulses considered by us, with their position in $\psi_N$ used to colour-code the points; the colour in (c) and (d) shows the number of pulses where profiles cross each bin in the $\left( \RLNe, \RLTe \right)$ space. All data points lie above the slab-ETG stability threshold of $\RLTe = 0.8 \RLNe$ \citep{Jenko_lin}, depicted as the lower blue dash-dotted line in (a,b). The upper blue line represents $\RLTe = 1.2 \RLNe$, as explained in \cref{ssec:gradientLitReview}. }}
	\label{fig:RLTeRLNe}
\end{figure}

In this section, we describe the qualitative behaviour of the temperature and density gradients across our database. These are expected to be the essential parameters of the local plasma equilibrium: local gradients represent a source of free energy, which causes microinstabilities. In the local picture, these microinstabilities saturate into a turbulent state that in turn determines the transport of heat (and particles) at each location. The object of theory and modelling is the relationship between the resulting transport coefficients and the underlying equilibrium gradients and the consequent equilibrium profiles in a plasma carrying a known heat flux determined on the device scale.

\cref{fig:RLTeRLNe} shows the normalised gradients in the pedestal region for all pulses, separated based on their position relative to $\psitop{T_e}$, $\psitop{n_e}$, and $\psisep$. This separation gives rise to a threshold that is effectively described by $\RLNe > \mathcal{O}(20)$ in the steep-gradient region; concomitantly, $\RLTe > \mathcal{O}(100)$ for the majority of pulses in the steep-gradient region. For both the density-top and the steep-gradient regions, it is apparent that the pulses reside in a regime roughly characterised by $\LNe / \LTe > 1$. In the density-top region, this difference in magnitude between $\RLTe$ and $\RLNe$ is consistent with the temperature pedestal's inwards shift \citep{Frassinetti_Database}, and is a contrivance of the split at $\psitop{n_e}$. In contrast, in the steep-gradient region, it is believed to represent an important physical feature originating from the instability of electron-scale modes, as we will see below.

\subsection{Instabilities and Electron-Temperature-Gradient Turbulence}\label{ssec:gradientLitReview}

Besides the density and temperature gradients, which can trigger instabilities, the pedestal region features a large $E \times B$ flow shear. This has a suppressive effect on microinstabilities \citep{Zhang_shear_1992,Hahm_Burrell_Stabilisation_1995,Terry_shear_2000}, quenching those of them that have growth rates smaller than the shear rate. Amongst the instabilities that are fast enough to overcome this suppression, the most relevant ones are believed to be electron-scale modes, which cause heat transport in the electron channel \citep{Told_gyrokinetic_2008,Jenko_gyro_2009,Hatch_gyro_2015,Hatch_microtearing_2016,Hatch_JETgyrokin,Kotschenreuther_ped_transp_2017,Kotschenreuther_transportCulprits}. Namely, these are micro-tearing modes \citep[MTMs; see][]{Drake_tearing_1977,Drake_microtearing_1980,Applegate_microtearing_2007,Guttenfelder_MTM_2012,Zocco_kinetic_2015,Larakers_MTM_conductivity_2020,Larakers_global_2021,Giacomin_MTM_2023}, which are electromagnetic instabilities relying on finite resistivity, and electron-temperature-gradient modes \citep[ETG; see][]{Horton_toroidalETG_1988,Lee_collisionlessETG_1989,Jenko_ETG_2000,Adkins_ETG,Adkins_ScaleInvariance}, which are electrostatic modes. Both of these are driven by the equilibrium electron-temperature gradient. 


While gyrokinetic simulations of pedestal-regime plasmas recover various contributions to heat transport from MTMs, ETG modes, or other instabilities \citep[e.g.,][]{Leppin_Complex,Chapman-Oplopoiou_2022,Guttenfelder_2021}, the consensus appears to be that ETG modes are responsible for the majority of the turbulent heat transport in the pedestal \citep[see][and references therein]{Ren_ETGreview_2024}. A recent study of two JET pulses (one JET-C and one JET-ILW) confirming the dominance of ETG modes was given by \cite{Hatch_JetTransp2019}. Therefore, it makes sense to discuss the implications of ETG turbulence for the pulses in our database.

The ETG instability has two branches, one mediated by the magnetic-field curvature, called toroidal ETG, and one independent of the curvature and mediated by fluctuating parallel electron flow (current), called slab ETG. The slab-ETG instability is driven by the electron-temperature gradient but stabilised by the electron-density gradient. A measure of their competition is the gradient ratio
\begin{equation}\label{eq:etaDefinition}
	\eta_e(R) = \frac{\LNe}{\LTe},
\end{equation}
already introduced in \cref{sec:intro} as a key (local) physical parameter.

Fluid models \citep{Adkins_ETG} and linear GENE simulations \citep{Jenko_lin} all result in linear stability thresholds characterised by a near-unity $\etacr$ such that, for $\eta_e>\etacr$, slab-ETG modes are unstable. The numerical stability scans performed by \citet{Jenko_lin} yielded
\begin{equation} \label{eq:JenkoLimit}
	\left(\frac{R}{\LTe}\right)_{\mathrm{crit}}= \mathrm{max}\left\{ 0.8 \ \frac{R}{\LNe} , (1+\tau)(1-1.5\varepsilon)\left(1.33 + 1.91 \frac{\hat{s}}{q}\right)  \left( 1+0.3 \varepsilon \der {\kappa}{\varepsilon} \right)\right\} ,
\end{equation}
where $\tau = Z_{\mathrm{eff}} T_e/T_i$ is the temperature ratio, $\varepsilon  = a/R $ is the aspect ratio, $q$ is the magnetic safety factor, $\hat{s}$ is the magnetic shear, and $\kappa$ is the plasma elongation. This formula contains the linear thresholds for the slab-ETG branch (left) and the toroidal-ETG branch (right). Typically, the second term in \cref{eq:JenkoLimit} has order-unity values and is, therefore, much smaller than $0.8 \RLNe$. 

The linear gyrokinetic stability analyses of pedestal plasmas by \citet{Jenko_gyro_2009} and \citet{Guttenfelder_2021} find a value of $\etacr$ in the range $0.9-1.2$. This variation in the critical threshold is attributed to the effect of magnetic geometry and toroidal modes on the instability \citep{Jenko_gyro_2009}. The region of $\etacr \in [0.8,1.2]$ is marked in \cref{fig:RLTeRLNe}, bounded by the two blue dash-dotted lines. These order-unity values of $\etacr$ are situated well below the experimental values of the gradients; only near-separatrix values considerably overlap with this region, albeit they are subject to large uncertainties. This means that throughout the pedestal, the slab-ETG modes are linearly unstable, to the best of existing theoretical knowledge. 

This discussion prepares the analysis of the gradient ratio $\eta_e$, and the outcomes of reconstructing $T_e$ profiles based on the assumption that the pedestal gradients are \quotemarks{pinned} to a specific value of $\eta_e$ in \cref{sec:etae}. This is based on the possibility that the system simply pins itself locally to a marginal value of $\eta_e$ (although whether that marginality is with respect to the traditional linear stability threshold is to be discussed). In contrast to this approach, in \cref{sec:heatFluxModels}, we will present the results of reconstructing pedestals using scalings of turbulent heat transport: we will relate the separatrix loss power $\Psep$ to the background gradients of temperature and density, allowing $\eta_e$ to vary.

\section{Gradient Ratio \ensuremath{\eta_e} and Pedestal Reconstruction}\label{sec:etae}

A simple way to use a physics-inspired approach for pedestal reconstructions is to make use of $\eta_e = \LNe / \LTe$. First, we will understand the disposition of this gradient ratio in relation to $\RLTe$, $\RLNe$, and the pedestal location $\psirenorm$. Once we understand the trends of the experimental values of $\eta_e$ in \cref{ssec:etaDist}, we will use our own model fits to reconstruct the temperature-pedestal profiles in \cref{ssec:etaPredictions}.

\subsection{Distributions of the Gradient Ratio \ensuremath{\eta_e} } \label{ssec:etaDist}

\begin{figure}
	\centering
	\includegraphics[width=\textwidth]{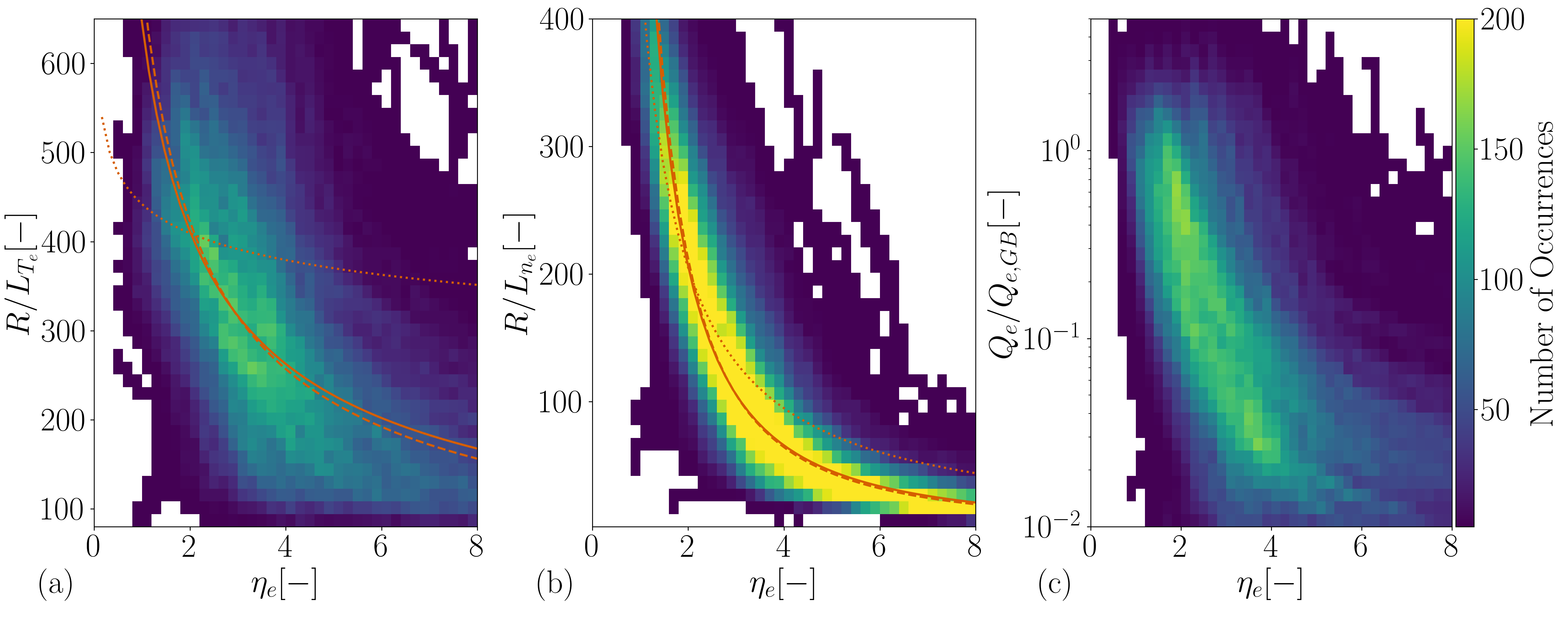}
	\caption{\centering Distributions of the gradient ratio $\eta_e$, defined in \cref{eq:etaDefinition}, across the database conditioned on (a) $\RLTe$, (b) $\RLNe$, (c) $Q/\Qegb$. All data is obtained from the pedestal region, between $\psitop{T_e}$ and $\psi_{Sep}$. Panel (b) suggests that  $\RLNe\propto\eta_e  ^\zeta$ describes the pedestal data well, as will be discussed in \cref{sec:alpha}. The orange lines in panels (a) and (b) represent three fit values of $\zeta$ to: the maxima of the $\RLNe$ distribution (dashed), the full $\RLNe$ distribution (dotted), the steep-gradient-region prediction across the database discussed in \cref{ssec:alphaPredictions} (solid). }
	\label{fig:etaDistributions}
\end{figure}

In \Cref{fig:etaDistributions}(a) and (b), the data from \cref{fig:RLTeRLNe} is recast to show an important property of the pedestals: the steep-gradient region ($\RLNe \gtrsim 20$) displays a strong tendency of \nolinebreak$\eta_e$ to maintain values of around $2$ across this region. It is also apparent that there is a power-law relationship between $\eta_e$ and $\RLNe$, which will be discussed in \cref{sec:alpha}. There is no such relationship between $\eta_e$ and $\RLTe$ evident in \cref{fig:etaDistributions}(a), as there is far more variance in $\RLTe$ for any given $\eta_e$. Furthermore, the gyro-Bohm-normalised heat flux $Q/\Qegb$ appears to increase as $\eta_e$ approaches the stability threshold (or a different $\mathcal{O}(1)$ value). This confirms the intuition of \citet{Field_Stiff} regarding the role of the linear threshold, although with the notable difference that the heat flux must decrease further away from it.

\begin{figure}
	\centering
	\includegraphics[width=\textwidth]{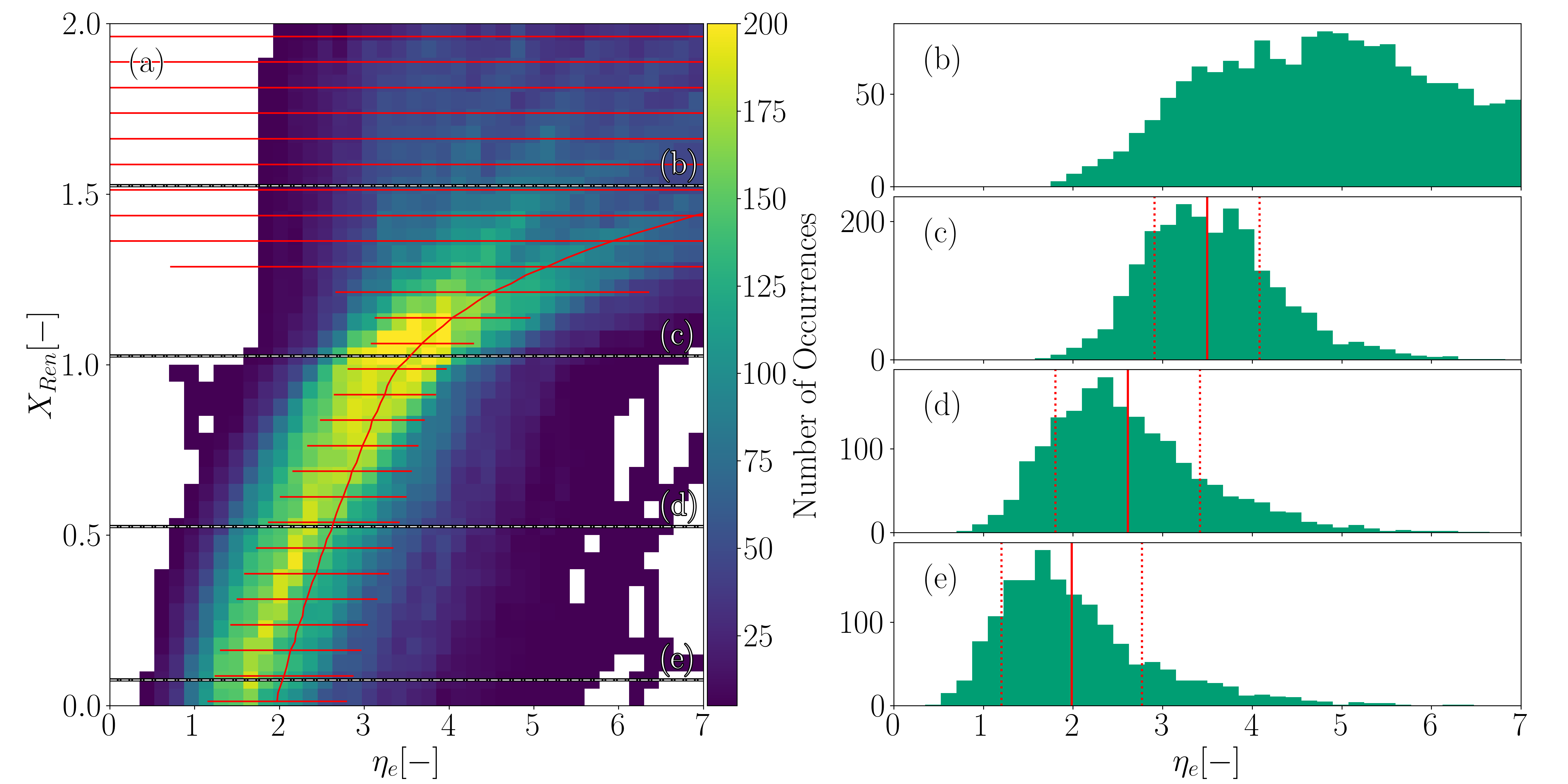}
	\caption{\centering {
			 (a) Distribution of $\eta_e$ [as defined in \cref{eq:etaDefinition}] conditioned on the renormalised profile position $\psirenorm$, defined in \cref{eq:psirenorm}. The red lines represent the mean and variance of $\eta_e$ at each $\psirenorm$. (b-e) Cross-sections of the histogram, at (b) $(\psitop{T_e}+\psitop{n_e}) / 2$, (c) $\psitop{n_e}$, (d) $\psisteep$, and (e) $\psisep$, corresponding to the white dash-dotted lines in (a). In each of these cross-sections, the continuous red lines represent the mean value of $\eta_e$, and the dotted red lines represent the values one standard deviation away from the mean. Note that these lines are outside of the range of $\eta_e$ captured in panel (b).  }}
	\label{fig:etaslice}
\end{figure}

In \cref{eq:psirenorm}, we defined a radial coordinate $\psirenorm$ in order to exploit the geometry of each pulse and to match $\psisep$, $\psitop{n_e}$, and $\psitop{T_e}$ across the database. In \cref{fig:etaslice}, we provide the explicit dependence of $\eta_e$ on the radial location relative to these pedestal landmarks. An important pedestal property highlighted by \cref{fig:etaslice} is that the most probable values of $\eta_e$ range between $1.5$ and $3.5$ across the steep-gradient region, increasing from the separatrix (\cref{fig:etaslice}e) towards $\psitop{n_e}$ (\cref{fig:etaslice}c). Since the stability threshold of ETG modes is expected to be at $\eta_e \approx 1$ (see \cref{sec:gradients}), this implies a measurable deviation of the turbulence from the linear threshold. Consequently, the turbulence resides in an order-unity supercritical regime where $\eta_e$ is closer to $2$ \citep[which has also been reported in several ASDEX experiments, e.g., ][]{Neuhauser_eta2}. The distribution of $\eta_e$ does overlap with the linear threshold in the vicinity of the separatrix, although only just.

We note that $\eta_e$ increases rapidly once the profiles cross $\psirenorm = 1$, which is a consequence of the inward shift of the electron-temperature pedestal: the temperature pedestal maintains large gradients irrespective of the reduction in the density gradient. This could be indicative of a difference between the saturated turbulent states in the steep-gradient and density-top regions. This qualitative difference between the two regions can be interpreted to imply that, in the steep-gradient region, the density gradient plays an essential role in characterising the turbulence, and the turbulent state resides in a critical regime where the temperature gradient is (almost) uniquely determined by the density gradient. In contrast, inside the density-pedestal top, the density gradient does not uniquely determine the temperature gradient, and $\RLTe$ depends more strongly on other plasma parameters, as can be inferred from the comparison of \cref{fig:etaDistributions}(a) and (b). This difference could be attributed to the physically expected prevalence of slab-ETG modes in the steep-gradient region, with the density gradient being the dominant stabilising effect on these modes. At small $\RLNe$, toroidal-ETG modes become unstable before the slab-ETG ones \citep{Jenko_lin}, and the addition of such sub-dominant modes can alter the heat transport \citep{Parisi_3dTurb_2022}. This might be the relevant physics in the density-top region. The inclusion of electromagnetic effects into this picture complicates matters further, as we briefly discussed in \cref{ssec:gradientLitReview}.

\subsection{Electron-Temperature Reconstruction Using Constant Gradient Ratio} \label{ssec:etaPredictions}

\begin{figure}
	\centering
	\includegraphics[width=\textwidth]{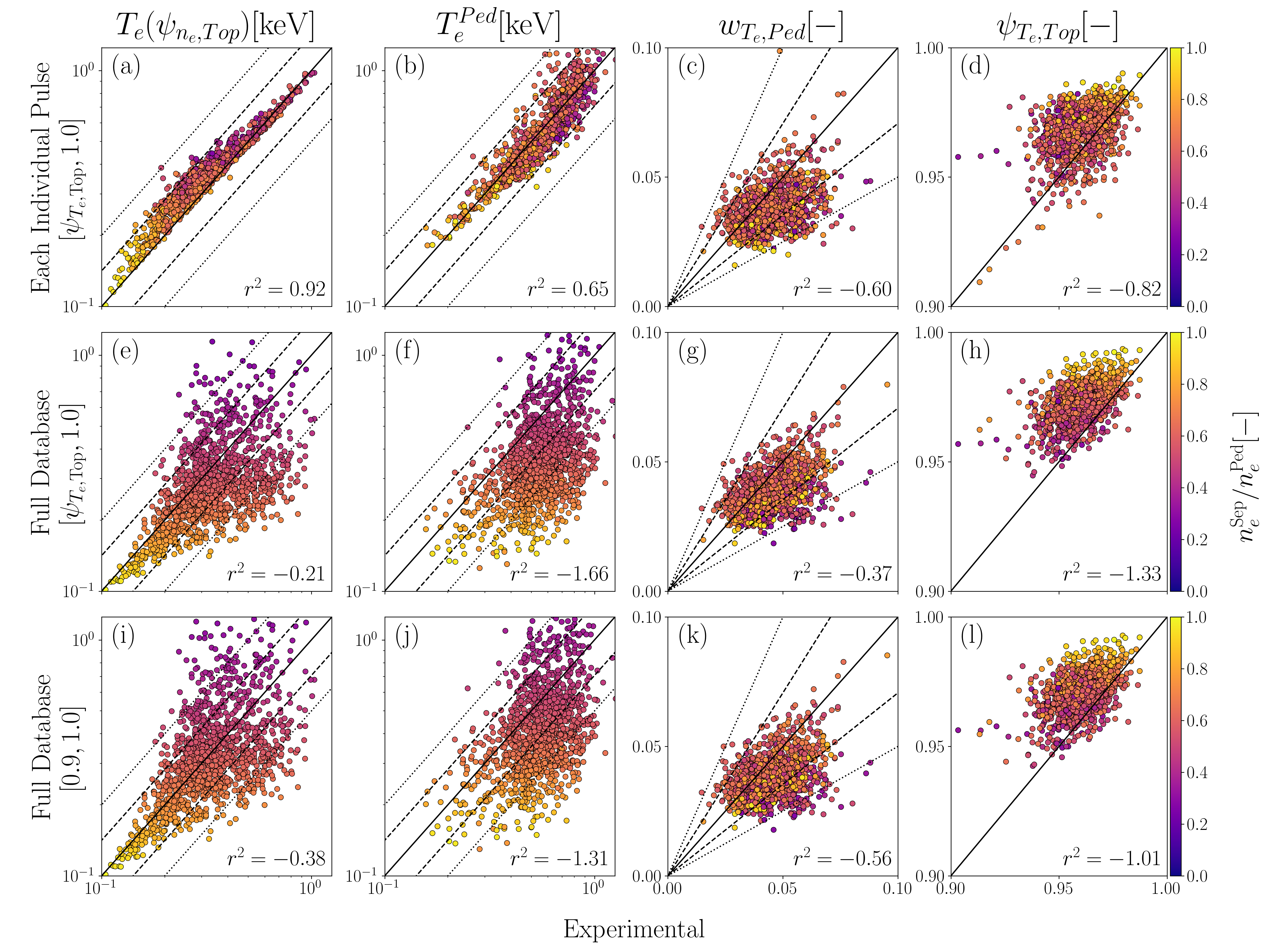}
	\caption{\centering {  Database predictions obtained by integrating $1 / \LTe = \eta_e / \LNe$ inwards from the separatrix, (a-d) with the best-fit $\eta_e$ for each pulse over the pedestal region, (e-h) with the best-fit $\eta_e=1.90$ across the database over the pedestal region, and (i-l) with the best-fit $\eta_e=2.19$ across the database over the region $\psi_N \in[0.9,1.0]$. The colour bars represent the ratio between the separatrix density $\Nesep$ and the density-pedestal height $\Netop$. The plots follow the same layout as in \cref{fig:mlFullProfPred}.}}
	\label{fig:etaResults}
\end{figure}

Using the tendency of $\eta_e$ to have order-unity values in the pedestal region, we will proceed to predict the electron-temperature profiles assuming there is a simple functional relationship between the normalised gradients $\RLTe$ and $\RLNe$. The simplest such relationship, proposed by \citet{Field_Exhaust}, would be a set value of $\eta_e$ describing the turbulence across the whole profile. This is, in itself, an assumption of infinitely \quotemarks{stiff} transport limited by a critical $\eta_e$, maintained across the pedestal: the heat flux diverges as soon as this value of $\eta_e$ is crossed, reducing the local temperature gradient and consequently bringing the heat flux back to a finite value. Based on the results of \cref{ssec:etaDist}, we expect this imposed value of $\eta_e$ to be in the vicinity of $2$. We note that the experimental value of $\eta_e \approx 2$ across the pedestal was initially measured by \citet{Neuhauser_eta2} on ASDEX, but has been observed to characterise H-mode pedestals correctly on other devices as well \citep{Ren_ETGreview_2024}. \citet{Horton_eta2} used the assumption of $\eta_e=2$ to reproduce correctly an electron-temperature pedestal profile on ASDEX Upgrade.

In order to find the values of $\eta_e$ that best describe the pulses in our database, we reconstruct the electron-temperature profiles by first assuming that there is a constant $\eta_e$ in \cref{eq:etaDefinition}, and then integrating inwards from the boundary at $\Rsep$ (as detailed in \cref{appendix:predictionMethods}). We then identify the best \quotemarks{nominal} $\eta_e$ such that the difference between the $\fit$ and the reconstructed pedestal is minimised (as explained in \cref{sappendix:bestFit}).

In \cref{fig:etaResults}, we present three sets of results obtained using this method of integration and parameter optimisation over different domains. These three database reconstructions are calculated by maximising the accuracy of the $T_e$ profile across:
\begin{predictions}
	\item[(a-d): ] the pedestal region, $\psi_N \in \left[\psitop{T_e}, 1.0\right]$, with a different $\eta_e$ for each pulse;
	\item[(e-h): ] the pedestal region, $\psi_N \in \left[\psitop{T_e}, 1.0\right]$, with the best-fit $\eta_e = 1.90$ across the database;
	\item[(i-l): ] the region including a part of the core, $\psi_N \in \left[0.9, 1.0\right]$, with the best-fit $\eta_e = 2.19$ across the database.
\end{predictions}

From the pedestal parameter comparison in \cref{fig:etaResults}, it is apparent that reconstructed temperature-pedestal profiles have temperature-top locations pushed further outwards than the experimental location. The inflexibility of mandating a constant value of $\eta_e$ yields predicted values of $\psitop{T_e}$ that follow $\psitop{n_e}$, irrespective of whether $\eta_e$ is fixed across the database or allowed a different value for each pulse. Furthermore, for the database-wide $\eta_e$ predictions in \Cref{fig:etaResults}(e-h), we find that the temperature-pedestal height, width, and top locations are strongly correlated to the ratio between the separatrix density $\Nesep$ and the pedestal density $\Netop$, supporting this observation. Overall, setting a constant value of $\eta_e\approx 2$ produces poor predictions of the pedestal-top regions. The performance of the models using database-wide $\eta_e$ are presented in entries 31-34 of \cref{table:mlPars}. In particular, the ones presented in \Cref{fig:etaResults}(e-h) and (i-l) correspond to Entries 32 and 34, respectively.

\begin{figure}
	\centering
	\includegraphics[width=\textwidth]{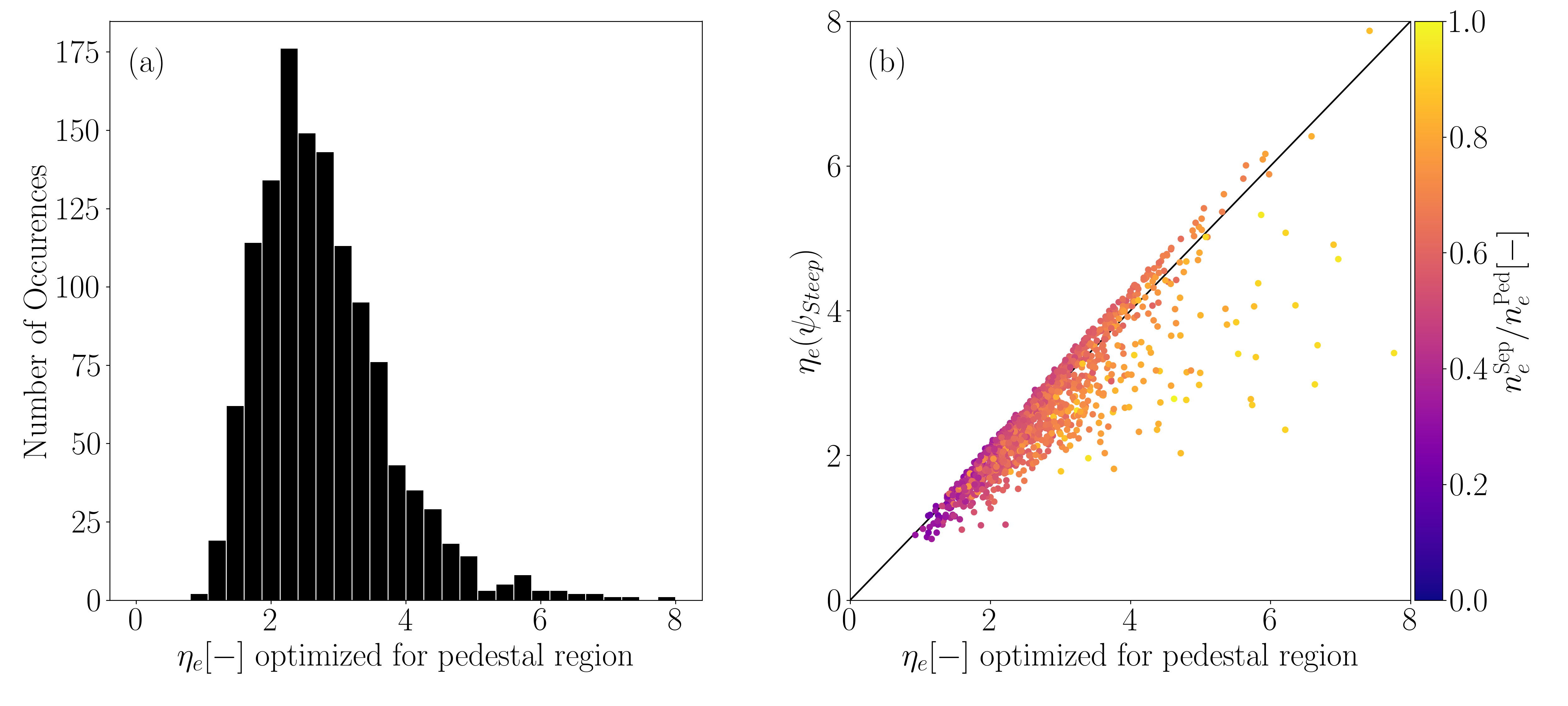}
	\caption{\centering { Values of $\eta_e$ that best predict the $T_e$ profiles across the pedestal region. (a) The distribution of values of fit $\eta_e$. (b) Comparison between the fit values of $\eta_e$ and the database values of $\eta_e$ at $\psisteep$ (as shown in \cref{fig:etaslice}d).}}
	\label{fig:etaBest}
\end{figure}

Thus, allowing the value of $\eta_e$ to vary across the database, i.e., assuming that the turbulence in each individual pedestal can be described by a single parameter that is constant across the profile, yields reasonably good predictions for the steep-gradient region, while over-predicting $T_e$ at $\psitop{n_e}$ and under-predicting $\Tetop$ within a factor of $\sqrt{2}$ of the experimental value. The distribution of the best-fit $\eta_e$ across the steep-gradient region is shown in \cref{fig:etaBest}. We find a considerable overlap between the best-fit values of $\eta_e$ from the pedestal predictions of \Cref{fig:etaResults}(a-d) and the experimental values of $\eta_e(\psisteep)$ at the steep-gradient point (shown in \cref{fig:etaslice}d). The largest discrepancies between the experimental $\eta_e(\psisteep)$ and the fit $\eta_e$ are correlated with the ratio $\Nesep/\Netop$: outwards shifted $n_e$ pedestals require a higher optimal $\eta_e$ than the experimental $\eta_e$ in order for the $T_e$ pedestal to rise enough. In general, pedestals with a significant shift have a poor prediction accuracy because of the inability of the constant-$\eta_e$ models to reproduce large $\RLTe$ beyond the density-top point.

Applying symbolic regression models \citep{Cranmer_PySR} to estimate the optimised $\eta_e$ for each pulse as a function of database parameters shows that the engineering and magnetic-field parameters lack any correlation with the obtained $\eta_e$. The inclusion of $\Nesep/\Netop$ alongside these parameters is essential for the correct estimation of $\eta_e$, but the scalings fit to the optimised $\eta_e$ are extremely complex and unlikely to have physical meaning. The importance of $\Nesep/\Netop$ could not have been elicited in \cref{sec:neuralnets} as the location of the separatrix (and hence the value of $\Nesep$) was omitted from the training information of the neural networks. This importance of $\Nesep/\Netop$ must complement the relationship this ratio bears with the pedestal pressure \citep{Frassinetti_2021}.

\subsection{Physical Implications of $\eta_e$ Results}\label{ssec:etaDiscussions}

The distribution of experimental values of $\eta_e$ (\cref{fig:etaDistributions}) over the steep-gradient region and their agreement with the best-fit $\eta_e$ values over the pedestal region in \cref{fig:etaBest}(b) represent an important finding. These best-fit values of $\eta_e$ are nominally higher than $2$, with a considerable spread. 

One can interpret the results of this section as providing physical insight into the steep-gradient-region turbulence in several alternative ways:
\begin{concl}
	\item [(i) ] the turbulence can be described by a quasilinear model strongly dominated by the linear picture of electrostatic and electromagnetic modes, but the true linear threshold $\etacr$ differs from the solution of \citet{Jenko_lin} by an order-unity value;
	\item [(ii) ] the turbulence saturates in a non-linear \quotemarks{Dimits}-like state \citep{Dimits_simulation_2000,Ivanov_Dimits,Colyer_DimitsETG_2017}, and the transport level increases abruptly once some non-linear threshold $\eta_{e,\mathrm{Dimits}}$ is surpassed; based on our results, we expect that $\eta_{e,\mathrm{Dimits}}$ would be higher than the linear $\etacr$ by an $\mathcal{O}(1)$ amount;
	\item [(iii) ] the turbulence causes a moderately (rather than near-infinitely) \quotemarks{stiff} heat transport \citep{Field_Stiff}, and, given the separatrix loss power $\Psep$ in JET-ILW pulses, it results in a value of $\eta_e$ that is greater than $\etacr$ by an $\mathcal{O}(1)$ amount.
\end{concl}  

In \cref{ssec:etaDist}, we also identified and discussed the stark difference between the steep-gradient and density-top regions. Explaining this difference demands a theoretical model that accounts for qualitative changes in the turbulent behaviour with the change of the density gradient $\RLNe$. Such changes could be the result of the contribution of toroidal-ETG modes to the turbulent transport, as recently found by \citet{Chapman_toroidal_2024} and \citet{Krutkin_toroidal_2025}. We will discuss gyrokinetics-inspired models \citep{Guttenfelder_2021,Chapman-Oplopoiou_2022,Hatch_RedModels} that use the turbulent heat flux to reconstruct the $T_e$ profiles in \cref{sec:heatFluxModels}, but before that, in \cref{sec:alpha}, we will construct and assess a simpler scaling prescription that allows for $\eta_e$ to adjust over the pedestal region.

	\section{Scaling Exponent $\alpha$ and Pedestal Reconstruction}\label{sec:alpha}

 The relationship between $\eta_e$ and $\RLNe$ shown by the dashed orange line in \cref{fig:etaDistributions}(b) suggests that a power-law scaling between the gradients can effectively describe the pedestal. Furthermore, a clear characteristic of the plasma gradients in \cref{fig:RLTeRLNe} is that $\RLTe$ monotonically increases as $\RLNe$ does. This increase can perhaps be described by a power-law relationship of the form 
 \begin{equation} \label{eq:polyPred} 
 	\frac{R}{\LTe}  = A \left(\frac{R}{\LNe}\right)^\alpha,
 \end{equation}
where $\alpha$ and $A$ are parameters to be measured (and theorised about). To analyse the behaviour of the exponent $\alpha$ at different locations in the pedestal, we formally define 
\begin{equation}\label{eq:alphaDefinition}
	\alpha(R) = \der {\ln{(\RLTe)}}{\ln{(\RLNe)}},
\end{equation}
where $R$ is the radial location at which the gradients are evaluated. Following the $\RLTe$ vs. $\RLNe$ loci for any given pulse in \cref{fig:RLTeRLNe}, we find that $\alpha(R)$ decreases from $\psitop{T_e}$ (low-gradient region of \cref{fig:RLTeRLNe}a) to $\psisep$ (dark-red lines in \cref{fig:RLTeRLNe}b). 

In \cref{ssec:alphaDist}, this local $\alpha(R)$ will be calculated for each pulse in the database. Then, in \cref{ssec:alphaPredictions}, we will fit $\alpha$-based models to the database and reconstruct the electron-temperature pedestals, similar to our analysis of $\eta_e$ in \cref{sec:etae}.

\subsection{Distributions of the Scaling Exponent \ensuremath{\alpha}} \label{ssec:alphaDist}

\begin{figure}
	\centering
	\includegraphics[width=\textwidth]{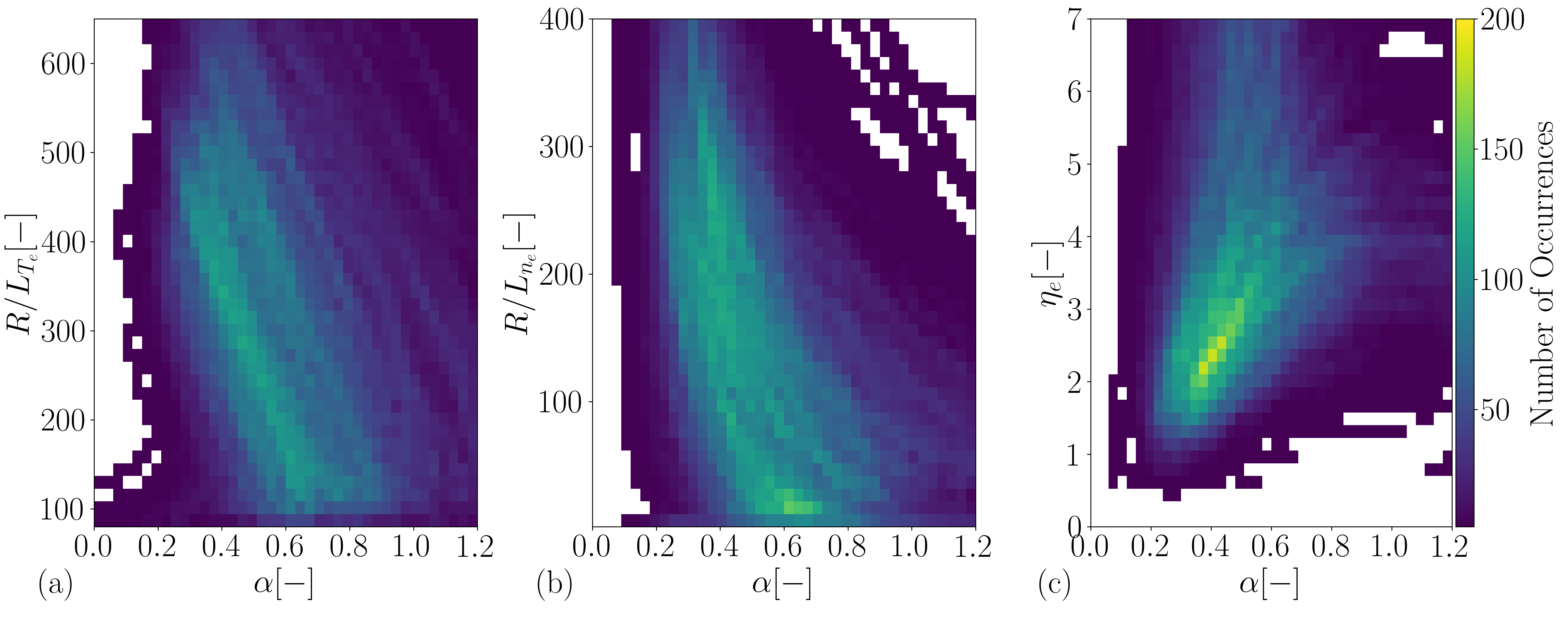}
	\caption{\centering {Distribution of the exponent $\alpha$, defined in \cref{eq:alphaDefinition}, across the database as a function of (a) $\RLTe$, (b) $\RLNe$, (c) $\eta_e$. All data is obtained from the pedestal region, between $\psitop{T_e}$ and $\psi_{Sep}$.}}	\label{fig:alphaDistributions}
\end{figure}

\begin{figure}
	\centering
	\includegraphics[width=\textwidth]{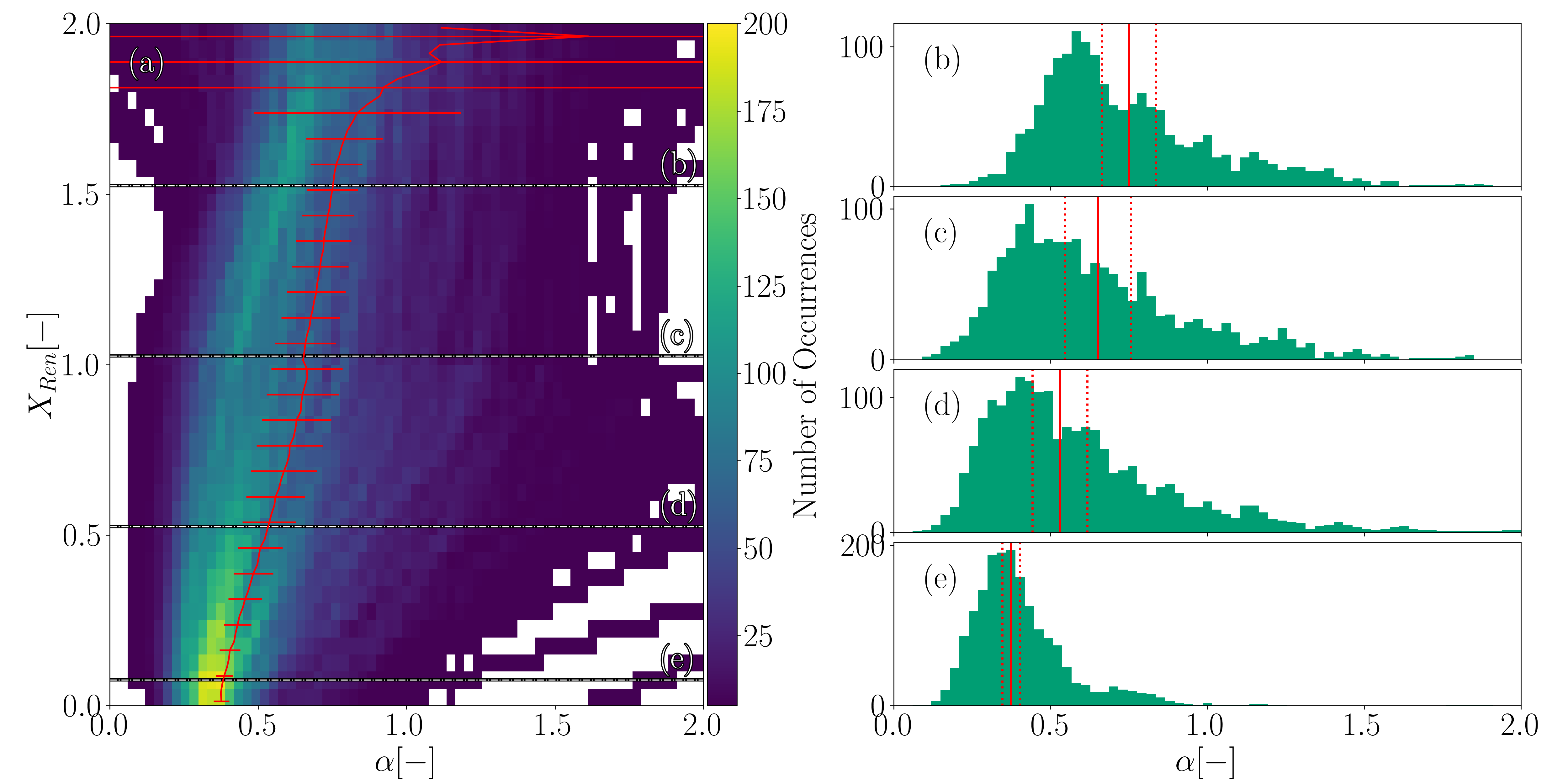}
	\caption{\centering { (a) Distribution of $\alpha$, as defined in \cref{eq:alphaDefinition}, conditioned on $\psirenorm$. The red lines represent the mean and variance of $\alpha$ at each $\psirenorm$. (b-e) Cross-sections of the histogram, at \mbox{(b)  $(\psitop{T_e}+\psitop{n_e}) / 2$}, (c) $\psitop{n_e}$, (d) $\psisteep$, and (e) $\psisep$, corresponding to white dash-dotted lines in panel (a). In each of these cross-sections, the continuous red lines represent the mean value of $\alpha$, and the dotted red lines represent the values one standard deviation away from the mean. }}
	\label{fig:alphaslice}
\end{figure}

In \cref{fig:alphaDistributions}, we show the distribution of the exponent $\alpha$ conditioned on the gradients $\RLTe$, $\RLNe$, and their ratio $\eta_e$. The values of $\alpha$ at the separatrix are clustered around $\approx 0.33$ (as seen in \cref{fig:alphaDistributions}b at large $\RLNe$). Overall, the range of values taken by $\alpha$ is narrow, indicating that a power-law relationship between $\RLTe$ and $\RLNe$ with a constant exponent might indeed represent the profiles in the steep-gradient region correctly. The relationship between $\alpha$ and $\eta_e$ is shown in \cref{fig:alphaDistributions}(c), and it implies a clear correlation between increasing $\eta_e$ and increasing $\alpha$.

Rearranging the values of $\alpha$ using the renormalised radial coordinate $\psirenorm$ given by \cref{eq:psirenorm} gives us \cref{fig:alphaslice}. Its most important feature is the convergence of all profiles to $\alpha \approx 0.33$ at the separatrix ($\psirenorm$ approaching $0$ in \cref{fig:alphaslice}a). This coincides with the highest values of the normalised gradients $\RLTe$ (\cref{fig:alphaDistributions}a) and $\RLNe$ (\cref{fig:alphaDistributions}b). The exponent $\alpha$ commonly lies in the range of $0.3 - 0.5$ in the steep-gradient region, with a large spread towards $\psitop{n_e}$. The values of $\alpha$ remain order unity in the density-top region.

\subsection{Electron Temperature Reconstructions Using Constant Exponent $\alpha$} \label{ssec:alphaPredictions}

Because of the small variance of $\alpha$ in the pedestal region, it makes sense to explore the consequences of asserting that $\alpha$ is constant across the pedestal. This is conceptually similar to the analysis carried out in \cref{ssec:etaPredictions}, where $\alpha=1$ and $A\equiv \eta_e$, but now $\alpha \neq 1$ allows the physically relevant parameter $\eta_e$ to adjust over the pedestal (as a function of either of the gradients). The case for such an adjustment of $\eta_e$ across the pedestal is clear in \Cref{fig:etaDistributions}(a,b), where the orange lines represent fits of $\alpha$ to the distributions of $\eta_e$ vs. $\RLNe$ -- in the figure caption, these fits are presented as $\RLNe \propto \eta_e^\zeta$, where $ \zeta = (\alpha-1)^{-1} $. The dotted line with $\alpha \approx 0.1$ is the fit to the whole distribution of $\eta_e$ in \cref{fig:etaDistributions}(b), while the dashed line with $\alpha = 0.41$ (obstructed by the continuous orange line with $\alpha = 0.39$ obtained below as a best-fit for database reconstructions) is the fit to the highest-likelihood $\eta_e$ values in the histogram. It is clear that $\alpha\approx 0.1$ fails to reproduce the trend of pedestal profiles, whereas $\alpha = 0.41$ meaningfully matches the relationship between $\eta_e$ and $\RLNe$. Hence, using a constant $\alpha$ to describe a pedestal should lend enough flexibility to the relationship between $\RLTe$ and $\RLNe$. 

\begin{figure}
	\centering
	\includegraphics[width=\textwidth]{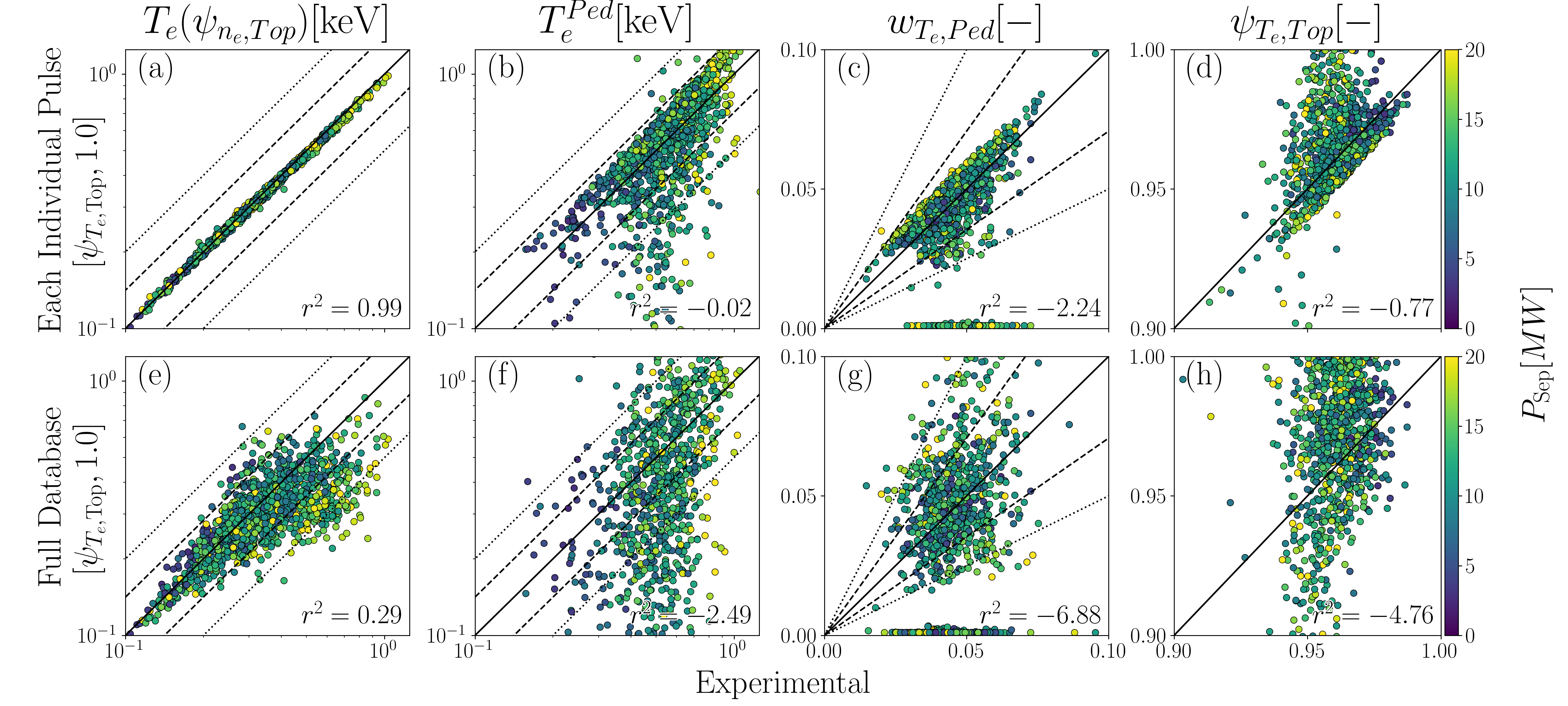}
	\caption{\centering { Database predictions obtained by integrating $\RLTe = A\left(\RLNe\right)^\alpha$ inwards from the separatrix, with (a-d) the best-fit $A$ and $\alpha$ for each pulse, and (e-h) best-fit $A = 50.6$ and $\alpha=0.39$ across the database. The fits are done over the pedestal region. The colour bar represents the separatrix loss power $\Psep$. The layout of the figure is the same as in \cref{fig:mlFullProfPred}.}}
	\label{fig:alphaResults}
\end{figure}

\begin{figure}
	\centering
	\includegraphics[width=\textwidth]{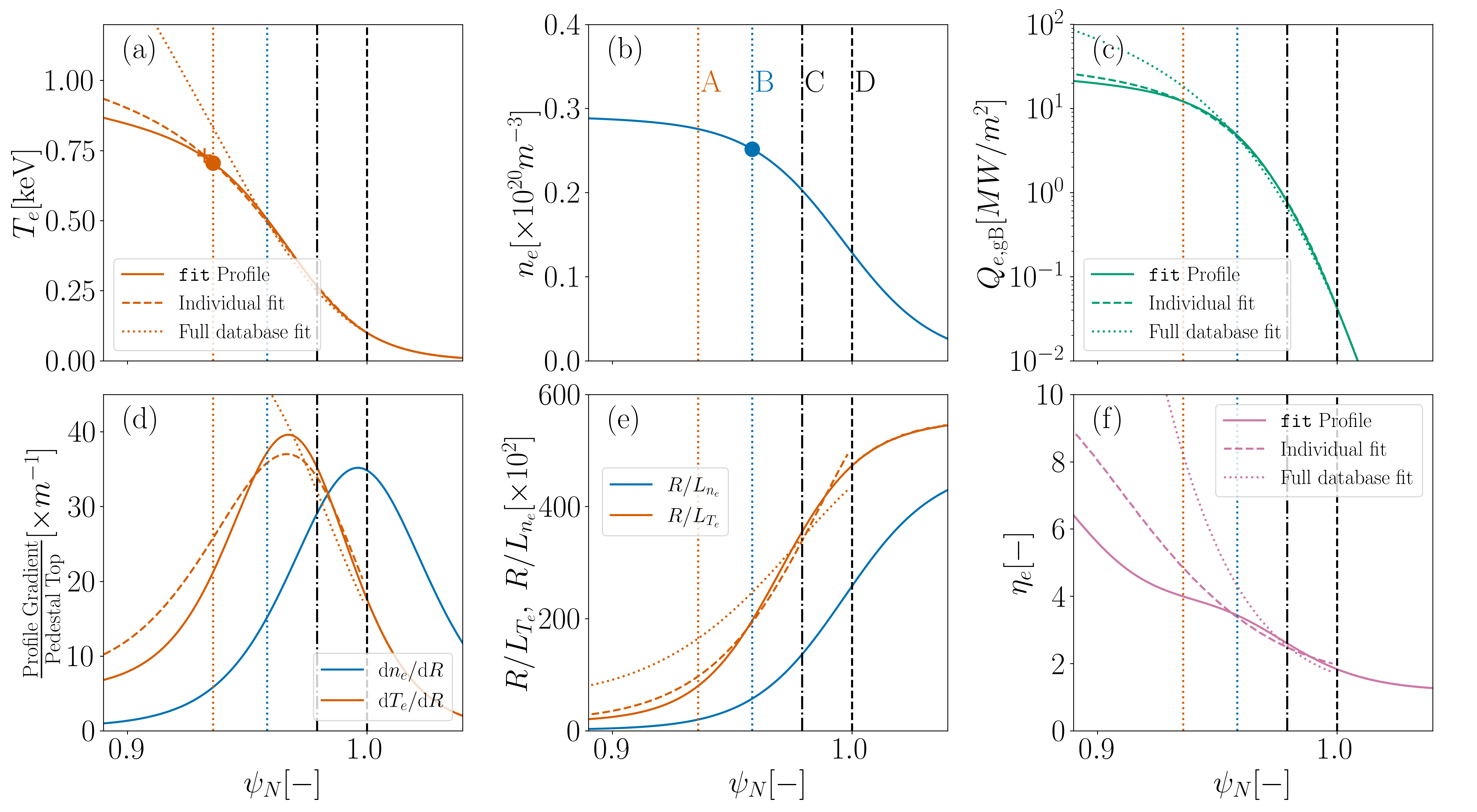}
	\caption{\centering {
			Pulse $\examplepulse$ $\fit$ profiles compared with predictions using the scaling \nolinebreak\cref{eq:polyPred}. The layout is the same as in \cref{fig:Gradients}. The $\fit$ profiles are shown as solid lines, while the predictions using the $A$ and $\alpha$ optimised individually for pulse $\examplepulse$ are shown as dashed liens dashed lines, and predictions using the $A$ and $\alpha$ optimised for the whole database are shown as dotted lines. Both fits are done over the interval $[\psitop{T_e}, \psisep]$. The vertical lines and dots represent the landmarks of the $\fit$ profiles (see \cref{ssec:locations,fig:Gradients}), while the crosses represent the pedestal top of the $T_e$ predictions.}}
	\label{fig:exampleVsAlpha}
\end{figure}

Thus, similar to our approach in \cref{ssec:etaPredictions}, we attempt to reproduce the pedestals using \cref{eq:polyPred} as a fitting formula, seeking the best values of $A$ and $\alpha$. \cref{fig:alphaResults} shows two database reconstructions that were calculated using the scaling \cref{eq:polyPred}. These reconstructions maximise the accuracy of the $T_e$ profile across the pedestal region, $\psi_N \in \left[\psitop{T_e}, \psisep=1.0\right]$, with:
\begin{predictions}
	\item[(a-d): ] $A$ and $\alpha$ fit individually for each pulse;
	\item[(e-h): ] $A=50.6$ and $\alpha=0.39$ fit across the database.
\end{predictions}
For these reconstructions in \cref{fig:alphaResults}, the values of $A$ and $\alpha$ are the best-fit values obtained using the optimisation method described in \cref{sappendix:bestFit}. Note that the value of $\alpha = 0.39$ in this method agrees with the fit to the highest-likelihood values of $\eta_e$ of \cref{fig:etaDistributions}(b).

\begin{figure}
	\centering
	\includegraphics[width=\textwidth]{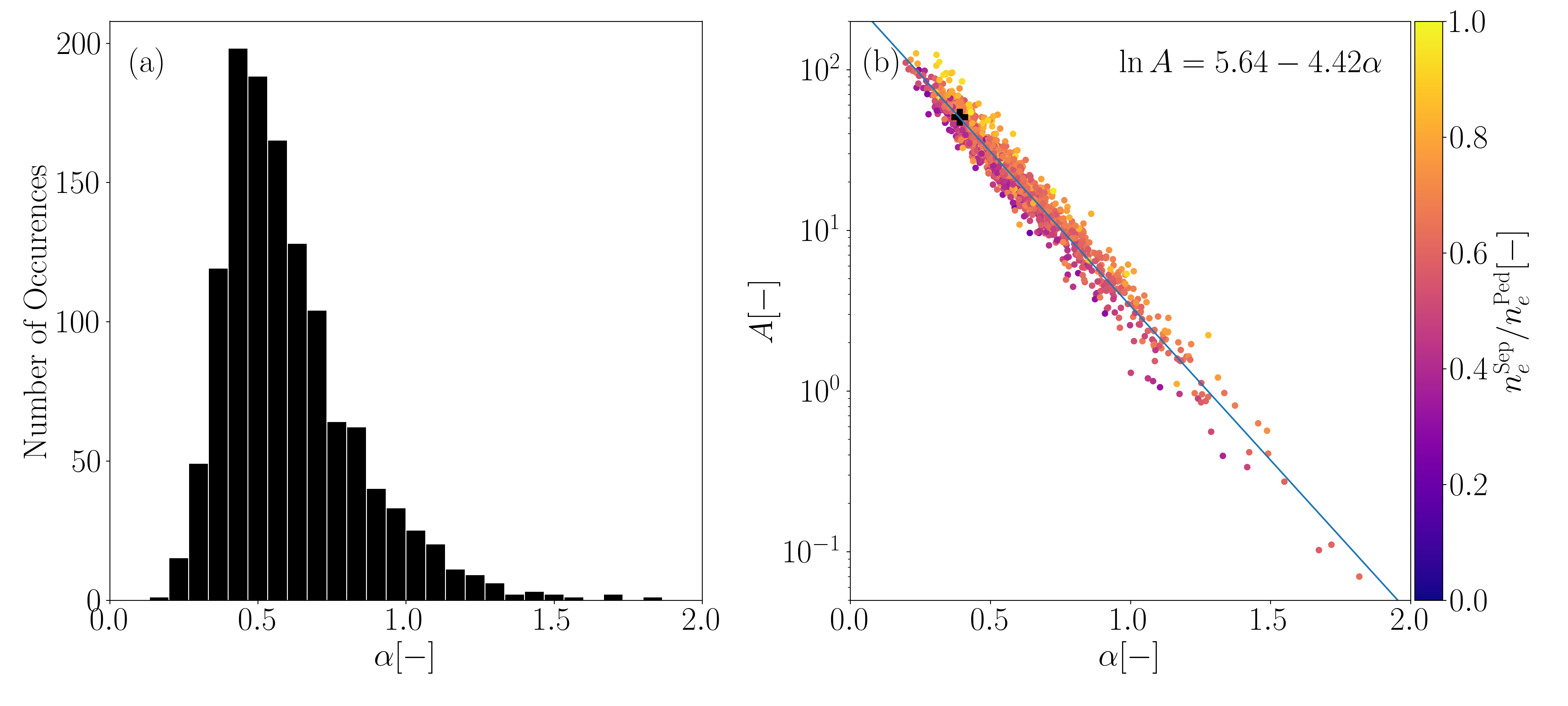}
	\caption{\centering { Values of $\alpha$ and $A$ that best predict the $T_e$ profiles individually across the pedestal region: (a) the distribution of values of fit $\alpha$, (b) the fit values of $\alpha$ vs $A$. The black cross represents the $A$ and $\alpha$ fit for the whole database (shown in \Cref{fig:alphaResults}e-h). The linear fit between $\ln A$ and $\alpha$ is shown as a blue line. }}
	\label{fig:alphaBest}
\end{figure}

The predictions obtained by allowing $\alpha$ and $A$ to be fit individually for each pulse (\Cref{fig:alphaResults}a-d) are very accurate in reproducing the pedestal region overall, as seen in \cref{fig:alphaResults}(a). Furthermore, the pedestal shapes are reconstructed well for a large proportion of the database, although for some pulses, the $T_e$ profiles continue increasing exponentially towards the core, so $\mtanh$ parameters do not describe these profiles well. This can be seen in \Cref{fig:alphaResults}(b-d), where a subset of the $\mtanh$ parameters is very far from the correct prediction.

This issue of unbounded $T_e$ profiles is more pronounced for the reproductions using a single combination of $A = 50.6$ and $\alpha = 0.39$ across the database. Since many reconstructed profiles do not resemble an $\mtanh$, the corresponding pedestal parameters are obtained poorly (\Cref{fig:alphaResults}f-h). However, the accuracy of these fits across the steep-gradient region (\cref{fig:alphaResults}e) is much better than the best-fitting $\eta_e$ predictions of \cref{ssec:etaPredictions} (\Cref{fig:etaResults}e,i), meaning that such an approach is more effective for the steep-gradient region than the methods in \cref{sec:etae}. This combination of $A$ and $\alpha$ is shown in \Cref{fig:etaDistributions}(a,b) as the continuous orange lines, recovering the fit performed directly to the distribution of $\RLNe$ and $\eta_e$ shown by the dashed line.

If $A$ and $\alpha$ are instead optimised  over the interval $\psi_N \in [0.9,1.0]$  for a database-wide fit, we obtain $A = 31.7$ and $\alpha =  0.46$ (not shown). This combination produces slightly degraded pedestals over the $\psi_N \in \left[\psitop{T_e}, 1.0\right]$ interval when compared to the combination in \Cref{fig:alphaResults}(e-h), and it does not solve the general issue of the exponential divergence of $T_e$ profiles. The combinations of $A$ and $\alpha$ optimised over the intervals $\left[\psitop{T_e}, 1.0\right]$ and $\left[0.9, 1.0\right]$ correspond to entries 36 and 38 of \cref{table:FitCoeff} in \cref{appendix:parameters}, where their accuracies over different ranges are compared with our other attempts at $T_e$-pedestal reconstructions in this study.

Examples of the profiles that constitute \cref{fig:alphaResults} are given in \cref{fig:exampleVsAlpha} for the pulse $\examplepulse$. The database-optimised fit (dotted lines) has a lower value of $\alpha$ than the optimal value for pulse $\examplepulse$, causing $\RLTe$ to vary less with $\RLNe$. Because of this, the database-optimised values give a large $\RLTe$ beyond the density-pedestal top and a diverging $T_e$ profile without a $\psitop{T_e}$ within the bounds shown in \cref{fig:exampleVsAlpha}. The optimal combination of $A$ and $\alpha$ for this individual pulse (dashed lines) results in a good $T_e$ pedestal and reproduces $\eta_e$ correctly between $\psitop{T_e}$ and $\psisep$.

The distributions of $A$ and $\alpha$ fit for each pedestal (corresponding to \Cref{fig:alphaResults}a-d) are shown in \cref{fig:alphaBest}. The key finding, manifest in \cref{fig:alphaBest}(b), is that, while we found that a two-parameter fit \cref{eq:polyPred} is much better than the one-parameter fit ($\alpha = 1$, $A = \eta_e$ in \cref{sec:etae}), the two parameters appear to be highly correlated. Their relationship is given by  $\ln A = 5.64 \pm 0.05 - (4.42 \pm 0.08) \alpha$, which allows for \cref{eq:polyPred} to be recast as
\begin{equation}\label{eq:oneParameterExponentFit}
	\frac{R}{\LTe} = (281 \pm 15) \left( \frac{1}{84 \pm 7} \frac{R}{\LNe} \right) ^ \alpha.
\end{equation}
The ranges for the numerical constants represent a $95\%$ confidence interval. This means that the fits performed in this Section in fact intrinsically have only one degree of freedom. We have found that a similar locus of fit parameters $A$ and $\alpha$, given by $ \ln A = 5.73 \pm 0.09 - (5.18 \pm 0.12) \alpha $, describes a subset of $273$ JET-C pulses from the EUROfusion Pedestal Database, selected based on criteria similar to those described in \cref{ssec:dataSelection}. These JET-C pulses are not otherwise included in this study.




Finding a relationship between $\alpha$ and physical parameters would represent a major success in modelling the pedestal steep-gradient region. However, we anticipate that a model like \cref{eq:oneParameterExponentFit} might be much harder to justify physically than other alternatives used in \cref{sec:etae,sec:heatFluxModels}. We also note that we have failed to find any correlation between the distribution of $\alpha$ and database parameters.

	\section {Heat-Flux Models}\label{sec:heatFluxModels}

As we saw throughout the parameter-combination scans of \cref{ssec:mlImportantPars}, the separatrix loss power $\Psep$ is strongly correlated with the $T_e$ pedestals. As also highlighted by the correlation matrix in \cref{fig:correlationMatrix}, the values of the $T_e$ pedestal profile at the pedestal-top locations $\Tetop$ and $\NetopTe$ are positively correlated with $\Psep$. We will now attempt to exploit this dependence of $T_e$ on $\Psep$ using models of heat transport to reconstruct $T_e$ pedestals.

This section shows the outcomes of $T_e$-pedestal-profile reconstruction using models that relate the gradients $\RLNe$ and $\RLTe$ to the turbulent heat flux transported in the electron channel. First, in \cref{ssec:etgTurb}, we build upon the discussion of linear instabilities in the pedestal given in \cref{ssec:gradientLitReview} and present the properties of the turbulent transport caused by ETG modes and other instabilities. This will provide context to the rest of our analysis. \Cref{ssec:heatFluxAssumptions} describes the calculation of the turbulent heat flux and its normalisation using database parameters. Using the models explained in \cref{ssec:etgTurb}, in \cref{ssec:heatFluxPredictions}, we reconstruct electron-temperature pedestals and assess the quality of these predictions. Our reconstructions use the parameter-optimisation methods described in \cref{appendix:predictionMethods}, and the results in this section represent an illustrative sub-set of the full set of models we assessed in \cref{appendix:heatFluxFits}.

\subsection{Electron-Scale Turbulence}\label{ssec:etgTurb}

The non-linear dynamics of ETG-driven turbulence in the steep-gradient region of pedestals are strongly influenced by the complex interactions between micro-instabilities that emerge as a result of the underlying plasma equilibria. Here, we briefly review the key findings from non-linear gyrokinetic simulations and their implications for pedestal turbulence.

\citet{Chapman-Oplopoiou_2022} carried out gyrokinetic simulations using the nominal gradients $\RLTe$ and $\RLNe$ evaluated at $\psisteep$ for four JET-ILW pedestals, which are also present in this study. Varying these gradients within the experimental confidence intervals showed that the electron-channel turbulent heat flux $\Qeturb$ increased with increasing $\RLTe$ and decreased with increasing $\RLNe$. This is consistent with the linear picture of the slab-ETG instability, where electron-temperature gradients drive the instability, while electron-density gradients suppress it. \citet{Chapman-Oplopoiou_2022} showed that a good fit to their results was the relationship
\begin{equation} \label{eq:chapmanModel}
 	\frac{\Qeturb}{\Qegb} = A  \left(\frac{R}{\LTe}\right)^2 \left(\eta_e -\etacr\right)^{\beta},
\end{equation}
where, as defined in \cref{ssec:exampleGradients}, $\Qegb = n_e T_e \vthe \rho_e^2 / R^2 $, and the best fit parameters were $A=0.85$, $\etacr = 1.28$, and $\beta = 1.32$. The simulated heat fluxes were also consistent with the fit found in simulations of DIII-D pedestals \citep{Guttenfelder_2021}: $A=1.5$, $\etacr = 1.4$, and $\beta =1$. Models of this form can also reproduce results such as those from the $\RLTe$ scans in \citet{Parisi_3dTurb_2022}: keeping $\eta_e$ constant yields $\Qeturb/\Qegb \propto (\RLTe)^2$, whereas keeping $\RLNe$ constant gives  $\Qeturb/\Qegb \propto (\RLTe)^4$. This behaviour would be matched by \cref{eq:chapmanModel} with $\beta=2$ when $\eta_e \gg \etacr$. 
 
The behaviour of the turbulent heat flux $\Qeturb$ parametrised by \cref{eq:chapmanModel} can be justified by the physical picture proposed by \citet{Field_Stiff}. In their model, the $(\RLTe)^2$ term is justified by replacing the radial size $R$ of the tokamak with the locally-relevant length scale $\LTe$ in the gyro-Bohm normalisation of the heat flux:
\begin{equation} \label{eq:QeGBrescale}
 	\Qegb = n_e T_e \vthe \frac{\rho_e^2}{R^2} \;\;\; \rightarrow \;\;\; \Qegb^* = n_e T_e \vthe \frac{\rho_e^2}{\LTe^2},
\end{equation} 
i.e., the relevant scale associated with free-energy injection is $\LTe$, rather than $R$. This is a natural choice if turbulent transport is controlled by the slab-ETG instability. With the redefinition \cref{eq:QeGBrescale}, \cref{eq:chapmanModel} becomes 
\begin{equation} \label{eq:chapmanModel2}
 	\frac{\Qeturb}{\Qegb^*} =A \left(\eta_e - \etacr\right)^{\beta},
\end{equation}
and therefore, for $\beta>0$, the normalised turbulent heat flux is a function of how much the slab-ETG drive (expressed through $\eta_e$) exceeds the (non-linear) threshold $\etacr$. This is an essential characteristic that sets this model of turbulent transport apart from the models using a fixed $\eta_e$ (\cref{sec:etae}) or a power law for the relationship between $\RLTe$ and $\RLNe$ (\cref{sec:alpha}). Those models amounted to assuming that the turbulence was marginal and, therefore, \quotemarks{pinned} to a specific value of $\etacr$, constant across the pedestal or varying in a prescribed manner. The model \cref{eq:chapmanModel} allows the profiles to reach beyond this marginal regime.

This type of modelling of turbulent transport is common in the context of ITG turbulence in core plasmas and has been reported to reproduce measurements correctly \citep[e.g.,][]{Mantica_ITG,Garbet_ITGcoeff}. In the case of ITG turbulence, the critical value of the ion temperature gradient is usually expected to be larger than the linear stability threshold because the non-linear saturation just above the linear threshold is in the \quotemarks{Dimits} regime, in which strong, non-linearly saturated zonal flows suppress heat transport \citep{Ivanov_Dimits,Dimits_simulation_2000}. A similar behaviour is sometimes found in simulations of ETG turbulence, where the long-term onset of zonal flows suppresses the heat transport caused by ETG \citep{Colyer_DimitsETG_2017}. Thus, it is conceivable (although, at this stage, far from certain) that $\etacr$ is a non-linear threshold.

Note also that in the limit of strong drive ($\eta_e \gg \etacr$), \cref{eq:chapmanModel} with $\beta = 1$ leads to the scaling $Q / \Qegb \propto (R/\LTe)^3$, which results from a simple scaling theory \citep{Barnes_CritBal,Adkins_ScaleInvariance}\footnote{\citet{Barnes_CritBal} proposed this scaling for ITG turbulence, but its validity has recently been questioned by \citet{Nies_ZFsaturation_2024}. It is not clear whether a similar objection exists for ETG turbulence.}.

ETG turbulence in the pedestal is subject to further complexity \citep[see the review by][]{Ren_ETGreview_2024}. The structure of toroidal-ETG modes along the field lines \citep{Parisi_ETGdominance,Jenko_gyro_2009} leads to  non-linear coupling between ion-scale ($k_y \rho_i \sim 1$) and electron-scale ($k_y \rho_e \sim 1$) modes. This can lead to quenching of ETG turbulence under certain conditions \citep{Parisi_3dTurb_2022}. Other studies of the effects of toroidal-ETG on pedestal turbulence suggest that the contribution of toroidal-ETG amplifies transport \citep{Chapman_toroidal_2024,Krutkin_toroidal_2025}.

The multi-scale study of pedestal transport by \citet{Leppin_Complex} suggests that the modes that constitute the turbulence are different between the pedestal-top and the steep-gradient region. At the top, ion-scale trapped-electron modes (TEM) dominate the transport, whereas, in the steep-gradient region, they are suppressed as a result of high $\mathbf{E} \times \mathbf{B}$ shear \citep{Hahm_Burrell_Stabilisation_1995}. Steep-gradient transport is, therefore, dominated by slab-ETG and micro-tearing modes, as also confirmed by \citet{Kotschenreuther_transportCulprits} and \citet{Chapman-Oplopoiou_2022}. Electromagnetic effects were also found to be important for the saturation of pedestal turbulence in the comparison carried out by \citet{Dicorato_complexTurb_2024}. In spite of this, they found that turbulent transport had a much higher contribution from ETG modes for the Type-I ELMy pedestal when compared to the small-ELM pedestals, supporting our development of models centred  around ETG transport.


In what follows, we extend our considerations to other heat-flux models that have been obtained by analysing a large dataset of gyrokinetic simulations by \citet{Hatch_RedModels}. Their Table 1 contains several symbolic-regression fits to the heat transport in those simulations. These relationships are explicitly stated [see \cref{eq:Hatch123,eq:Hatch5}] and tested in \cref{sappendix:equations}. Optimised versions of these relationships will also be used in our analysis below in the forms of \cref{eq:ha1_22,eq:ha5_30}.
 
Further extrapolations obtained from simulation datasets are those found by \citet{Farcas_transportEq_2024},
\begin{equation}\label{eq:farcasModel}
	\frac{\Qeturb}{\Qegb } \propto \sqrt\frac{m_e}{m_i} \left(\frac{R}{\LTe}\right) ^{1.40} \left(\eta_e - 1\right)^{0.79} q^{-0.51} \beta_e^{-0.87} \lambda_D^{-0.51},
\end{equation}
and by \citet{Hatch_modeling_2024},
\begin{equation}\label{eq:hatch2024Model}
	\frac{\Qeturb}{\Qegb } \propto \sqrt\frac{m_e}{m_i} \left(\der{\ln T_e}{\psi}\right) ^{2} \left(\eta_e - 1\right) \eta_e ^{1.57} \tau_e ^ {-0.5} \lambda_D^{-0.4}.
\end{equation}
Here $\lambda_D$ is the electron Debye length normalised with respect to the electron gyroradius \nolinebreak $\rho_e$ in \citet{Farcas_transportEq_2024} and to the ion sound radius $\rho_s = \sqrt{m_i T_e }/eB$ in \citet{Hatch_modeling_2024}. These are relationships where the properties of the magnetic field appear via the safety factor $q$, the electron beta $\beta_e$, and the normalised temperature gradient with respect to the poloidal flux, $1/\tilde{L}_{T_e} \equiv \dertxt{\ln T_e}{\psi}$. Because we do not have a good magnetic-equilibrium reconstruction for our pedestals, we omit these models from our study.

\subsection{Assumptions for the Turbulent Heat Flux}\label{ssec:heatFluxAssumptions}

The non-linear dynamics of ETG turbulence provide a pathway to connect the separatrix loss power $\Psep$ with pedestal profiles. Using the scalings derived from gyrokinetic simulations, $T_e$ pedestal profiles will be reconstructed in \cref{ssec:heatFluxPredictions} by assuming that the heat flux is dominated by the electron-channel turbulent transport, $Q(R) = \Qeturb $, and that $Q$ matches the experimentally determined separatrix loss power $\Psep$:
\begin{equation}
	Q(\Rsep) = \frac{ \Psep }{ S_{\mathrm{Sep} }} ,
\end{equation}
where $S_{\mathrm{Sep}} \approx 135 \mathrm{m}^2$ is the surface area of the separatrix. Further to this, we argue that, given the small width of the pedestal $w_\mathrm{Ped} = \mathcal{O}(\mathrm{cm})$ compared to the device length scales $R$ or $a = \mathcal {O}(\mathrm{m})$, the variation in the area of the flux surface as a function of $R$ is negligible. We also assume that the plasma profiles are in steady state \citep[see \cref{sec:geometry} and][]{Frassinetti_Database}, so the rate of change of the thermal energy over the small extent of the pedestal is small. Therefore, we approximate the turbulent heat flux as constant over the pedestal:
\begin{equation}
	\Qeturb(R) \approx \frac{ \Psep }{ S_{\mathrm{Sep} }} .
\end{equation}

To calculate the gyro-Bohm normalisation $\Qegb$ of the turbulent heat flux [see \cref{eq:QeGBrescale}], we use a locally approximated value of the magnetic field: 
\begin{equation}
	B(R) = \sqrt{B_\theta(R)^2 + B_\phi(R)^2} \approx B_\mathrm{Vac} \frac{R_0}{R} \sqrt{1 - \frac{1}{q_{95}^2}} \approx 0.66 B_\mathrm{Vac},
\end{equation}
where $B_\theta$ and $B_\phi$ are, respectively, the poloidal and toroidal components of the magnetic field, $B_\mathrm{Vac}$ is the vacuum magnetic field on the axis of JET-ILW (shown as $B$ in \cref{fig:ParamHists,table:EngPars}), $q_{95}$ is the safety factor evaluated at $\psi_N = 0.95$, and $R_0 = 0.92 \; \mathrm{m}$ is the radius of the axis of JET-ILW. This should remain a good approximation over the extent of the pedestal, irrespective of the divergence of the safety factor at the separatrix, $q \rightarrow \infty$ as $R \rightarrow \Rsep$. This approximation also agrees well with what one would obtain by using $\Iplasma$ and Ampere's law to estimate $B_\theta(R)^2$. These approximations fall short of the precision of the magnetic field calculated via the MHD equilibrium, but they are good enough for the purpose of gyro-Bohm normalisation.

Because of the approximations that we use, uncertainties remain as to the exact contribution of additional transport mechanisms such as neoclassical transport \citep{Guttenfelder_2021} and ELM-induced losses \citep{Field_Exhaust}. Order-unity errors of the heat flux do not significantly affect the reconstruction of pedestals because of the \quotemarks{stiffness} of pedestal transport described in \cref{ssec:etgTurb}. This property means that small changes in $\RLTe$ in the models introduced below, \crefrange{eq:ch2_8}{eq:ha5_30}, can lead to large changes in heat flux. Conversely, a large inaccuracy in the heat flux results in a comparatively small change in $\RLTe$, meaning that pedestal reconstructions are robust. This is illustrated in \cref{fig:chapmanParamScan} (\cref{appendix:chapmanScan}), where an order-unity change in $\Qeturb / \Qegb$ [represented in \cref{fig:chapmanParamScan}(d) by a change in $A$] has only a small effect on $\RLTe$ and the reconstructed $T_e$ profile.

\subsection{Electron-Temperature Reconstruction Using Heat-Flux Models}\label{ssec:heatFluxPredictions}

Using the methods outlined in \cref{appendix:predictionMethods}, we optimise the free parameters in \cref{eq:chapmanModel} and the general forms of the heat-flux models from \citet{Hatch_RedModels} given in \cref{eq:Hatch123,eq:Hatch5}. With these, we then reconstruct the $T_e$ profiles.

\begin{figure}
	\centering
	\includegraphics[width=\textwidth]{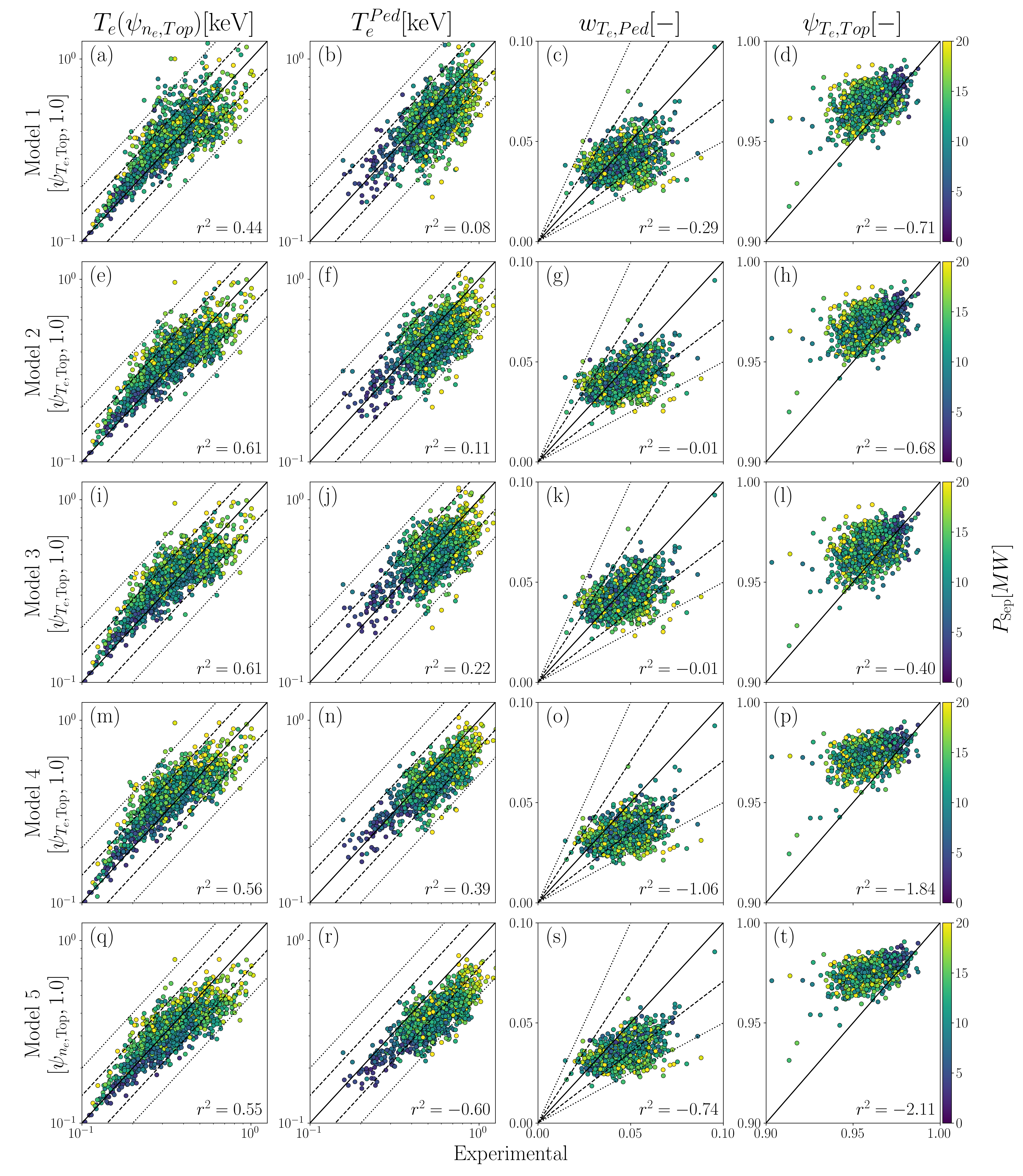}
	\caption{\centering { Database predictions obtained by inverting heat-flux models for $\RLTe$ and integrating inwards from the separatrix. The models use best-fit parameters over the pedestal region (a-p) [models \crefrange{eq:ch2_8}{eq:ha1_22}] and over the steep-gradient region (q-t) [model \cref{eq:ha5_30}]. The colour bar represents the separatrix loss power $\Psep$. The plots follow the same layout as in \cref{fig:mlFullProfPred}.}}
	\label{fig:heatFluxResults}
\end{figure}

\cref{fig:heatFluxResults} shows the \mtanh parameters that describe such reconstructions using models optimised over the pedestal region $\psi_N \in [\psitop{T_e},1]$ [panels (a-p)] and the steep-gradient region $\psi_N \in [\psitop{n_e},1]$ [panels (q-t)]. The exact models that we used correspond to the panels of \cref{fig:heatFluxResults} as follows: 
\begin{predictions}
	\item[(a-d): ] model \cref{eq:chapmanModel} with fixed $\beta=1$ and $\etacr=1.4$, and the value of $A=0.95$ optimised over the pedestal region $\psi_N \in [\psitop{T_e},1]$; this corresponds to entry 8 in \cref{table:FitCoeff}:
	\begin{equation} \label{eq:ch2_8}
		\mathrm{Model \; 1}: \;\; \frac{\Qeturb}{\Qegb} = 0.95 \left(\frac{R}{\LTe}\right) ^ 2 \left( \eta_e - 1.4 \right) ;
	\end{equation}

	\item[(e-h): ] model \cref{eq:chapmanModel} with fixed $\beta=1$ and the values of \quotemarks{$\etacr$}$=-0.55$ and $A=0.29$ optimised over the pedestal region $\psi_N \in [\psitop{T_e},1]$; this corresponds to entry 17 in \cref{table:FitCoeff}:
	\begin{equation} \label{eq:ch2_17}
		\mathrm{Model \; 2}: \;\; \frac{\Qeturb}{\Qegb} = 0.29 \left(\frac{R}{\LTe}\right) ^ 2 \left( \eta_e + 0.55 \right) ;
	\end{equation}

	\item[(i-l): ] model \cref{eq:chapmanModel} with the values of \quotemarks{$\etacr$}$=-8.6$, $\beta=3.75$, and $A=10^{-3}$ optimised over the pedestal region $\psi_N \in [\psitop{T_e},1]$; this corresponds to entry 14 in \cref{table:FitCoeff}:
	\begin{equation} \label{eq:ch1_14}
		\mathrm{Model \; 3}: \;\; \frac{\Qeturb}{\Qegb} = 10^{-3} \left(\frac{R}{\LTe}\right) ^ 2 \left( \eta_e + 8.6 \right) ^ {3.75} ;
	\end{equation}

	\item[(m-p): ] model \cref{eq:Hatch123} from \citet{Hatch_RedModels} with the values of the numerical constants optimised over the pedestal region $\psi_N \in [\psitop{T_e},1]$; this corresponds to entry 22 in \cref{table:FitCoeff}:
	\begin{equation} \label{eq:ha1_22}
		\mathrm{Model \; 4}: \;\; \frac{\Qeturb}{\Qegb} = 9.64 \frac{R}{\LTe}  \left( \eta_e ^ {2.92} + 16.02 \right) ;
	\end{equation} 

	\item[(q-t): ] model \cref{eq:Hatch5} from \citet{Hatch_RedModels} with the values of the numerical constants optimised over the steep-gradient region $\psi_N \in [\psitop{n_e},1]$; this corresponds to entry 30 in \cref{table:FitCoeff}:
	\begin{equation} \label{eq:ha5_30}
		\mathrm{Model \; 5}: \;\; \frac{\Qeturb}{\Qegb} = 0.01 \frac{R}{\LTe}  \left(   \eta_e  + 8.86 \right) ^{4.77} .
	\end{equation}

\end{predictions}

\begin{figure}
	\centering
	\includegraphics[width=\textwidth]{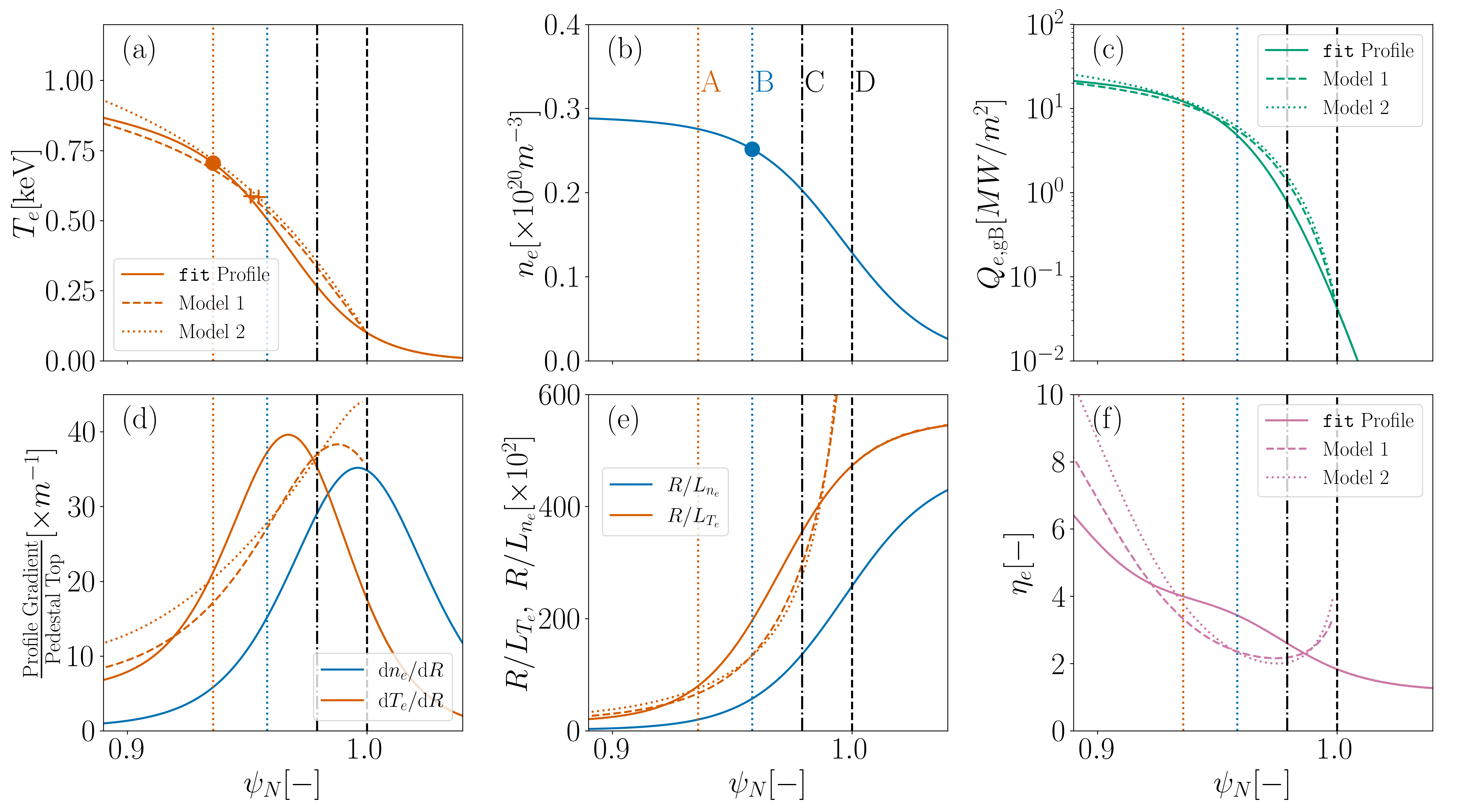}
	\caption{\centering {
			Pulse $\examplepulse$ $\fit$ profiles compared with predictions using heat-flux models 1 \cref{eq:ch2_8} and 2 \cref{eq:ch2_17}. 
			The layout is the same as in \cref{fig:Gradients}. 
			The $\;\fit$ profiles are shown as solid lines, with the the predictions using the heat-flux models are shown as dashed [model \cref{eq:ch1_14}] and dotted [model \cref{eq:ch2_8}] lines.
			The fit is done over the interval $[\psitop{T_e}$, $\psisep]$. The vertical lines and dots represent the landmarks of the $\fit$ profiles (see \cref{ssec:locations,fig:Gradients}), while the crosses represent the pedestal top of the $T_e$ predictions.}}
	\label{fig:exampleVsChapman}
\end{figure}

\begin{figure}
	\centering
	\includegraphics[width=\textwidth]{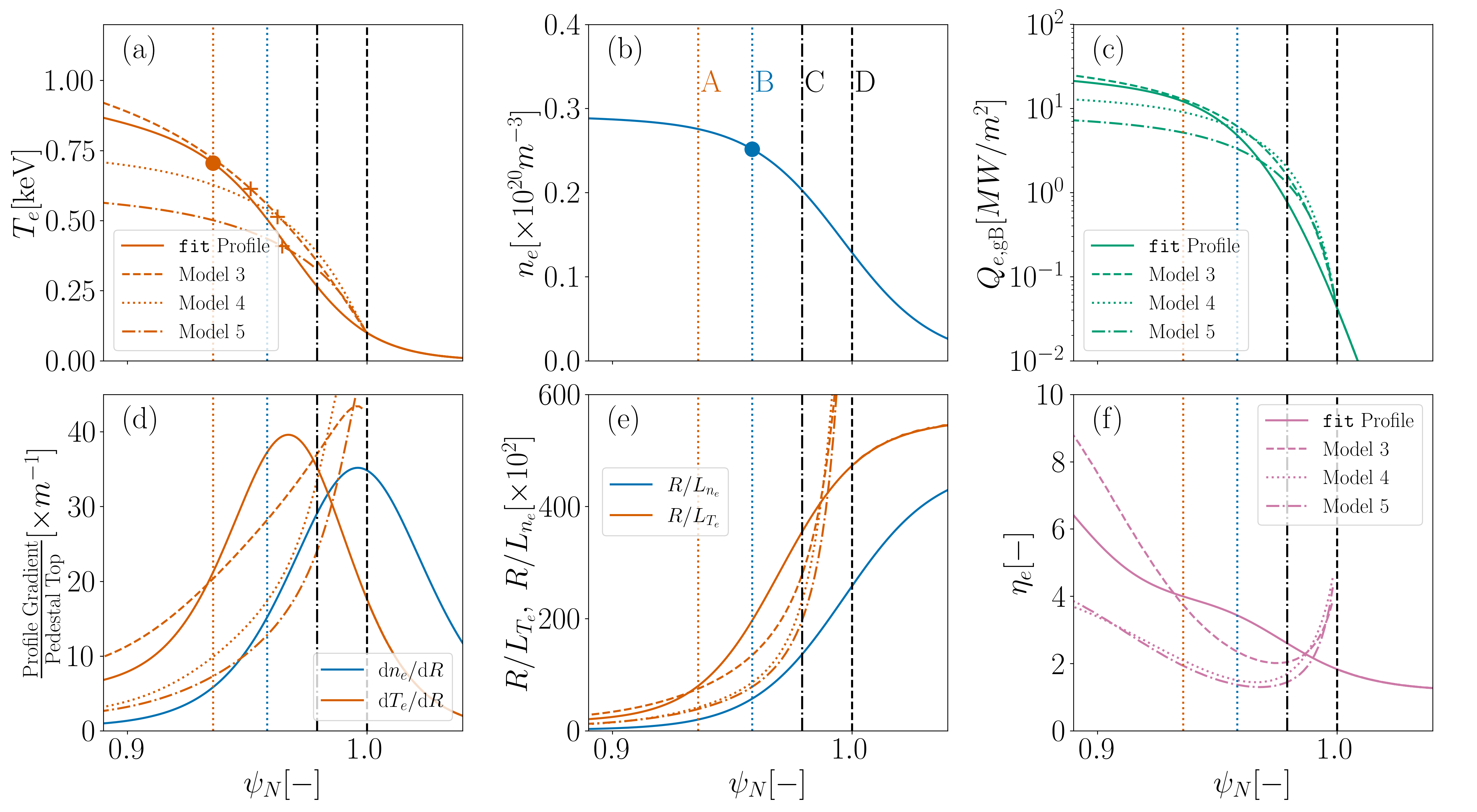}
	\caption{\centering {
			Same as \cref{fig:exampleVsChapman}, but for predictions using heat-flux models 3 \cref{eq:ch1_14}, 4 \cref{eq:ha1_22}, and 5 \cref{eq:ha5_30}. 
			The $\;\fit$ profiles are shown as solid lines, while the predictions using different heat-flux models are shown as dashed [model \cref{eq:ch2_17}],  dotted [model \cref{eq:ha1_22}], and  dash-dotted [model \cref{eq:ha5_30}] lines.
			The fit is done over the intervals $[\psitop{T_e}$, $\psisep]$ for models \cref{eq:ch1_14,eq:ha1_22}, and $[\psitop{n_e}$, \nolinebreak $\psisep]$ for model \cref{eq:ha5_30}.}}
	\label{fig:exampleVsHeatFluxOptimised}
\end{figure}

A more comprehensive list of accuracy metrics for various heat-flux models is given in \cref{table:FitCoeff} of \cref{appendix:heatFluxFits}. There, we present the performance of the exact models from \citet{Chapman-Oplopoiou_2022} and \citet{Hatch_RedModels}: those given by \crefrange{eq:ch2_8}{eq:ha5_30} and their variations that minimise the mean-squared difference between the reconstructed and the \fit $T_e$ profiles. We also compare them to the results of \cref{sec:neuralnets,sec:etae,sec:alpha}. 

We note that the optimised prefactor $A = 0.95$ of \cref{eq:chapmanModel} over the steep-gradient region is in good agreement with the value of $A = 0.85$ proposed by \citet{Chapman-Oplopoiou_2022}. 

Model \cref{eq:ch2_8} maintains the physical intuition of ETG-based turbulent transport outlined in \cref{ssec:etgTurb} and manages to capture the shapes of pedestals much better than the simplistic models of \cref{sec:etae,sec:alpha}. Model \cref{eq:ch2_17} is a further improvement to all metrics of pedestal shape, although the optimisation of $\etacr$ yields a negative value over the ranges of the pedestal or steep-gradient region (see entries 12, 14, 16, and 17 of \cref{table:FitCoeff}). The improvement in reconstruction accuracy by setting this value to be negative casts doubt on the idea of a turbulent threshold $\etacr$. Models \cref{eq:ch2_8,eq:ch2_17} share two systematic limitations: $\RLTe$ is over-predicted close to the separatrix and the reconstructed $\psitop{T_e}$ tracks the experimental $\psitop{n_e}$. The inaccuracy in $\RLTe$ comes as a result of the following process, illustrated in \cref{fig:exampleVsChapman}: the large value of $\Qeturb / \Qegb$ near the separatrix yields a large $\RLTe$ (panel d), and this then quickly increases the value of the $T_e$ profile as the integration goes on, reducing the value of $\Qeturb / \Qegb$ faster than the experimental profile suggests (panel c). This means that in the steep-gradient region of the $T_e$ pedestal, the temperature gradient is always over-predicted, with an over-estimated $\RLTe$ at the separatrix. Further, the increase of $n_e$ away from the separatrix increases $\Qegb$, which reduces $\Qeturb / \Qegb$. This produces the shape of the $T_e$ pedestal, by switching to low values of $\RLTe$ away from the separatrix, and, in the process, causing $\psitop{T_e}$ to match $\psitop{n_e}$ [see \Cref{fig:heatFluxResults}(d,h,l) and \cref{fig:exampleVsChapman}(a)]. 

Seeking optimal values of the exponent $\beta$ in \cref{eq:chapmanModel} yields large exponents for the term containing the gradient ratio $\eta_e$ (entries 13 and 14 of \cref{table:FitCoeff}). In spite of this, the parameter scan shown in \cref{fig:chapmanParamScan} (in \cref{appendix:chapmanScan}) suggests that reconstructions using \nolinebreak \cref{eq:chapmanModel} are not very sensitive to the exponent $\beta$ and that the corresponding reduction of the predicted values of $\RLTe$ are compensated by the decrease in the prefactor \nolinebreak $A$ \nolinebreak to \nolinebreak $10^{-3}$ (effectively increasing the turbulent heat flux). This insensitivity of the qualitative shape of predictions to the exponent $\beta$ is also present for models \cref{eq:ha1_22,eq:ha5_30}. 

The predictions using models \cref{eq:ha1_22,eq:ha5_30} \citep[derived from ][]{Hatch_RedModels} give improved $r^2$ values of the predictions of $\NetopTe$ and $\Tetop$ [see \Cref{fig:heatFluxResults}(m,n,q,r)] when compared to those obtained from model \cref{eq:ch2_8}. However, the shape of pedestals suggests that there is little variation in $\psiwid{T_e}$ and $\psitop{T_e}$ [see \Cref{fig:heatFluxResults}(o,p,s,t)], and that the reconstructed pedestals are very thin and pushed towards the separatrix. This signals the importance of a larger exponent of the temperature gradient in such models [such as $(\RLTe)^2$ in \cref{eq:chapmanModel}], since the use of $(\RLTe)^1$ in \cref{eq:ha1_22,eq:ha5_30} [corresponding to the models \cref{eq:Hatch123,eq:Hatch5} further discussed in \cref{appendix:heatFluxFits}] demands a large value of the temperature gradient in order to balance the magnitude of $\Qegb / \Qeturb$ near the separatrix. This temperature gradient produces outwards-shifted $T_e$ pedestals, as illustrated in \cref{fig:exampleVsHeatFluxOptimised}. The sheer magnitude of $\RLTe$ near the separatrix also means that there is little sensitivity of the \mtanh parameters to the $n_e$ profile for models \cref{eq:ha1_22,eq:ha5_30}.

Another important common characteristic of our fit parameters is the embarrassing tendency of $\etacr$ to become negative. This is manifest in \cref{eq:ch1_14,eq:ch2_17,eq:ha1_22,eq:ha5_30} and in the majority of optimisations in \cref{table:FitCoeff}, and we discuss this at length in \cref{appendix:heatFluxFits}. While the sign of $\etacr$ does not have a dramatic effect on the reconstructed $T_e$ profiles [see \Cref{fig:heatFluxResults}(a-h) and \cref{fig:exampleVsChapman}], we interpret this result as an indication that either the forms of the models for $\Qeturb$ are inaccurate for the data, or that the physical argument for $\Qeturb \propto \eta_e - \etacr$ for some \textit{physical} threshold $\etacr$ is not supported by the data.

\subsection{Discussion of Heat-Flux Model Results}
In this section, we have obtained proof that the models of \citet{Guttenfelder_2021} and \citet{Chapman-Oplopoiou_2022} perform well in reproducing the $\mtanh$ parameters of pedestals (\cref{fig:heatFluxResults}), alongside matching the experimental profile well over the regions with steep gradients (\cref{table:FitCoeff}). 

A thorough analysis of the best gyrokinetics-inspired turbulent transport models has confirmed the necessity of $\RLTe$ appearing with an exponent larger than unity in the expression for $\Qeturb/\Qegb$, as also recognised by models \cref{eq:hatch2024Model,eq:farcasModel}. We have also found a unanimously negative sign of the constant $\etacr$, however, this physically dubious feature is not demanded strongly by the predictions: forcing $\etacr$ to be positive does not severely impact prediction performance.

We note that all of our pedestal reconstructions using optimised heat-flux models generally produce similar $\mtanh$ parameters. The minimisation of the sum of squared discrepancies between prediction and experiment also yields comparable reconstruction accuracy between fits done across the same range of the pedestal (entries 7-30 of \cref{table:FitCoeff}). These two outcomes mean that a significant part of our assessment of which transport models are best comes from the qualitative shape of the profiles such as those in \cref{fig:exampleVsChapman,fig:exampleVsHeatFluxOptimised} (as we discuss at the end of \cref{ssec:heatFluxPredictions}). 

Overall, the inclusion of the turbulent heat flux into our pedestal-reconstruction methods addresses to a noticeable degree the shortcomings of the \quotemarks{marginal} models that used database-wide parameters in \cref{sec:etae,sec:alpha}. This said, the fact that the reconstructed $T_e$ pedestals inherit the shape of the $n_e$ pedestals is a sign that such models require more complexity in order to capture the density-top region correctly, and, hence, the shift between the $T_e$ and $n_e$ pedestals \citep{Frassinetti_Database}. This is also supported by the significantly better performance of our neural-network models in \cref{ssec:mlLocalVal}, which successfully learned this difference in turbulent particle and energy transport.

It is also worth noting that the absence in the database of the exact profiles of the magnetic field, e.g., $q(R), \; B(R), \; \hat s (R)$, limits the range of models that can be tested without including the results of numerical magnetic-field reconstruction. 

	\section{Summary and Conclusions} \label{sec:conclusions}

In this paper, we have analysed the JET-ILW pedestal database \citep{Frassinetti_Database} and reconstructed pedestal temperature profiles using density profiles and other engineering parameters. 

In \cref{sec:geometry}, we described the pedestal profiles used in this study and defined the pedestal parameters via $\mtanh$ fits. We showed the distributions and correlations of database parameters in \cref{sec:database,appendix:parameters}. This overview helped frame \cref{sec:neuralnets}, which navigated this parameter-space using machine-learning methods. Our objective was to use neural networks to predict $T_e$ profiles. Different parameter sets were tested, and, in \cref{ssec:mlFullProf}, we found that using the full $n_e$ profiles yielded excellent results for reconstructing $T_e$ pedestals, in particular when all of the available information was included. This was shown both in terms of $\mtanh$ metrics such as pedestal height and location (\cref{fig:mlFullProfPred}), and in terms of mean-squared difference between prediction and experiment over the steep-gradient region (\cref{table:FitCoeff}). In \cref{ssec:mlLocalVal}, local predictions of the $T_e$ profile showed good results for the pedestal heights, although there were inaccuracies in predicting the pedestal width and location (\cref{fig:mlBigCropPred}).

The ability to reconstruct pedestals with excellent quality using neural networks suggests that the predicting $T_e$ pedestals using $n_e$ profiles and other database parameters is a well-posed problem, both because of the underlying physics being governed by reproducible phenomena, and because the dataset is statistically robust. The success of local predictions indicates that this task can be achieved using models that take into account the local amplitudes and gradients of the pedestal. This said, we have observed that the inclusion of the radial location $R$ of the local prediction resulted in improved performance. This could be explained by the fact that, during training, the networks developed a general model of the large-scale structure of pedestals, and the inclusion of the radial location indirectly gave an indication of global physics, improving the prediction. 
In \cref{ssec:mlImportantPars}, we explicitly identified the most important parameters for accurate $T_e$ prediction in the global and local cases: the separatrix loss power $\Psep$, the plasma current $\Iplasma$, the fuelling rate $\fuelrate$, and the strike-point configuration. 

In \cref{sec:gradients}, we discussed the importance of the normalised temperature and density gradients $\RLTe$ and $\RLNe$ in the context of known linear ETG mode stability thresholds (\cref{fig:RLTeRLNe}). We found that all steep-gradient region pedestals were in an order-unity-supercritical regime, which suggested that the gradient ratio $\eta_e$ was an important parameter. In \cref{sec:etae}, our analysis of $\eta_e$ showed that it is not a good descriptor of the $T_e$ and $n_e$ inside of the density-top location because the values of $\eta_e$ range over several orders of magnitude. We found that $\eta_e \approx 2$ across the steep-gradient region (\cref{fig:etaslice}), confirming across a large database the previous experimental observations of values of $\eta_e$ in the pedestal. This result could be interpreted as pointing to the existence of a non-linear threshold $\etacr$ in pedestals, or as an indication that the turbulence saturates in an order-unity supercritical regime to match the transport necessary to maintain a steady state.

Next, we assumed a simple prescription for the relationship between the pedestal gradients $\RLTe$ and $\RLNe$: a constant $\eta_e$ across the pedestal. This was empirically motivated by the clustering of $\eta_e$ around $2$ in the steep-gradient region, and is equivalent to assuming that turbulence is marginal, close to a (in general, non-linearly) established threshold. 
Reconstructing $T_e$ pedestals using a given $\eta_e$ value gave reasonable but not very accurate pedestal predictions (\cref{fig:etaResults}). Using an optimisation algorithm to find a nominal value of $\eta_e$ across the database yielded values between $1.9$ and $2.1$. Pulse-by-pulse fits of $\eta_e$ performed much better at reproducing the pedestal region [\Cref{fig:etaResults}(a-d)], although $T_e$ pedestals where shifted outwards towards the separatrix, in line with the $n_e$ pedestals. The pulse-by-pulse optimisation yielded a distribution of fit values of $\eta_e$ that matched the experimental $\eta_e$ in the middle of the steep-gradient region (\cref{fig:etaBest}), indicating that the value of $\eta_e$ is a reliable indicator of the transport processes in the pedestal.

The results shown in \cref{fig:etaDistributions} hinted at a power-law relationship between $\RLTe$ and $\RLNe$ of the form $\RLTe = A(\RLNe)^\alpha$, a possibility that was explored further in \cref{sec:alpha}. There, we examined the distribution of the exponent $\alpha$ of such a hypothetical power-law relationship between the temperature and density gradients. While this lacked physical grounding, it was a simple model for our data.
In \cref{ssec:alphaPredictions}, we found that the best database-wide value of the exponent was $\alpha \approx0.4$ for the pedestal region [\Cref{fig:alphaResults}(e-h)]. Fitting $\alpha$ and $A$ independently for each pulse gave excellent agreement over the steep-gradient region [\cref{fig:alphaResults}(a)]. The resulting pairs of $A$ and $\alpha$ all turned out to lie on the similarity line $\ln A = 5.64 - 4.42 \alpha$ (\cref{fig:alphaBest}), giving the relationship \nolinebreak \cref{eq:oneParameterExponentFit} to describe all pedestals our JET-ILW subset of the database. Remarkably, a large set of JET-C pulses also obeyed a similar linear relationship between the gradients. This was one of the clearest modelling insights that we obtained in this study, even though we could not find a relationship between the pedestal-wide $\alpha$ and other database parameters. Constructing a model for $\alpha$ would complete the predictive model of the steep-gradient region, but the lack of physical intuition for such a power-law relationship makes this task difficult.

Finally, in \cref{sec:heatFluxModels}, we explored the role of the separatrix loss power $\Psep$ in relating temperature and density gradients. We focused our analysis on models inspired by gyrokinetic simulations \citep{Guttenfelder_2021,Chapman-Oplopoiou_2022,Hatch_RedModels}. In \cref{ssec:heatFluxPredictions,appendix:heatFluxFits}, we computed the best-fit parameters for these models and discussed the quality of the resulting $T_e$ profile reconstructions.

The models of the form \cref{eq:chapmanModel} \citep{Chapman-Oplopoiou_2022,Guttenfelder_2021} produced good reconstructions of the $T_e$ profile without major adjustments. Such models have a theoretical root in slab-ETG-driven turbulent transport, as described in \cref{ssec:heatFluxAssumptions}. Reconstructions of pedestal profiles using parameters optimised for these models and those found by \citet{Hatch_RedModels} provided excellent values for the $T_e$ profile values at the top of the $n_e$ and $T_e$ pedestals as a result of our optimisation process, although the pedestal shapes were subject to systematic inaccuracies (\cref{fig:heatFluxResults}).
The temperature gradients were wildly over-estimated near the separatrix when the formulae used for the heat flux included a prefactor of $(\RLTe)^1$ (\cref{fig:exampleVsHeatFluxOptimised}), whereas models with $(\RLTe)^2$ resulted in much better reconstructions (\cref{fig:exampleVsChapman}). This is a confirmation that the locally relevant length scale $\LTe$ must be used for the gyro-Bohm estimate of the heat flux $\Qegb$. Optimising for the critical value of the gradient ratio $\etacr$ almost always resulted in negative values (\cref{table:FitCoeff}), undermining the physical picture that we described in \cref{ssec:heatFluxAssumptions}. We interpret this as a sign that the models that we analysed are missing key elements of the underlying physics of the pedestal.

In spite of the shortcomings of our heat-flux models, the results of \cref{sec:heatFluxModels} confirm that we have successfully accounted for the separatrix loss power in order to find the electron-temperature gradients. These results were a major improvement over the predictions that used database-wide fitting parameters in \cref{sec:etae,sec:alpha}. However, the much better quality of the local-neural-network reconstructions of \cref{ssec:mlLocalVal} suggests that the upper bound of prediction accuracy has not yet been reached by the \quotemarks{physics-based} models. It appears that, in order to obtain a more accurate model of pedestal transport, a clearer theoretical understanding of the underlying gyrokinetic turbulence is required, and also more clarity regarding whether the pedestal-shape parameters, such as the width and position of the pedestal, are determined by local turbulence or by device-scale physics. Extending our investigation to other devices could give the key for creating a simple generalised model for pedestal transport.

\subsection{Acknowledgements}

This work has been carried out within the framework of the EUROfusion Consortium, funded by the European Union via the Euratom Research and Training Programme (grant number 101052200 - EUROfusion) and by the EPSRC Energy Programme (grant number EP/W006839/1). Views and opinions expressed herein are, however, those of the authors only and do not necessarily reflect those of the European Union or the European Commission. Neither the European Union nor the European Commission can be held responsible for them.

The work of L.-P. T. was supported by an EPSRC CASE studentship (grant number EP/W524311/1) in partnership with UKAEA. The work of A. A. S. was supported in part by EPSRC (grant number EPR034737/1), the STFC (grant number STW000903/1), and by the Simons Foundation via a Simons Investigator Award. The work of L. F. was supported by Vetenskapsrådet (grant number 2023-04895).
L.-P. T. would like to thank M. Abazorius, G. Acton, B. Chapman-Oplopoiou, W. Clarke, R. Greif, D. R. Hatch, P. G. Ivanov, S. Newton, P. Reichherzer, J. Ruiz Ruiz, M. Senstius, and F. C. Spangle for helpful advice and discussions.

	\addtocontents{toc}{{Appendices}\par} 
	\appendix 
	\renewcommand{\thesection}{\Alph{section}} 
	\renewcommand{\thesubsection}{\thesection.\arabic{subsection}} 
	
	\let\oldsection\section
	\renewcommand{\section}[1]{
		\refstepcounter{section} 
		\addcontentsline{toc}{section}{\thesection. #1} 
		\oldsection*{\thesection. #1} 
	}
	

\section{Pulse Parameters and Correlations}\label{appendix:parameters}

In this appendix, we list the definitions of each plasma parameter included in our study of the pedestal database. The distributions of these parameters across the database are shown in \cref{fig:ParamHists}. Their measurement units, as well as their minimum, maximum, mean, and median values are given in \cref{table:EngPars}. This is relevant context for carrying out profile predictions using neural networks in \cref{sec:neuralnets}. We loosely distinguish two groups of parameters, based on their origin: engineering parameters and magnetic-equilibrium parameters.

In \cref{table:EngPars} and \cref{fig:ProfValsHists}, we also present the distributions of the $\mtanh$ fit parameters, which describe the $T_e$ and $n_e$ profiles in our databases, alongside other relevant parameters such as the separatrix density $\Nesep$, the temperature at the density-top location $T_e(\psitop{n_e})$, and the ratio between the separatrix density and the pedestal-top density $\Nesep/\Netop$. The correlation matrix for all the parameters presented in \cref{table:EngPars} is presented in \cref{fig:correlationMatrix}. This is similar to Figure 1 of \citet{Kit_ML_database}, adapted to our dataset of $1251$ pulses.

\begin{table}
	\def~{\hphantom{0}}
	\centering\begin{tabular}{c c c c c c c}
		\textbf{Parameter} & \textbf{Units} & \textbf{Min} & \textbf{Max} & \textbf{Mean} & \textbf{Median} & \\ 
		\hline
		$\Psep$ 				 & $[MW]$ 					& 2.37~ & 27.0~~ & 11.9~~ & 11.7~~ & \\
		$\Gamma_{D} $ 			 & $[e^-/s\times 10^{22}]$ 	& 0.0~~ & 22.3~~ & ~2.08~ & ~1.58~ & \\
		$\Iplasma$ 				 & $[MA]$ 					& 0.972 & ~3.97~ & ~2.12~ & ~1.99~ & \\
		$B$ 		 			 & $[T]$ 					& 0.967 & ~3.68~ & ~2.37~ & ~2.38~ & \\
		$\overline{n_e}$ 		 & $[m^{-2}\times 10^{20}]$ & 0.0~~ & ~0.907 & ~0.434 & ~0.430 & \\
		\hline
		$\kappa$ 				 & $[-]$ 					& 1.58~ & ~1.82~ & ~1.68~ & ~1.68~ & \\
		$\langle \delta \rangle$ & $[-]$ 					& 0.153 & ~0.456 & ~0.265 & ~0.262 & \\
		$\delta_\mathrm{Lower}$	 & $[-]$ 	 				& 0.232 & ~0.492 & ~0.327 & ~0.332 & \\
		$\delta_\mathrm{Upper}$	 & $[-]$ 	 				& 0.021 & ~0.461 & ~0.2~~ & ~0.182 & \\
		$q_{95}$				 & $[-]$					& 2.42~ & ~5.44~ & ~3.46~ & ~3.37~ & \\
		$\beta_{\theta} $ 		 & $[-]$ 					& 0.226 & ~1.41~ & ~0.615 & ~0.551 & \\
		$\nu*$ 					 & $[-]$ 					& 0.002 & ~0.133 & ~0.021 & ~0.017 & \\
		$\ngw$ 					 & $[m^{-3}\times 10^{20}]$ & 0.361 & ~1.43~ & ~0.787 & ~0.768 & \\
		$\fgw$ 					 & $[-]$ 					& 0.29~ & ~1.05~ & ~0.729 & ~0.734 & \\ 		
		\hline 
		$\Tetop$				 & $[keV]$ 					& 0.157 & ~1.38~ & ~0.573 & ~0.556 & \\
		$\psiwid{T_e}$			 & $[-]$ 					& 0.014 & ~0.144 & ~0.046 & ~0.045 & \\
		$\psitop{T_e}$			 & $[-]$ 					& 0.899 & ~0.987 & ~0.96~ & ~0.961 & \\
		$\Netop$				 & $[m^{-3}\times 10^{20}]$ & 0.179 & ~1.08~ & ~0.457 & ~0.44~ & \\
		$\psitop{n_e}$			 & $[-]$ 					& 0.923 & ~0.999 & ~0.979 & ~0.98~ & \\
		$\psiwid{n_e}$			 & $[-]$ 					& 0.016 & ~0.231 & ~0.045 & ~0.041 & \\
		$T_e(\psitop{n_e}) $	 & $[keV]$					& 0.104 & ~1.032 & ~0.366 & ~0.325 & \\
		$\Nesep	$				 & $[m^{-3}\times 10^{20}]$ & 0.058 & ~0.658 & ~0.273 & ~0.254 & \\
		$\Nesep / \Netop$		 & $[-]$ 					& 0.212 & ~0.988 & ~0.596 & ~0.597 & \\
	\end{tabular}
	\caption{\centering {Relevant engineering parameters, magnetic-equilibrium parameters, and profile values in the database, also depicted in \cref{fig:ParamHists,fig:ProfValsHists}. Each group of parameters is separated by a horizontal bar. The measurement units, ranges, mean values, and median values of the parameters are displayed for the subset of the database analysed in this paper.}}
	\label{table:EngPars}
\end{table}

\subsection{Pulse-Parameter Definitions} \label{sappendix:params}

The engineering parameters used in this study are:
\begin{itemize}[noitemsep,topsep=0pt,parsep=0pt,partopsep=0pt]
	\item[-- ] the separatrix loss power $\Psep $;
	\item[-- ] the deuterium gas fuelling rate $\fuelrate$;
	\item[-- ] the line-integrated electron density following one of the interferometer chord diagnostics of JET through the plasma edge $\overline{n_e}$;
	\item[-- ] the plasma current $\Iplasma$;
	\item[-- ] the strike-point configuration, defined by which of the divertor plates the footprints of the open field lines rest upon, with several basic configurations described below.
	
\end{itemize}

\begin{figure}
	\centering
	\includegraphics[width=\textwidth]{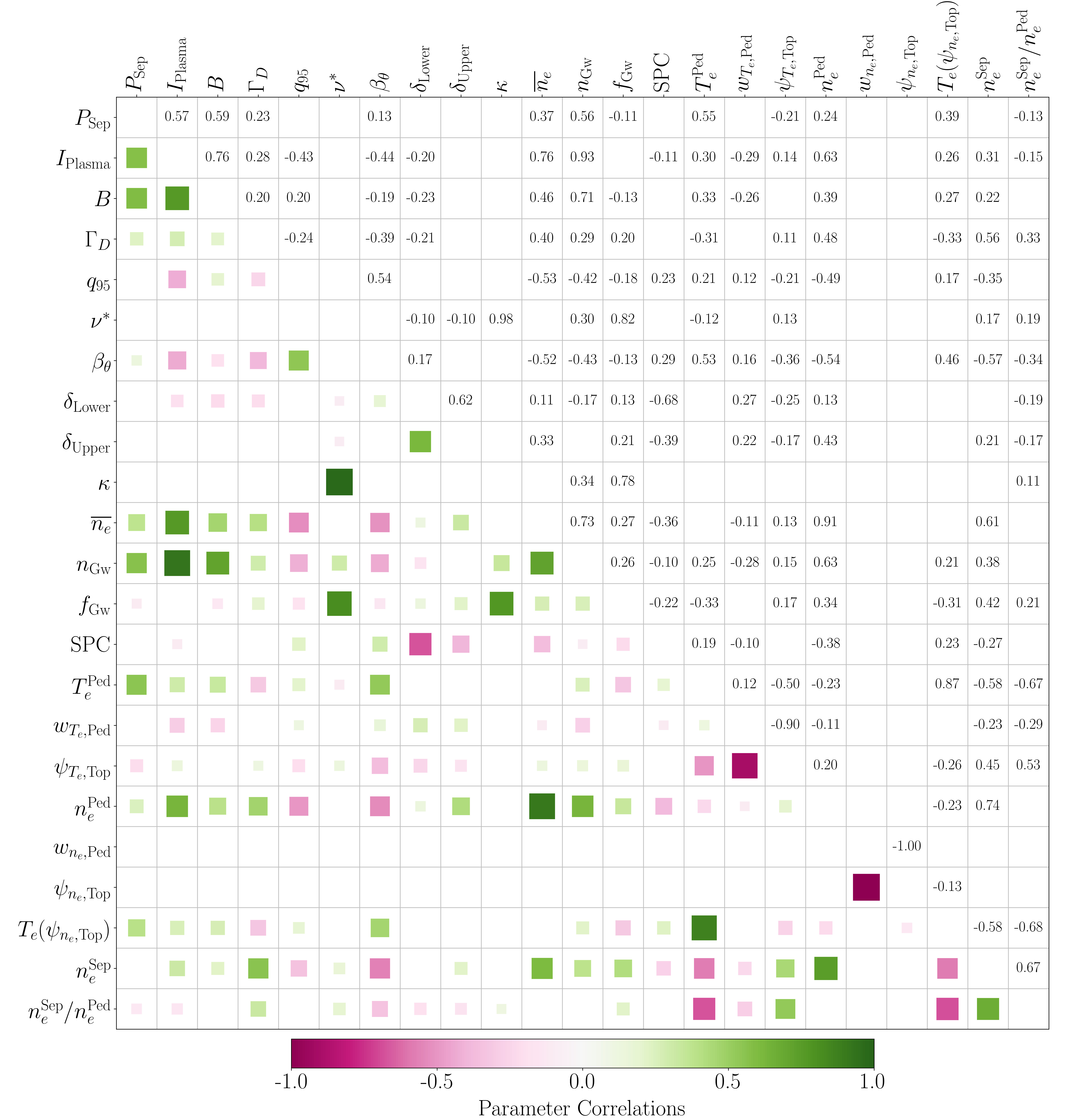}
	\caption{\centering {Matrix of Pearson correlation coefficients computed for engineering parameters, magnetic-equilibrium parameters, and relevant profile parameters across the database. SPC refers to the strike-point configurations, as denoted in \cref{sappendix:params}. The upper triangle contains the numerical values of the correlation coefficients, the lower triangle represents the same values pictorially. Values that are smaller in magnitude than $0.10$ are omitted.}}
	\label{fig:correlationMatrix}
\end{figure}

The separatrix loss power is computed from the power-balance of the plasma, viz.,
\begin{equation}
	\Psep = P_\mathrm{Ohm} + P_\mathrm{NBI} + P_\mathrm{ICRH} - P_\mathrm{Shine} - P_\mathrm{Rad},
\end{equation}  
where the different power components are the Ohmic heating power $P_\mathrm{Ohm}$, the neutral-beam injection power $P_\mathrm{NBI}$, the ion-cyclotron-resonance heating power $P_\mathrm{ICRH}$, the shine-through power $P_\mathrm{Shine}$ (the power injected but not absorbed by the plasma), and the power radiated by the bulk of the plasma $P_\mathrm{Rad}$.

The strike-point configurations are formatted as \quotemarks{inner strike point / outer strike point} in \cref{fig:ParamHists}. The different ways to contact the divertor plates are: V for vertical, C for corner, and H for horizontal. A depiction of these strike point configurations in JET-ILW can be found in Figure 5 of \citet{Frassinetti_Database}. For the purpose of computing parameter correlations in \cref{fig:correlationMatrix}, the strike point configurations (denoted as SPC in the Figure) are encoded by integer values: V/H $\rightarrow$ 0, C/H $\rightarrow$ 1, V/V $\rightarrow$ 2, C/V $\rightarrow$ 3, V/C $\rightarrow$ 4, and C/C $\rightarrow$ 5.

The magnetic-equilibrium parameters used in this study are:
\begin{itemize}[noitemsep,topsep=0pt,parsep=0pt,partopsep=0pt]
	\item[-- ] the magnetic field on the axis $B$;
	\item[-- ] the safety factor $q_{95}=\dertxt{\Phi}{\psi}$ evaluated at $\psi_N = 0.95$ (here, $\Phi$ and $\psi$ are the toroidal and poloidal magnetic fluxes, respectively);
	\item[-- ] the $\delta_{\mathrm{Upper}}$ and $\delta_{\mathrm{Lower}}$ plasma triangularities, used  in the Miller equilibrium parametrisation \citep{Miller_eq};
	\item[-- ] the plasma elongation, Miller equilibrium index  $\kappa$; 
	\item[-- ] the kinetic pressure normalised by the poloidal magnetic energy density, $\betapol= 2\mu_0p /B_\theta^2 $;
	\item[-- ] the electron-electron collisionality at the pedestal top $\nu^*$, normalised to the thermal ion bounce frequency, as per Equation (7) of \citet{Frassinetti_Database};
	\item[-- ] the Greenwald density limit  $\ngw$, a measure of the maximum line-averaged density at the edge of the plasma, defined as $\ngw = \Iplasma/\pi a^2$, where $a$ is the minor radius of the device;
	\item[-- ] the Greenwald fraction $\fgw$, the ratio between the measured line-averaged density and $\ngw$.
	
\end{itemize}

\section{Neural Networks and Methods}\label{appendix:neuralnets}

This appendix details the machine-learning methods used to reconstruct $T_e$ in our database. We will describe our methods of handling data, including selection, processing, and augmentation, alongside our choice of training strategies and network architecture.

\subsection{Data Pre-processing} \label{sappendix:preprocessing}

The database that we use includes parameters associated with each pulse, which can have discrete values (e.g., the strike-point configuration) or continuous ones (e.g., the database parameters or the profiles). The aim of our data pre-processing is to translate this data into a series of scalars that can be assembled into structured vectors of labels (inputs) and features (outputs) to describe each pulse. These sets of labels and features will then be used to reconstruct $T_e$ profiles using a neural network.

In order to be able to make use of the strike-point configurations through our numerical method, the discrete-valued strike-point configurations (which can take $6$ possible values) are \quotemarks{one-hot} encoded to a $6$-dimensional vector. Consequently, any strike-point configuration is represented by a vector that has all components equal to $0$, with the exception of one component that is $1$. The location of the non-zero component uniquely maps to each strike-point configuration.

The profiles of $n_e$, $T_e$, $\RLNe$, and $R$ are sampled from \fit profiles at $128$ (\cref{ssec:mlFullProf,ssec:mlLocalVal}) or $64$ (\cref{ssec:mlImportantPars}) equally spaced radial locations. The radial locations are chosen in an interval between $R=3.7\;\mathrm{m}$ and $R=3.85\;\mathrm{m}$. This range encompasses all points located between $\psitop{T_e}$ and $\psisep$. The use of $R$ as the radial coordinate (as opposed to $\psi_N$ or $\psirenorm$) is made in order to avoid the inclusion of data that contains a priori information about the $T_e$ pedestal -- as $\psi_N$ contains information obtained from the equilibrium reconstruction of the plasma profiles, and $\psirenorm$ makes use of the pedestal-top locations.

Where available, experimental uncertainties are included. These can be associated with pulse parameters or profiles. For \fit profiles, we include the \raw uncertainties. The experimental errors included in \raw profiles are sampled at the same $128$ or $64$ radial locations using the following procedure: the \raw values of the experimental uncertainties of the $n_e$ and $T_e$ profiles are convolved with a Gaussian kernel of full-width-half-maximum matching the instrument function of the HRTS diagnostic \citep{Frassinetti_MeasurementMethods}. Then, a low-pass filter is applied to obtain a smooth numerical interpolation of these convolved errors. The convolved and filtered experimental errors offer an excellent estimate for the mean uncertainties at each location in the pedestal. One such set of uncertainties is displayed in \cref{fig:randomCrop} for pulse $\examplepulse$. These values will be used in data augmentation, as will be explained next.

\subsection{Augmentation} \label{sappendix:augmentation}

Augmentation is essential for the stability and convergence of neural networks, and plays an important role in transforming our limited database of $1251$ pulses into a statistically meaningful training set. For robust interpolation, it is also important to introduce awareness of the experimental uncertainties into the neural network. Therefore, we indirectly weight each pulse based on how sparse the neighbouring parameter space is. We will describe below our measure of parameter-space sparsity and how we use it to decide the amount of augmentation (duplication) that each piece of data must be subjected to. We also add random Gaussian noise to the augmented data in order to model the experimental uncertainties.

\begin{figure}
	\centering
	\includegraphics[width=\textwidth]{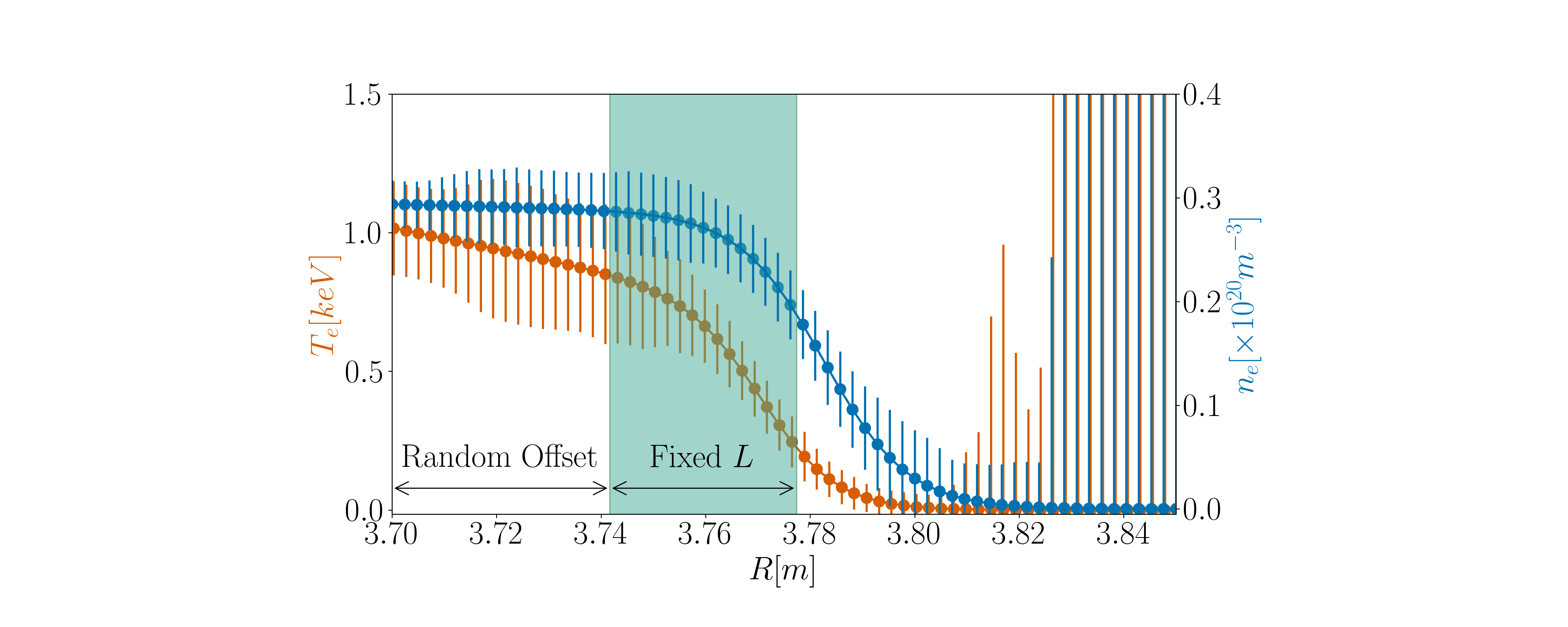}
	\caption{\centering  {The $n_e$ (blue) and $T_e$ (orange) profiles for pulse $\examplepulse$, with $64$ numerical samples at equally spaced radial locations. The error bars represent the experimental uncertainties, obtained as described in \cref{sappendix:preprocessing}. The green box represents a subset of $L$ consecutive numerical values used in the training of neural networks. The beginning position of the subset of length $L$ is chosen randomly as part of the augmentation process described in \cref{sappendix:augmentation}.}}
	\label{fig:randomCrop}
\end{figure}

For the purpose of assembling the labels, each separate set of them consists of $N$ different parameters, not including profiles. These pulse parameters are therefore treated as populating an $N$-dimensional space, for which we first calculate the parameter-space density. Consider one of the possible parameters $j$, whose value for each pulse is $j(\mathrm{Pulse})$. We calculate the parameter-space density using a Gaussian-kernel density estimator as follows:
\begin{equation}\label{eq:augDensity}
	\rho_j (x) \propto \sum_\mathrm{Pulse} \exp \left\{ - \frac{  [j(\mathrm{Pulse}) - x]^2}{2 h^2}\right\},
\end{equation}
where $x$ is the value of $j$ where the parameter-space density is evaluated, and $h$ is the full-width half-maximum of the Gaussian kernel. We set $h = [\max(j) - \min (j)]/10$ in order to discern distributions that are not unimodal. Any overall pre-factor in \cref{eq:augDensity}  is irrelevant in this method, as will be seen below.

Using the single-parameter density evaluated for each individual pulse, $\rho_j (\mathrm{Pulse}) \equiv \rho_j [ j (\mathrm{Pulse})]$, we calculate the number of times each pulse must be duplicated (augmented) for the dataset. First we define the \quotemarks{augmentation score} for each pulse, 
\begin{equation} \label{eq:augScore}
	S(\mathrm{Pulse}) = \sum_{j} \exp \left[\frac{d+1}{d+{\rho_j(\mathrm{Pulse})}/{\mathrm{max}(\rho_j)}}\right],
\end{equation}
where $d$ is a positive constant and the sum is over all parameters. The definition \cref{eq:augScore} assigns the highest scores to the least represented pulses in parameter space. Using a value of $d$ of around $0.2 - 0.4$ results in an approximately uniform distribution of $S$. The augmentation number of each pulse, defined as the number of times this pulse will appear in the training set as a result of the augmentation process, is a linear re-scaling of the augmentation score: 
\begin{equation}
	A(\mathrm{Pulse}) = A_{\min} + (A_{\max} - A_{\min}) \frac{S(\mathrm{Pulse}) - \min(S)}{\max(S)-\min(S)} ,
\end{equation}
where $A_{\min}$ and $A_{\max}$ are chosen as follows: 1200 and 2500, respectively, for full-profile predictions (\cref{ssec:mlFullProf}); 2000 and 3500 for local predictions (\cref{ssec:mlLocalVal}); 500 and 5000 for the parameter scans (\cref{ssec:mlImportantPars}). This means that for the pulses that sample high-density regions of the parameter space, $A(\mathrm{Pulse})$ will be smaller than for those in sparser regions. Consequently, the pulses in dense regions will appear fewer times in the training set of our neural network to compensate for the density. This effectively balances our data set. 

The choice of the augmentation score \cref{eq:augScore} was made so as to avoid the behaviour of some alternatives, e.g., $S(\mathrm{Pulse}) = \left[\prod_{j} \rho_{j}(\mathrm{Pulse}) \right]^{-1/N}$, that caused $> 95\%$ of pulses to end up with the maximum augmentation score $A(\mathrm{Pulse})\approx A_{\max}$ when $N$ was large, which would have defeated the purpose of the augmentation procedure.

Our augmentation procedure includes a random cropping of the profiles of $T_e$, $n_e$, and $\RLNe$, if they are included in the information used in the network. The same cropping is also applied to the values of the radial location $R$ where the profiles are sampled numerically. This size reduction also includes selecting randomly located subsets of $L$ consecutive values from each profile, as illustrated in \cref{fig:randomCrop}, and discarding the rest of the profile. For full-profile predictions, this value of $L$ is set to 80\% of the number of numerical samples of each profile so as to allow the data to be subject to radial \quotemarks{jitter}. For local-value predictions, $L=3$ for each profile, effectively selecting an extremely narrow (one-point) sample. The random positioning of the subsets helps the network match the features of the profiles to the prediction, avoiding an inflexible memorisation based on the exact radial location of each numerical value. The inclusion of $R$ in the input data is used to test the importance of the location for the prediction.

So far, each pulse is represented in the dataset by $A(\mathrm{Pulse})$ vectors of labels and $A(\mathrm{Pulse})$ vectors of features. Each of the label vectors has length $N + M L$, being composed of the $N$ pulse parameters and $M$ cropped profiles of length $L$. The feature vectors have length $L$, representing the cropped $T_e$ profile.
 
All resulting entries in the augmented dataset (parameters and profiles) also include Gaussian random noise matching the experimental uncertainties in each individual value. This noise is added to each pulse parameter:
\begin{equation}
	j(\mathrm{Pulse}) \rightarrow j(\mathrm{Pulse}) + \mathcal F [\sigma_j (\mathrm{Pulse})],
\end{equation}
where $\sigma_j (\mathrm{Pulse})$ is the experimental uncertainty of the parameter $j$ in a given pulse, and $\mathcal F [\sigma]$ is a random sample from a normal distribution with standard deviation $\sigma$ and zero mean. Similarly, each point in the cropped profiles of $T_e$, $n_e$, and $\RLNe$ is subjected to random noise corresponding to the experimental uncertainty evaluated at the point's location.

\subsection{Normalisation and Loss} \label{sappendix:normAndLoss}

The augmentation procedure described above results in a dataset comprised of vectors of labels of length $N+ML$ and vectors of features of length $L$, containing the concatenated parameters and profile values. For numerical stability, all of the values in these vectors are normalised after augmentation. Depending on the case, we map each label and each feature independently to a distribution with mean zero and standard deviation $1$. These have no effect on the results, but they are sometimes necessary to avoid divergence during training.

In order to train the neural network, a mean-squared loss function between the target features and the predicted features is minimised. This loss function is defined to be
\begin{equation}
	\mathcal L  = \frac{1}{L A_\mathrm{total}} \sum_{k=1}^{A_\mathrm {total}}\sum_{i=1}^{L} [(y_{k,i})_\mathrm{true}- (y_{k,i})_\mathrm{pred}]^2 ,
\end{equation}
where $A_\mathrm {total}= \sum_{\mathrm{Pulse}}A(\mathrm{Pulse})$ is the total number of label-feature pairs of vectors, $(y)_\mathrm{true}$ are the target values (in our case, the normalised and cropped values of $T_e$), and  $(y)_\mathrm{pred}$ are the features predicted by the neural network. We denote by $y_{k,i}$ the $i$'th value of the $k$'th feature vector. 

Additionally, $L_2$ regularisation is used in the optimisation of the model, i.e., each linear transformation layer in the architecture of the network (explained in \cref{sappendix:modelArchitecture}) contributes proportionally to its $L_2$ norm to the loss function. This means that the optimiser causes the linear transformations in the neural network to have few singular values in their singular-value decompositions, reducing over-fitting.

\subsection{Train-Test Split} \label{sappendix:trainTestSplit}

We use a randomized $80\%-20\%$ train-test split, i.e., $1000$ pulses are used for the training of the neural networks and $251$ pulses are used for testing the accuracy of predictions, without being seen during training. For the augmentation process, parameter-space densities and normalisations are calculated using only the $1000$ training pulses. These parameter-space densities and normalisations are then applied in the augmentation of both the training $1000$ pulses and the $251$ test pulses. 

A prediction of the $T_e$ profile for a pulse is made by using the trained network to make predictions using all the corresponding augmented entries in the dataset. Then, the final $T_e$ profile is computed by averaging over all predictions corresponding to augmented entries.

\subsection{Neural-Network Architecture} \label{sappendix:modelArchitecture}

\begin{figure}
	\centering
	\includegraphics[width=\textwidth]{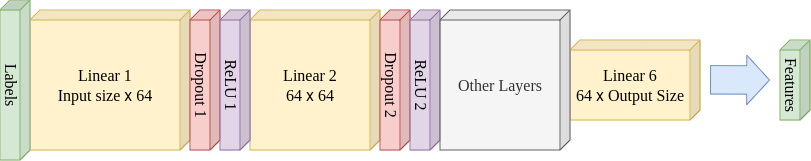}
	\caption{\centering {Feed-forward neural-network architecture described in \cref{sappendix:modelArchitecture}. The layers are shown in the order in which they are applied to the labels. The internal dimensions of the linear layers are $64 \times 64$, whereas the first and the last layers match the dimensions of the label and the feature vectors, respectively. Dropout layers are included only for local-value predictions. The \quotemarks{other layers} block represents a sequence of $64\times 64$ linear layers, dropout layers, and ReLU layers.}}
	\label{fig:nnArchitecture}
\end{figure}

Our predictions are carried out using neural networks that consist of several transformations that are applied sequentially. We refer to these transformations as layers, and because of the sequential application of the layers (as per the network architecture shown in \cref{fig:nnArchitecture}), these networks are called \quotemarks{feed-forward}. Our feed-forward neural-network architecture consists of a sequence of $6$ linear layers, interposed with $5$ rectified-linear-unit (ReLU) layers. If appropriate, dropout layers are also included before the ReLU layers. These will be explained below.

Each linear layer represents an affine transformation of a vector $\mathbf x$, of the form $\mathbf x \mapsto W \mathbf x + \mathbf b$, where $W$ is a weight matrix and $\mathbf b$ is a bias vector, both of which are \quotemarks{learned} during training (i.e., they are subject to the optimisation algorithm that reduces the loss and regularisation terms, introduced in \cref{sappendix:normAndLoss}). The internal dimensions of the affine transformations are set to $64$, whereas the input and output dimensions are matched to the profiles and parameter combinations in the labels and features, respectively, as described in \cref{sappendix:augmentation}. 

The ReLU layers apply a piecewise linear transformation to every component of a vector $\mathbf x$, viz., $\mathrm{ReLU}(x_i) = \max(0, x_i)$. This introduces non-linearity into the network and allows for complex behaviour. Without such a source of non-linearity, the network would simply amount to a single affine transformation.

To improve generalisability and stability, particularly in models with small $L$ for local predictions, where overfitting may be more pronounced, we optionally include dropout layers. These layers perform a transformation on a vector $\mathbf x$ that randomly chooses a subset of components $x_i$ and sets them to $0$ when applied during training. Outside of training, the probability of \quotemarks{dropping out} any $x_i$ is $0$, and the network is fully used. This method reduces over-reliance on specific labels and helps prevent overfitting.

Further details regarding the numerical methods described in this Appendix can be found in the documentation of the PyTorch package\footnote{The documentation of the PyTorch package can be found at \url{https://pytorch.org/docs/}.} \citep{PyTorch_2019}, which was used to obtain our results.

\section{Method for Integrating Temperature Profiles}\label{appendix:predictionMethods}

The results obtained in \cref{ssec:etaPredictions,ssec:alphaPredictions,sec:heatFluxModels,} will, in general, begin by stating an explicit or implicit functional relationship between $\RLTe$ and $\RLNe$. Therefore, the starting point of our profile predictions is a differential equation of the form $\RLTe = f (*)$, or, equivalently:
\begin{equation}\label{eq:dtdr}
	\der{T_e}{R} = - \frac{T_e}{R} f(*),
\end{equation}
where the $*$ represents the database parameters and the numerical constants that are used for numerical integration, including $R$.

Following \citet{Simpson_Separatrix}, the boundary condition for the integration is set as $T_e=100\;\mathrm{eV}$ at $\Rsep$, the radial location where $\psi_N = 1.0$ (\cref{ssec:radialcoord}). Thus, we match the starting point of the integration to where the \fit $T_e$ profile has the value of $100\;\mathrm{eV}$. The numerical integration is then carried out inwards from $\Rsep$, spanning the extent of the pedestal. Once a predicted profile has been computed for each pulse, a new $\mtanh$ profile is fit to the prediction, and the resulting parameters are compared to the parameters that are contained in the original database, i.e., to the $\mtanh$ parameters of the $\fit$ profile.

\subsection{Finding Best-Fit Parameters}\label{sappendix:bestFit}
Determination of the best-fit parameters for a certain scaling, e.g., identifying an optimal $\eta_e$ in \cref{eq:etaDefinition}, is carried out by minimising the absolute difference between the $\fit$ $T_e$ profile and the reconstructed $T_e$ profile over some radial domain $\mathcal{D}$. The optimisation and integration are done in real space, but, for the consistency of exposition, we use intervals of $\psi_N$ or $\psirenorm$ to denote locations. The domain $\mathcal{D}$ can represent a full-pedestal interval encompassing the core slope $\psi_N \in [0.85,1.0]$, or any region in the pedestals matching some subinterval of $\psirenorm \in [0,2]$. This means that we can find the optimal parameters that match the steep-gradient region, the pedestal region, or a larger part of the pedestal. 

Minimising the residual between the $\fit$ $T_e$ profile and the reconstructed $T_e$ profile,
\begin{equation} \label{eq:optimisationResidual}
	\Delta = \int_{\mathcal{D}} \mathrm{d} R \left[ T_{e,\fit}(R)-T_{e,\mathrm{Rec}}(R) \right]^2 \rightarrow \min,
\end{equation}
is equivalent to minimising a \quotemarks{weighted} difference between the $\fit$ $\RLTe$ and the reconstructed $\RLTe$. This weighting arises because any profile is given by
\begin{equation}
	T_e(R) = \Tesep \exp \left( - \int^R_{\Rsep}\mathrm{d} r \frac{ 1}{\LTe (r)} \right),
\end{equation}
and, therefore, an inaccuracy in $\RLTe$ close to the separatrix is compounded across the whole $T_{e,\mathrm{Rec}}$ profile. While it is true that if $(\RLTe)_\fit = (\RLTe)_\mathrm{Rec}$, then $T_{e,\fit} = T_{e,\mathrm{Rec}}$, we aim for a correct prediction of $T_{e,\mathrm{Rec}}$ and, therefore, we allow for more leeway for the prediction of $(\RLTe)_\mathrm{Rec}$.

\section{Heat-Flux-Model Optimisation and Comparison of $T_e$ Reconstructions}\label{appendix:heatFluxFits}

In this appendix, we give a general overview of the equations used to obtain the results of \cref{sec:heatFluxModels} and the outcomes of numerical optimisation of the constants determining these equations. In order to obtain these results, we optimised the free parameters of the transport models listed in \cref{sappendix:equations}, originally proposed by \citet{Chapman-Oplopoiou_2022} and \citet{Hatch_RedModels}. These optimised transport models are used in \cref{sec:heatFluxModels} to reconstruct $T_e$ profiles. A comprehensive set of optimised parameters, including those from \cref{sec:heatFluxModels}, and their accuracy, are documented in \cref{table:FitCoeff}, where we also give a comparison of the quality of predictions performed in \cref{ssec:mlFullProf,ssec:mlLocalVal,ssec:etaPredictions,ssec:alphaPredictions}. The metrics of comparison listed in \cref{table:FitCoeff} are explained in \cref{sappendix:bigComparison}.

\subsection{Explicit Expressions for Heat-Flux Models}\label{sappendix:equations}

For all of the heat-flux models described here, we maintain the usage of three free dimensionless parameters: $A$, $\fitcoeff$, and $\beta$.
The two models proposed by \citet{Chapman-Oplopoiou_2022} are of the form:
\begin{equation} \label{eq:Chapman12}
	\frac{\Qeturb}{\Qegb} = A \left(\frac{R}{\LTe}\right) ^ 2 \left( \eta_e - \fitcoeff \right) ^\beta,
\end{equation}
where the major distinction between the two is in the value of $\beta$. The case with a general value of $\beta$ is referred to in \cref{table:FitCoeff} as \quotemarks{Chap 1 models}; \quotemarks{Chap 2 models} have $\beta=1$, as originally posited by \citet{Guttenfelder_2021}. Their physical foundation is discussed by \citet{Field_Stiff}: the positive value of $\fitcoeff$ in \cref{eq:Chapman12} has the role of a threshold value of $\eta_e$, as described in \cref{ssec:etgTurb}. 

The five models obtained by fitting to a database of gyrokinetic simulations in Table \nolinebreak 1 of \citet{Hatch_RedModels} (referred to there as $Q_1$ to $Q_5$) are of three types. First, 
\begin{equation} \label{eq:Hatch123}
	\frac{\Qeturb}{\Qegb} = A \frac{R}{\LTe}  \left(  \eta_e ^\beta - \fitcoeff \right)  ,
\end{equation}
where $\beta=2$ for $Q_1$ (\quotemarks{Hatch 1}) and $\beta=4$ for $Q_2$ and $Q_3$ (\quotemarks{Hatch 2} and  \quotemarks{Hatch 3}). The distinction between the Hatch 2 and Hatch 3 models is in the constant $A$ and the inclusion of the temperature ratio $\tau = Z T_e / T_i$ as a parameter. Here, we take our deuterium plasmas to have the temperature ratio $\tau = 1$ throughout the pedestal profile as an extrapolation from measurements that suggest $T_e = T_i$ \citep[available only for a small subset of the database: see][]{Frassinetti_Database}.
The second model type is
\begin{equation} \label{eq:Hatch5}
	\frac{\Qeturb}{\Qegb} = A \frac{R}{\LTe}  \left( \eta_e  - \fitcoeff \right) ^\beta  ,
\end{equation}
where $\beta=4$ for $Q_5$ (\quotemarks{Hatch 5}).
Finally, $Q_4$ of \citet{Hatch_RedModels} is the $\beta=1$ version of \cref{eq:Chapman12} (\quotemarks{Chap 2}).

Below, we will assess the relative performance of all these models in predicting correct temperature profiles and also compare them with the models introduced in \cref{sec:etae,sec:alpha}. In this process, we also adjust the parameters of \cref{eq:Chapman12,eq:Hatch123,eq:Hatch5} for an optimal $T_e$ prediction. All of this information is contained in \cref{table:FitCoeff}.

\subsection{Comparison of Models}\label{sappendix:bigComparison}

All of the numerical optimisations in this study are performed using the method described in \cref{sappendix:bestFit}. Given that most of these heat-flux models were inspired by gyrokinetic simulations of large-gradient and high-shear region of the pedestal, we limit the radial ranges of optimisation to the physically relevant regions:
\begin{itemize}
	\item[-- ] the steep-gradient region, \quotemarks{Steep}, bounded by $\psisep$ and $\psitop{n_e}$;
	\item[-- ] the full pedestal region, \quotemarks{Ped}, bounded by $\psisep$ and $\psitop{T_e}$;
	\item[-- ] the density-top region, \quotemarks{Top}, bounded by $\psitop_{n_e}$ and $\psitop{T_e}$.
\end{itemize}
These definitions are consistent with our approach across this study, and the pedestal locations are as defined in \cref{ssec:locations}. We also include an extended range: 
\begin{itemize}
	\item[-- ] the pedestal profile including part of the core, \quotemarks{Full}, bounded by $\psisep$ and $ \psi_N = 0.85$.
\end{itemize}

\begin{table}
	\makebox[\textwidth]{ \begin{tabular}{llccccccccccc}			
		& \textbf{Model} & \multicolumn{3}{c}{\textbf{Parameters}} & \multicolumn{8}{c}{\textbf{Fit Metrics}} \\
		&  & $\fitcoeff$ & $\beta$ & $A$ & $M_\mathrm{Steep}$ & $r^2_\mathrm{Steep}$ & $M_\mathrm{Ped}$ & $r^2_\mathrm{Ped}$ & $M_\mathrm{Top}$ & $r^2_\mathrm{Top}$ & $M_\mathrm{Full}$ & $r^2_\mathrm{Full}$ \\
		[-1.2ex]\hline 
		1 & Chap 1 \cref{eq:Chapman12} & 1.28 & 1.43 & 0.85 & 0.60 & 0.55 & 1.11 & 0.64 & 1.88 & 0.33 & 237.73 & -16.94 \\
		2 & Chap 2 \cref{eq:Chapman12} & 1.40 & 1.00 & 1.50 & 0.49 & 0.64 & 1.05 & 0.66 & 1.95 & 0.30 & 207.90 & -14.69 \\
		3 & Hatch 1 \cref{eq:Hatch123} & 1.80 & 2.00 & 6.73 & 16.13 & -11.05 & 28.26 & -8.09 & 43.30 & -14.51 & 2590.88 & -194.49 \\
		4 & Hatch 2 \cref{eq:Hatch123} & -2.88 & 4.00 & 0.50 & 2.35 & -0.75 & 5.50 & -0.77 & 9.47 & -2.38 & 68.59 & -4.18 \\
		5 & Hatch 3 \cref{eq:Hatch123} & -5.12 & 4.00 & 0.63 & 1.14 & 0.15 & 2.72 & 0.13 & 4.67 & -0.66 & 47.81 & -2.61 \\
		6 & Hatch 5 \cref{eq:Hatch123} & -0.41 & 4.00 & 0.31 & 11.13 & -7.32 & 21.03 & -5.76 & 32.25 & -10.55 & 913.32 & -67.91 \\
		[-1.2ex]\hline
		7 & Chap 1 \cref{eq:Chapman12} & 1.28 & 1.43 & \textbf{0.82} & 0.56 & 0.58 & \textbf{1.06} & \textbf{0.66} & 1.81 & 0.35 & 1.17 & 0.91 \\
		8 & Chap 2 \cref{eq:Chapman12} & 1.40 & 1.00 & \textbf{0.95} & 0.54 & 0.60 & \textbf{0.98} & \textbf{0.68} & 1.67 & 0.40 & 0.92 & 0.93 \\
		9 & Hatch 1 \cref{eq:Hatch123} & 1.80 & 2.00 & \textbf{116.41} & 0.81 & 0.39 & \textbf{1.38} & \textbf{0.56} & 2.19 & 0.18 & 2.46 & 0.81 \\
		10 & Hatch 2 \cref{eq:Hatch123} & -2.88 & 4.00 & \textbf{3.09} & 0.47 & 0.65 & \textbf{1.10} & \textbf{0.65} & 1.97 & 0.30 & 2.16 & 0.84 \\
		11 & Hatch 3 \cref{eq:Hatch123} & -5.12 & 4.00 & \textbf{2.10} & 0.50 & 0.63 & \textbf{1.14} & \textbf{0.63} & 2.02 & 0.28 & 2.05 & 0.85 \\
		12 & Hatch 5 \cref{eq:Hatch123} & -0.41 & 4.00 & \textbf{6.40} & 0.74 & 0.45 & \textbf{1.20} & \textbf{0.61} & 1.79 & 0.30 & 2.39 & 0.82 \\
		[-1.2ex]\hline
		13 & Chap 1 \cref{eq:Chapman12} & \textbf{-9.96} & \textbf{3.71} & \textbf{1e-3} & \textbf{0.33} & \textbf{0.75} & 0.72 & 0.77 & 1.36 & 0.48 & 0.97 & 0.93 \\
		14 & Chap 1 \cref{eq:Chapman12} & \textbf{-8.60} & \textbf{3.75} & \textbf{1e-3} & 0.42 & 0.68 & \textbf{0.70} & \textbf{0.78} & 1.13 & 0.57 & 0.75 & 0.94 \\
		15 & Chap 1 \cref{eq:Chapman12} & \textbf{-0.24} & \textbf{1.44} & \textbf{0.20} & 0.47 & 0.65 & 0.75 & 0.76 & 1.19 & 0.55 & \textbf{0.72} & \textbf{0.95} \\
		16 & Chap 2 \cref{eq:Chapman12} & \textbf{-1.15} & 1.00 & \textbf{0.33} & \textbf{0.34} & \textbf{0.75} & 0.73 & 0.77 & 1.30 & 0.48 & 0.85 & 0.94 \\
		17 & Chap 2 \cref{eq:Chapman12} & \textbf{-0.55} & 1.00 & \textbf{0.29} & 0.43 & 0.68 & \textbf{0.71} & \textbf{0.77} & 1.09 & 0.57 & 0.81 & 0.94 \\
		18 & Chap 2 \cref{eq:Chapman12} & \textbf{0.63} & 1.00 & \textbf{0.50} & 0.45 & 0.66 & 0.75 & 0.76 & 1.22 & 0.54 & \textbf{0.73} & \textbf{0.94} \\
		19 & Hatch 1 \cref{eq:Hatch123} & \textbf{-4.98} & 2.00 & \textbf{50.32} & \textbf{0.44} & \textbf{0.67} & 1.06 & 0.66 & 1.27 & 0.23 & 3.30 & 0.75 \\
		20 & Hatch 1 \cref{eq:Hatch123} & \textbf{-4.91} & 2.00 & \textbf{28.34} & 0.82 & 0.39 & \textbf{0.88} & \textbf{0.72} & 0.70 & 0.59 & 1.70 & 0.87 \\
		21 & Hatch 1 \cref{eq:Hatch123} & \textbf{-7.17} & 2.00 & \textbf{12.64} & 2.19 & -0.63 & 2.07 & 0.33 & 1.18 & 0.24 & \textbf{0.85} & \textbf{0.94} \\
		22 & Hatch 1 \cref{eq:Hatch123} & \textbf{-16.02} & \textbf{2.92} & \textbf{9.64} & 0.77 & 0.42 & \textbf{0.86} & \textbf{0.72} & 0.76 & 0.61 & 1.71 & 0.87 \\
		23 & Hatch 2 \cref{eq:Hatch123} & \textbf{-1.04} & 4.00 & \textbf{12.41} & \textbf{0.44} & \textbf{0.67} & 1.31 & 0.58 & 2.59 & 0.07 & 3.64 & 0.73 \\
		24 & Hatch 2 \cref{eq:Hatch123} & \textbf{-1.31} & 4.00 & \textbf{5.05} & 0.48 & 0.64 & \textbf{1.08} & \textbf{0.65} & 1.92 & 0.32 & 2.28 & 0.83 \\
		25 & Hatch 2 \cref{eq:Hatch123} & \textbf{-1.22} & 4.00 & \textbf{0.54} &       &      &       &      & \textbf{0.26} & \textbf{0.91} & 0.60 & 0.95 \\
		26 & Hatch 2 \cref{eq:Hatch123} & \textbf{-1.73} & 4.00 & \textbf{1.75} & 0.92 & 0.31 & 1.81 & 0.42 & 2.88 & -0.03 & \textbf{1.43} & \textbf{0.89} \\
		27 & Hatch 5 \cref{eq:Hatch123} & \textbf{-7.12} & 4.00 & \textbf{0.07} & \textbf{0.44} & \textbf{0.67} & 1.07 & 0.66 & 1.27 & 0.22 & 3.31 & 0.75 \\
		28 & Hatch 5 \cref{eq:Hatch123} & \textbf{-6.67} & 4.00 & \textbf{0.05} & 0.82 & 0.38 & \textbf{0.89} & \textbf{0.71} & 0.71 & 0.59 & 1.70 & 0.87 \\
		29 & Hatch 5 \cref{eq:Hatch123} & \textbf{-8.81} & 4.00 & \textbf{0.01} & 2.21 & -0.65 & 2.09 & 0.33 & 1.10 & 0.27 & \textbf{0.86} & \textbf{0.94} \\
		30 & Hatch 5 \cref{eq:Hatch123} & \textbf{-8.86} & \textbf{4.77} & \textbf{0.01} & \textbf{0.44} & \textbf{0.67} & 1.06 & 0.66 & 1.28 & 0.22 & 3.30 & 0.75 \\
		[-1.2ex]\hline
		31 & \cref{sec:etae} \cref{eq:etaDefinition} & \multicolumn{3}{c}{$\eta_e=$ \textbf{1.73}} & \textbf{0.88} & \textbf{0.34} & 2.82 & 0.09 & 5.28 & -0.94 & 6.76 & 0.49 \\
		32 & \cref{sec:etae} \cref{eq:etaDefinition} & \multicolumn{3}{c}{$\eta_e=$ \textbf{1.90}} & 0.91 & 0.32 & \textbf{2.67} & \textbf{0.14} & 5.14 & -0.85 & 5.90 & 0.56 \\
		33 & \cref{sec:etae} \cref{eq:etaDefinition} & \multicolumn{3}{c}{$\eta_e=$ \textbf{3.10}} &       &      &      &      & \textbf{0.53} & \textbf{0.81} & 1.29 & 0.90 \\
		34 & \cref{sec:etae} \cref{eq:etaDefinition} & \multicolumn{3}{c}{$\eta_e=$ \textbf{2.19}} & 1.33 & 0.01 & 3.47 & -0.12 & 6.47 & -1.32 & \textbf{5.27} & \textbf{0.60} \\
		35 & \cref{sec:alpha} \cref{eq:polyPred} & \multicolumn{3}{c}{$\alpha=$ \textbf{0.32}, $A=$ \textbf{81.21}} & \textbf{0.45} & \textbf{0.67} & 1.80 & 0.42 & 1.19 & -0.04 & 69.61 & -4.25 \\
		36 & \cref{sec:alpha} \cref{eq:polyPred} & \multicolumn{3}{c}{$\alpha=$ \textbf{0.39}, $A=$ \textbf{50.67}} & 0.53 & 0.60 & \textbf{1.37} & \textbf{0.56} & 1.13 & 0.25 & 7.59 & 0.43 \\
		37 & \cref{sec:alpha} \cref{eq:polyPred} & \multicolumn{3}{c}{$\alpha=$ \textbf{0.42}, $A=$ \textbf{29.33}} &       &      &      &      & \textbf{0.40} & \textbf{0.81} & 1.08 & 0.92 \\
		38 & \cref{sec:alpha} \cref{eq:polyPred} & \multicolumn{3}{c}{$\alpha=$ \textbf{0.46}, $A=$ \textbf{31.77}} & 0.69 & 0.48 & 1.80 & 0.42 & 2.66 & -0.19 & \textbf{1.81} & \textbf{0.86} \\
		[-1.2ex]\hline
		39 & \multicolumn{4}{l}{Local-value \textbf{None} NN}          & 0.83 & 0.38 & 1.71 & 0.45 & 2.55 & 0.09 & 1.81 & 0.86 \\
		40 & \multicolumn{4}{l}{Local-value \textbf{No-EFIT} NN}       & 0.27 & 0.80 & 0.43 & 0.86 & 0.55 & 0.80 & 0.27 & 0.98 \\
		41 & \multicolumn{4}{l}{Local-value \textbf{No-}$\mathbf R$ NN}    & 0.18 & 0.86 & 0.29 & 0.91 & 0.38 & 0.86 & 0.22 & 0.98 \\
		42 & \multicolumn{4}{l}{Local-value \textbf{All} NN}         & 0.17 & 0.87 & 0.25 & 0.92 & 0.34 & 0.88 & 0.15 & 0.99 \\
		43 & \multicolumn{4}{l}{Full-pedestal \textbf{None} NN} & 0.48 & 0.64 & 0.79 & 0.74 & 1.19 & 0.58 & 0.74 & 0.94 \\
		44 & \multicolumn{4}{l}{Full-pedestal \textbf{No-EFIT} NN}  & 0.22 & 0.83 & 0.32 & 0.90 & 0.45 & 0.84 & 0.21 & 0.98 \\
		45 & \multicolumn{4}{l}{Full-pedestal \textbf{EFIT} NN} & 0.16 & 0.88 & 0.22 & 0.93 & 0.30 & 0.89 & 0.20 & 0.99 \\
		46 & \multicolumn{4}{l}{Full-pedestal \textbf{All} NN}& 0.12 & 0.91 & 0.17 & 0.94 & 0.26 & 0.91 & 0.11 & 0.99 \\
		
	\end{tabular}}
	
	\caption{\centering {Parameters for the models defined in \cref{sappendix:equations} and metrics assessing the quality of $T_e$ reconstructions as detailed in \cref{sappendix:bigComparison}. Bold values in the \quotemarks{Parameters} columns denote optimised parameters, while, in the \quotemarks{Fit Metrics} columns, they indicate the corresponding optimisation ranges. Reconstructions using neural-network (NN) models discussed in \cref{sec:neuralnets} are included for comparison. \quotemarks{Chap} is an abbreviation referring to models from \citet{Chapman-Oplopoiou_2022}, and \quotemarks{Hatch} to models from \citet{Hatch_RedModels}.  }}
	\label{table:FitCoeff}
\end{table}

\Cref{table:FitCoeff} presents several sets of parameters and metrics of the quality of the pedestal reconstruction corresponding to each set. The reconstructions are carried out using \cref{eq:etaDefinition}, \cref{eq:polyPred}, and \cref{eq:Chapman12,eq:Hatch123,eq:Hatch5}. \Cref{table:FitCoeff} also includes the quality metrics of neural-network reconstructions carried out in \cref{ssec:mlFullProf,ssec:mlLocalVal}. 

Our metrics of reconstruction quality are all defined over a certain \quotemarks{Range}$\in\{$Steep, Ped, Top, Full$\}$. The first of them is $M_\mathrm{Range}$, the averaged sum of squared discrepancies between the experimental $T_{e,\fit} (R_i)$ and the reconstructed $T_{e,\mathrm{Rec}}(R_i)$, sampled at $100$ equally spaced radial locations $R_i$. Similarly to \cref{eq:optimisationResidual}, it is defined as
\begin{equation}
	M_\mathrm{Range} = \frac{1}{1251} \sum _\mathrm{Pulse} \sum_{i=1}^{100}\left[\frac{T_{e,\fit} (R_i)- T_{e,\mathrm{Rec}}(R_i)}{\max_j T_{e,\fit} (R_j) }\right]^2.
\end{equation}
$M_\mathrm{Range}$ is normalized for each pulse to the maximum value of the $\fit$ profile over this range. 
The second metric is given the coefficient of determination between $T_{e,\fit} (R_i)$ and $T_{e,\mathrm{Rec}}(R_i)$,  defined the usual way:
\begin{equation}
	r^2_\mathrm{Range} = 1 - \frac{\sum _\mathrm{Pulse} \sum_{i=1}^{100} [T_{e,\fit} (R_i)- T_{e,\mathrm{Rec}}(R_i)] ^2 }{\sum _\mathrm{Pulse} \sum_{i=1}^{100} [T_{e,\fit} (R_i)- \left\langle T_{e,\fit} \right\rangle] ^2},
\end{equation}
where $ \left\langle T_{e,\fit} \right\rangle = \sum _\mathrm{Pulse} \sum_{i=1}^{100} T_{e,\fit} (R_i) / 125100$ is the average value of $T_e$ across all data in the considered \quotemarks{Range}.

The bold values in the table indicate which parameters are optimised and which metrics correspond to the range over which the optimisation is carried out. Some optimisations do not converge when the exponent $\beta$ is used as a fit parameter, and they are therefore omitted. 

Entries 1-6 in \cref{table:FitCoeff} suggest that the exact scalings from \citet{Chapman-Oplopoiou_2022} result in generally passable reconstructions over the pedestal regions, while those from \citet{Hatch_RedModels} perform more poorly. Entries 7-12, which have the prefactor $A$ fit over the \quotemarks{Ped} region, result in vast improvements in all metrics for the Hatch models, but, notably, the change in the Chapman 1 model is minimal, whereas the Chapman 2 model requires an adjustment of $A$ of less than $3\%$. The Hatch 2 and 3 models optimised in this fashion reproduce the \quotemarks{Steep} region better than the Chapman models. It is worth noting that our renditions of the Hatch models in \cref{eq:Hatch123,eq:Hatch5} omit a square-root-mass-ratio factor $\sqrt{m_i / m_e}$ compared to the original Equations ($Q_1$) to ($Q_5$) of Table \nolinebreak 1 in \citet{Hatch_RedModels}. However, the adjustment of $A$ that we obtain in entries 9-12 of \cref{table:FitCoeff} cannot be explained by this omission. 

Further optimisations (entries 13-30) reveal that all of these models are superior to the constant-$\eta_e$ and the constant-$\alpha$ reconstructions of \cref{sec:etae,sec:alpha} [entries 31-34 \nolinebreak for \nolinebreak \cref{eq:etaDefinition} and 35-38 for \cref{eq:polyPred}]. The neural-network reconstructions (entries 39-46, presented in \cref{fig:mlFullProfPred,fig:mlBigCropPred}) perform far better than all other methods, with the exception of the local model that uses no input parameters besides $n_e$ [entry 41 and \Cref{fig:mlBigCropPred}(a-d)].

A somewhat problematic result of the $\fitcoeff$ optimisations is that the Chapman models yield $\fitcoeff<0$ over the \quotemarks{Steep} and \quotemarks{Ped} regions, clashing with the theoretical picture where $\fitcoeff$ is either the linear stability threshold or a larger non-linear one, and should certainly be positive. The values of $\fitcoeff$ that we obtain for the Hatch models are consistent with this (entries 19-30), and, consequently, all of our optimisations indicate that, in general, terms the form $f(\eta_e) + C$ are required for the correct modelling of pedestal heat transport, with $C$ a positive constant and $f$ a strictly increasing and positive function.

Finally, all optimisations over the \quotemarks{Full} range result in excellent $r^2$ coefficients, but the resulting pedestal parameters are very poorly recovered. This conclusion, in conjunction with the other results in \cref{table:FitCoeff}, seems to favour mixed models that have different fitting parameters in the qualitatively different regions of the pedestal.

\begin{figure}
	\centering
	\includegraphics[width=\textwidth]{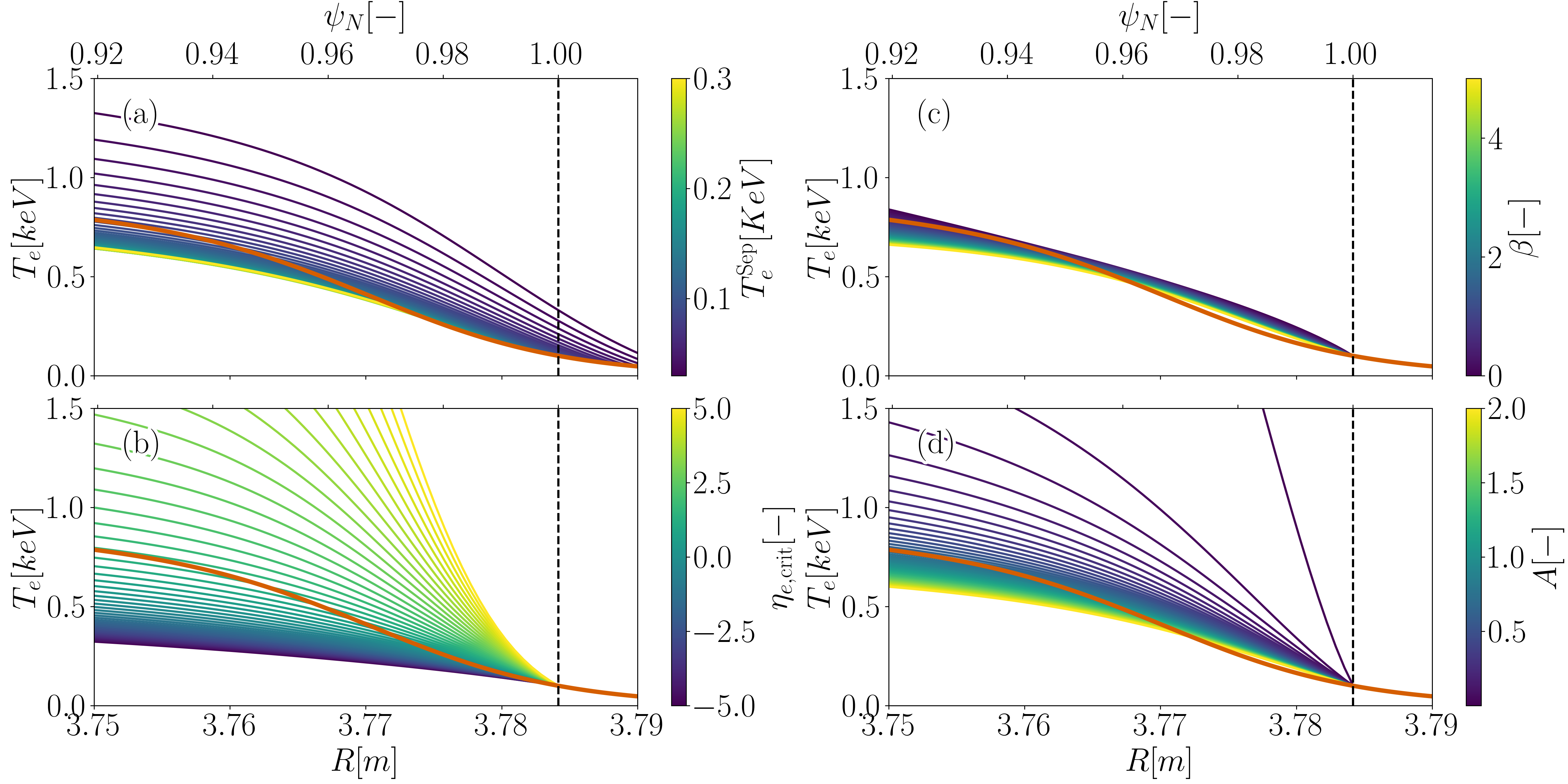}
	\caption{\centering  {The $\fit$ $T_e$ profiles for pulse $\examplepulse$ are shown in orange. Reconstructed profiles using the model \cref{eq:Chapman12} are shown in blue-yellow. These reconstructions are  integrated from an initial electron temperature $\Tesep$. Each panel shows the effect of varying one of the parameters of the model: (a) $\Tesep$, (b) $\etacr$, (c) $\beta$, (d) $A$; the other parameters are kept fixed at the values given in entry 1 in \cref{table:FitCoeff}. The vertical dashed lines represent the conventional separatrix location at $T_e=0.1\; \mathrm{KeV}$.}}
	\label{fig:chapmanParamScan}
\end{figure}

\subsection{Sensitivity of ETG-Heat-Flux Models to Parameters} \label{appendix:chapmanScan}

In \cref{sec:heatFluxModels}, we describe the results of adopting various heat-flux models motivated by gyrokinetic simulations of turbulence and, to some extent, physical considerations. In particular, we find \cref{eq:chapmanModel} [equivalently, \cref{eq:Chapman12}] to be a promising candidate for representing the turbulent transport in pedestals, giving good pedestal reconstructions. Optimising the parameters of this model yields two interesting results: the exponent $\beta$ becomes large if we attempt to match the region between $\psitop{T_e}$ and $\psisep$, and $\etacr$ takes negative values in this context. The latter is a physically important conclusion, however, the results of these optimisations are also a reflection of the numerical properties of the model \cref{eq:chapmanModel}.

\cref{fig:chapmanParamScan} shows parameter scans for the model \cref{eq:chapmanModel}. We take the \quotemarks{nominal} parameters to be $\etacr = 1.28$, $\beta =  1.43 $, and $A = 0.85$ (corresponding to entry 1 in \cref{table:FitCoeff}), and vary each of them individually over a range wide enough to encompass the results of all of our optimisations in \cref{appendix:heatFluxFits}. Alongside $\etacr$, $\beta$, and $A$, we also vary the implicit parameter $\Tesep$ (which is everywhere else set to $0.1 \; \mathrm {KeV}$) in order to show the effect that it has on the reconstructed $T_e$ profile. $\Rsep$ is consistently varied with $\Tesep$ so that the starting point of the integration lies on the $\fit$ profile of $\examplepulse$.

The predictions using the models \crefrange{eq:ch2_8}{eq:ch1_14} are very similar for the whole database. In order to understand the robustness of the prediction, it is therefore important to discuss the sensitivity of the prediction to the exact value of $\Tesep$ and the values of the heat-flux scaling parameters $\etacr$, $\beta$, and $A$. \cref{fig:chapmanParamScan} displays the results of scans in each of the free parameters. These models are very insensitive to changes of $\beta$, and $A$ requires order-of-magnitude variations to impact the resulting profile. Changing $A$ is equivalent to changing the local heat flux (or its gyro-Bohm normalisation), affecting the overall size of $\RLTe$. Changes in $\beta$ slightly adjust the position of the pedestal top, alongside affecting the profile's behaviour beyond it. High values of $\beta$ are equivalent to \quotemarks{pinning} $\RLTe$ to $ \etacr \times \RLNe$, whereas low values of $\beta$ reduce the importance of $\eta_e - \etacr$.
The profiles in this scan are insensitive to increases of $\Tesep$ as a result of the fact that, at nominal values of $\etacr$, $\beta$, and $A$, the prediction approximates the $\fit$ profile well. Large reductions of $\Tesep$ result in large displacements of $\Rsep$, causing high differences in the prediction.
The value of $\etacr$ changes the predictions most significantly. It introduces an effective minimal limit on $\RLTe$ in order to keep the term $\eta_e - \etacr$ positive, or, physically, if $\etacr > 0$, to maintain the turbulence in a supercritical regime. 

	\bibliographystyle{jpp}
	\bibliography{Bibliography}
\end{document}